\documentclass[11pt, a4paper, draft]{article}
\usepackage[a4paper,margin=25mm]{geometry}
\usepackage[T1]{fontenc}
\usepackage{lmodern,amsmath,amssymb,amsthm,mathtools,mathrsfs,booktabs,microtype,xcolor}
\usepackage{tikz,pgfplots}
\usetikzlibrary{arrows.meta,decorations.markings}
\usepgfplotslibrary{groupplots}
\pgfplotsset{compat=1.18}
\usepackage[colorlinks=true,linkcolor=blue!45!black,citecolor=blue!45!black,urlcolor=blue!45!black]{hyperref}
\numberwithin{equation}{section}
\newtheorem{theorem}{Theorem}[section]
\newtheorem{proposition}[theorem]{Proposition}
\newtheorem{lemma}[theorem]{Lemma}
\newtheorem{corollary}[theorem]{Corollary}
\theoremstyle{definition}
\newtheorem{assumption}{Assumption}[section]
\theoremstyle{remark}
\newcommand{\R}{\mathbb R}\newcommand{\C}{\mathbb C}
\newcommand{\dd}{\,\mathrm d}
\DeclareMathOperator{\Tr}{Tr}
\DeclareMathOperator{\Log}{Log}
\allowdisplaybreaks[2]
\title{Large $N$ limit of the ratio of Lukyanov integrals}
\author{Dong Wang\footnote{School of Mathematical Sciences, University of Chinese Academy of Sciences. \\ E-mail: \texttt{wangdong@wangd-math.xyz}.}} 

\begin{document}\maketitle
\begin{abstract}
We study the large-$N$ asymptotics of ratios of Lukyanov integrals
for fixed real coupling $b\geq1$. Our approach uses their
biorthogonal ensemble structure and vector Riemann--Hilbert
techniques. Under stated analytic hypotheses, we derive the
logarithmic ratio through its constant term, which is expressed
by a Fredholm determinant. The expansion is locally uniform in the
real parameter $\alpha$ and is formulated separately for even and
odd particle numbers. We verify the hypotheses for the correction
defined by the physical ground-state thermodynamic Bethe ansatz.
The relation to quantum-field-theoretic one-point functions is
discussed separately.
\end{abstract}

\section{Introduction}
\label{sec:introduction}

\subsection{Physical origin and the ratio problem}
The sinh-Gordon model is a massive integrable quantum field theory in
two dimensions, admitting a quantum inverse-scattering construction
and explicit form factors of local fields \cite{Sklyanin89,Fring-Mussardo-Simonetti93}.
Its finite-volume spectrum and thermodynamics are studied through
Baxter $Q$-functions, thermodynamic Bethe ansatz equations, and
quantum separation of variables \cite{Zamolodchikov06,Bytsko-Teschner06,Teschner08}. Lukyanov
\cite{Lukyanov01} proposed a representation of finite-volume vacuum expectation
values of exponential fields as renormalized limits of finite-dimensional
integrals. The quantity $\langle e^{a\varphi}\rangle_{R_{\rm circ}}$
is the normalized one-point function of the local exponential field
on an infinite Euclidean cylinder of circumference $R_{\rm circ}$.
Formally,
\[
 \bigl\langle e^{a\varphi}\bigr\rangle_{R_{\rm circ}}
 =\frac{\displaystyle\int\mathcal D\varphi\,
 e^{a\varphi(0)}e^{-\mathcal A_{\rm shG}[\varphi]}}
 {\displaystyle\int\mathcal D\varphi\,
 e^{-\mathcal A_{\rm shG}[\varphi]}},
\]
where $\mathcal A_{\rm shG}$ is the Euclidean sinh-Gordon action and
the exponential field is understood in a fixed renormalization
convention. It can be interpreted either as a thermal expectation
on the infinite line at temperature $R_{\rm circ}^{-1}$, or as a
ground-state expectation on a spatial circle of circumference
$R_{\rm circ}$; see \cite[Eqs.~(1)--(5)]{Lukyanov01}.

In Lukyanov's construction, separation
of variables gives a product of Baxter $Q$-functions, while the measure
contains two hyperbolic Vandermonde factors. Dividing the integral at
parameter $a$ by the one at $a=0$ removes parameter-independent normalization
factors. The remaining ratio still requires an ultraviolet power
renormalization.

To distinguish the cutoff from the number of integration variables,
write $I_M(R_{\rm circ},a)$ for Lukyanov's integral with $2M+1$
variables. With physical mass $m$ and $R=mR_{\rm circ}$, the
conjectural relation is \cite[Eq.~(23)]{Lukyanov01}
\begin{equation}
 \bigl\langle e^{a\varphi}\bigr\rangle_{R_{\rm circ}}
 =\kappa_a\lim_{M\to\infty}
 \left(\frac{4\pi M}{R}\right)^{-2a^2}
 \frac{I_M(R_{\rm circ},a)}{I_M(R_{\rm circ},0)}.
 \label{eq:intro-physical-cutoff}
\end{equation}
Here $\kappa_a$ is independent of the circumference and records the
normalization of the exponential field; see
\cite[Eqs.~(22)--(25)]{Lukyanov01}. It is not a parameter or a multiplicative
factor in either finite-dimensional integral or in their ratio.
With conformal field normalization, its value is fixed by the
infinite-volume expectation $\mathcal G_a$; the normalization implied
by our ratio asymptotics is stated in
Section~\ref{sec:field-normalization}. The original proposal was supported
by semiclassical and low-temperature calculations, rather than by a
large-$M$ asymptotic theorem.

Our convention is $N=2M+1$ and $\alpha=2a$, with separation variable
$\gamma=bQ\lambda$ and $Q=b+b^{-1}$.
Thus \eqref{eq:intro-physical-cutoff} predicts the term
$\frac12\alpha^2\log N$ in the logarithm of our ratio.
The physical cutoff uses odd $N$; the even subsequence is a
mathematical extension.

We call $\lambda$ the logarithmic, or rapidity, coordinate: the positive
coordinate used below is $x=e^{2B\lambda}$, with $B=1+b^{-2}$.
The standard physics term refers to the hyperbolic parametrization
$E=m\cosh t$, $p=m\sinh t$ of a massive particle's energy and momentum
in units with speed of light one. Lorentz boosts translate $t$.
Here the $\lambda_j$ are auxiliary separation variables; calling them
rapidity coordinates does not identify them with the rapidities of
$N$ physical particles. The rapidity plane means the complex
$\lambda$-plane.

The physical weight is determined by the ground-state thermodynamic
Bethe ansatz (TBA). Throughout, convolutions use Lebesgue measure.
With $\phi=\pi/(1+b^2)$, the normalized scattering kernel is
\[
 S_\phi(t)=\frac{\sin\phi\cosh t}
 {\pi(\sinh^2t+\sin^2\phi)},\qquad \int_\R S_\phi(t)\dd t=1.
\]
The pseudoenergy and the correction to the confining potential satisfy
\begin{equation}
 \epsilon=R\cosh-S_\phi*L,\qquad
 L=\log(1+e^{-\epsilon}),\qquad
 H(t)=\frac1\pi\int_\R\frac{L(v)}{\cosh(t-v)}\dd v.
 \label{eq:intro-TBA}
\end{equation}
The corresponding one-body factor is
$\exp[-r_*\cosh t+H(t)]$, where $r_*=R/\sin\phi$.
Analytic reconstruction from this TBA data is part of the established
$Q$-function theory; see \cite{Lukyanov01,Zamolodchikov06,Bytsko-Teschner06,Teschner08}.
The bounded-correction existence argument used below is due to
Fring--Korff--Schulz \cite{Fring-Korff-Schul99}. We specify both the measure and the
kernel because redistributing a factor of $2\pi$ between them is a
convention, whereas changing the integral operator changes the TBA.

\subsection{Previous mathematical results}
The large-$N$ analysis is complicated by the coexistence of a fixed
one-body weight and an increasing number of hyperbolically interacting
particles. The occupied interval in rapidity grows on the scale
$\log N$. Rescaling therefore produces an $N$-dependent equilibrium
problem, rather than the fixed regular potential occurring in the
usual one-cut setting. Moreover, the logarithm of the ratio is
much smaller than the individual logarithms. An error that is negligible
at the leading equilibrium scale can overwhelm the logarithmic term
or the finite part of the ratio.

Borot, Guionnet and Kozlowski \cite{Borot-Guionnet-Kozlowski16} developed a general approach
combining concentration of measure, equilibrium analysis, truncated
Wiener--Hopf operators, Riemann--Hilbert methods, and distributional
Schwinger--Dyson equations. Their expansion through $o(1)$ concerns a
rescaled sinh-model with scale $T_N=N^\delta$, $0<\delta<1/6$, and
a smooth, strictly convex, $N$-independent potential satisfying further
hypotheses. It supplies important methods for the present problem,
but does not directly give the physical Lukyanov ratio, whose rescaled
potential is $N$-dependent. Their discussion also distinguishes the
original Lukyanov prescription from the cutoff-dependent weights
suggested by lattice separation of variables \cite{Bytsko-Teschner06,Teschner08};
equality of the corresponding limits is an additional question.

Under their hypotheses, Dworaczek Guera and Kozlowski \cite{Dworaczek_Guera-Kozlowski25}
construct and analyze the
$N$-dependent equilibrium measure for the Lukyanov integral. Their
results include an integral representation for its density and
asymptotics for its support, obtained through the inversion of a
truncated Wiener--Hopf operator. These results support the predicted
power of $N$ in the ratio. The authors explicitly distinguish this
supporting calculation from a proof: their remainder estimates do
not control the ratio on the required logarithmic scale. In particular,
an explicit equilibrium measure alone does not evaluate the constant
term.

A complementary route uses the Muttalib--Borodin ensembles.
Muttalib~\cite{Muttalib95} introduced the additional pair interaction
in random-matrix models motivated by disordered conductors.
Borodin~\cite{Borodin99} developed the determinantal description and
computed microscopic limits for classical weights, involving Wright's
generalized Bessel functions. In the present problem, exponentiating
the rapidity variable changes the two hyperbolic Vandermonde factors
into $\prod_{i<j}(x_j-x_i)(x_j^\theta-x_i^\theta)$, with $\theta=b^2$.
The associated biorthogonal structure is therefore established,
although our weight and its equilibrium support depend on $N$.

The vector Riemann--Hilbert method avoids relying on a recurrence
of fixed length. Claeys and Wang~\cite{Claeys-Wang11} developed
two-component RH characterizations and steepest descent for an
equispaced external source, where the second polynomial family is
evaluated at an exponential variable. Claeys and Romano~\cite{Claeys-Romano14}
gave the corresponding $1\times2$ vector problems for the power
interaction and constructed one-cut equilibrium measures with and
without a hard edge. Their two components have different analytic
domains, with an identification of the sector edges for one component.
Their conformal map is the starting point for the map $J_N$ below.
The vector characterization holds for real $\theta\geq1$;
rationality is required for their finite recurrence and
Christoffel--Darboux formulas, not for the vector RH problem.

In subsequent steepest-descent analyses, the two components are
represented on a common auxiliary plane using the conformal map
and its inverse branches.
This produces scalar boundary relations that can involve values at
different points, often called shifted scalar RH problems.
Wang and Zhang~\cite{Wang-Zhang21} analyze two vector RH problems for
Laguerre-type weights $x^\beta e^{-nV(x)}$, $\beta>-1$, and prove
hard-edge asymptotics for positive integer $\theta$. Their
construction uses $(\theta+1)\times(\theta+1)$ Meijer-$G$ local
parametrices. Wang~\cite{Wang23a} extends the hard-edge analysis
to all real $\theta>0$ with a different local construction.
These results assume the corresponding regularity conditions on $V$.
There is also a square-matrix route through multiple orthogonality:
Kuijlaars and Molag~\cite{Kuijlaars-Molag19} treat $\theta=1/2$, and
Molag~\cite{Molag20} treats reciprocals of positive integers.
These restrictions on particular representations or local
constructions should be distinguished from the general vector
formulation.

For partition functions, Charlier~\cite{Charlier21} uses vector RH
steepest descent and its shifted scalar formulation to obtain
Muttalib--Borodin determinant asymptotics with Fisher--Hartwig
singularities for every real $\theta>0$. His integration interval
is fixed, $[a,b]$ with $0<a<b$, and both endpoints are hard edges.
The expansion includes an $n$-independent constant whose existence
is proved, but whose general value is not explicitly evaluated.
For correlation kernels, Claeys and Wang~\cite{Claeys-Wang22} derive bulk
and edge universality for the equispaced-external-source ensemble
directly from biorthogonal-polynomial asymptotics, without a
Christoffel--Darboux or double-contour formula. These works provide
the RH framework and precedents for determinant and kernel
asymptotics. Their conclusions do not directly supply the uniform
comparison needed here: the equilibrium scales move with $N$, and
the central lens jumps retain a nonzero limit. The analysis below
uses the established vector and conformal-map constructions while
treating those two features separately.

There are also independent physical descriptions of finite-temperature
one-point functions. Negro and Smirnov \cite{Negro-Smirnov13} propose determinant
formulas in a fermionic basis, involving a deformed TBA kernel and its
dressed resolvent. Their formulas give a natural comparison for a
finite part extracted from the Lukyanov integral; establishing that
comparison requires more than matching the coefficient of $\log N$.

\subsection{Method, scope, and organization}
Our object is
$F_N(\alpha)=\log[Z_{N,b}(\alpha;H)/Z_{N,b}(0;H)]$,
with the precise ensemble and analytic hypotheses stated in
Section~\ref{sec:conventions}. The expansion has the form
\[
 F_N(\alpha)=\frac{\alpha^2}{2}\log\frac{NC_b}{r_*}
 +\log\frac{\tau_\varepsilon(\alpha;H)}
 {\tau_\varepsilon(0;H)}+o(1),\qquad \varepsilon=(-1)^N,
\]
where
\[
 C_b=\frac{\Gamma\!\left(\frac1{2(1+b^2)}\right)
 \Gamma\!\left(\frac1{2(1+b^{-2})}\right)}{\sqrt\pi},
\]
and the trace-class determinant $\tau_\varepsilon$ is defined
in \eqref{eq:newtau}; its dependence on $b$ and $H$ is suppressed
in the notation there. This separates the equilibrium and outer
contribution from a central Fredholm determinant contribution that depends on $H$.

The main device is an exact identity for the mean resolvent on the
negative axis. It expresses the resolvent of the original ensemble
through the logarithmic derivative of one degree-$N$ polynomial for
the modified weight $w(x)/(x+c)$, with $c>0$. Integration over $c$
then gives the $\alpha$-derivative. Thus no finite recurrence in
$\theta$ is needed. In the steepest descent, the two endpoint models
are of Airy type, but the jumps in the central rapidity region retain
an order-one limit. They are included in the model problem and lead
to the Fredholm determinant above. The required error estimate is uniform in the insertion parameter
$c>0$ and integrable over its full range. The full scalar dressing identity applies to every lip, gap, and
Airy matching channel. We estimate the unmodified full operator in
scaled arclength spaces, with square-root jump weights on the omitted
channels. The resulting integrable comparison has error
$O(N^{-1}\log N)$; pointwise bulk asymptotics alone would not suffice.

Appendix~\ref{sec:tba-strip-proof} verifies Assumption~\ref{ass:H}
for the physical correction. Its bounded TBA existence proof is a
specialization of \cite[Section~5]{Fring-Korff-Schul99}, followed by regularity
estimates. The additional arguments establish automatic boundedness
within a specified real finite-convolution class and identify the
particular scalar amplitude required by the central RH problem.
The latter identification uses standard Fourier, Cauchy, and strip
Poisson formulas; no novelty is claimed for these general methods.

The asymptotic statement concerns the finite-dimensional ratio and
its Fredholm finite part. It does not determine the absolute
normalization of $Z_{N,b}(0;H)$, nor does it identify the Fredholm
expression with the independently normalized one-point function
in \eqref{eq:intro-physical-cutoff} or with the formulas of
\cite{Negro-Smirnov13}. Such an identification must retain the odd cutoff
and the field-normalization factor
$\kappa_a$. These distinctions are necessary when assessing the
physical conjecture. All asymptotics here have fixed $b$ and
$r_*>0$; no uniform ultraviolet or infinite-coupling limit is asserted.

Section~\ref{sec:biorthogonal-ensemble} derives the
Muttalib--Borodin representation and defines the monic biorthogonal
polynomials. Section~\ref{sec:newsetup} formulates the vector RH
problem and its central operator. Sections~\ref{sec:resolvent-identity}
and~\ref{sec:source-dressing} derive the resolvent identity and the effect of the weight modification. Sections~\ref{sec:newcorelimit}
and~\ref{sec:inverse-source-Hardy} treat the central limit and uniform
inversion; Section~\ref{sec:global-source-comparison} proves the full scalar dressing identity and the global
comparison, and Section~\ref{sec:ratio-completion} integrates the
$\alpha$-derivative. Appendix~\ref{sec:exact-general-equilibrium}
records the principal equilibrium construction, and
Appendix~\ref{sec:tba-strip-proof} treats the physical TBA correction.

\section{Conventions, analytic class, and ratio asymptotics}
\label{sec:conventions}
Fix $b\geq1$ and $r_*>0$, and write
\[
 \theta=b^2,\quad A=1+\theta,\quad B=1+\theta^{-1},\quad
 Q=b+b^{-1},\quad\phi=\pi/A,\quad\eta=1/(2B).
\]
Our normalization is
\begin{align}
 Z_{N,b}(\alpha;H)
 &=\int_{\R^N}\prod_{i<j}
 \sinh(A(\lambda_i-\lambda_j))\sinh(B(\lambda_i-\lambda_j))\notag\\
 &\hspace{15mm}\times\prod_{j=1}^N
 e^{-r_*\cosh\lambda_j+H(\lambda_j)+\alpha Q\lambda_j}\dd^N\lambda,
 \qquad F_N(\alpha)=\log\frac{Z_{N,b}(\alpha;H)}{Z_{N,b}(0;H)}.
 \label{eq:newZ}
\end{align}
\paragraph{Relation to Lukyanov's integral.}
Let $N=2M+1$, $\alpha=2a$, $R=mR_{\rm circ}$, and
$r_*=R/\sin\phi$. With the rapidity convention specified in
Appendix~\ref{sec:tba-strip-proof}, set $\gamma_j=bQ\lambda_j$.
The physical Baxter function in the normalization of
\cite[Eq.~(21)]{Lukyanov01} satisfies
\[
 \mathcal Q_R(bQ\lambda)^2
 =\exp[-r_*\cosh\lambda+H_R(\lambda)].
\]
We write $\mathcal Q_R$ as a function of the separation coordinate
$\gamma$, to distinguish it from the number $Q=b+b^{-1}$. In the symmetrized definition
\cite[Eq.~(13)]{Lukyanov01}, each differential is $d\gamma_j/b$ and each
pair contributes $2\sinh(\gamma_i-\gamma_j)
\sinh((\gamma_i-\gamma_j)/b^2)$. Therefore
\begin{equation}\label{eq:physical-integral-normalization}
 \boxed{\displaystyle
 I_M(R_{\rm circ},a)=
 \frac{2^{N(N-1)/2}Q^N}{N!}\,
 Z_{N,b}(2a;H_R),\qquad N=2M+1.}
\end{equation}
Indeed $d\gamma_j/b=Q\,d\lambda_j$,
$bQ=A$, $Q/b=B$, and $2a\gamma_j/b=2aQ\lambda_j$.
The prefactor in \eqref{eq:physical-integral-normalization} is
independent of $a$, so in particular
\begin{equation}\label{eq:physical-ratio-identification}
 \frac{I_M(R_{\rm circ},a)}{I_M(R_{\rm circ},0)}
 =\frac{Z_{2M+1,b}(2a;H_R)}{Z_{2M+1,b}(0;H_R)}.
\end{equation}
An overall $\lambda$-independent normalization of the Baxter function
would change the first identity by its $2N$th power but would leave
the ratio identity unchanged.

\begin{assumption}[Conditions on $H$]\label{ass:H}
For the fixed parameters $b\geq1$ and $r_*>0$, the function $H$
satisfies the following conditions.
\begin{enumerate}
\renewcommand{\labelenumi}{(\roman{enumi})}
\item \emph{Regularity and decay.}
$H$ is real and even on $\R$, holomorphic on a strip wider than
$|\Im\lambda|\leq\phi/2$, and exponentially decreasing there through
each fixed derivative.
\item \emph{Analytic central amplitude and strip bounds.}
Set
\[
 \mathcal R_\pm=e^{\pm i\phi}(0,\infty),\qquad
 \Omega_\phi=\C\setminus
 (\overline{\mathcal R_+}\cup\overline{\mathcal R_-}).
\]
For $\zeta\in\Omega_\phi$, define
\begin{equation}
 \mathcal C_H(\zeta)=\frac1{2\pi i}\int_\R H(v)
 \left[\frac{2e^{2v-i\phi}}{e^{2v-i\phi}-\zeta}
       -\frac{2e^{2v+i\phi}}{e^{2v+i\phi}-\zeta}\right]\dd v
 \label{eq:newCH}
\end{equation}
using the direct integral values. The rays $\mathcal R_\pm$ are traced
by $e^{2v\pm i\phi}$ as $v$ runs over $\R$, and are the loci where
the corresponding denominators can vanish. For real $t$, put
\begin{equation}
 \mathcal A_H(t)=\exp\left\{
 i r_*\sin\phi\sinh(t+i\phi/2)
 +\mathcal C_H(e^{2t})-\mathcal C_H(e^{2t+2i\phi})
 -H(t+i\phi/2)\right\}.
 \label{eq:newamp}
\end{equation}
Both Cauchy values in this formula are direct values of
\eqref{eq:newCH}; the second is not obtained by continuing the first
across its ray jump. The arguments $e^{2t}$ and $e^{2t+2i\phi}$ have
angles $0$ and $2\phi$, so they belong to $\Omega_\phi$ for
$0<\phi\leq\pi/2$. At $b=1$ the excluded rays are the two imaginary
half-axes and the evaluation points lie on the two real half-axes.
There exists a function $g$ such that
$\mathcal A_H(t)=g(t+i\phi/2)$, where $g$ is bounded and holomorphic
on $0<\Im z<\pi/2$, has boundary values satisfying
\begin{equation}
 |g(t)|\leq1,\qquad |g(t+i\pi/2)|\leq\rho<1,
 \label{eq:newschur}
\end{equation}
and has double-exponential decay, uniformly on each closed interior
substrip used in the contour deformation.
\end{enumerate}
\end{assumption}

Assumption~\ref{ass:H} specifies the weight and its central analytic
class; it does not assume a finite-$N$ RH error estimate or a
partition-function limit. For $H=0$ it holds with
$g(z)=\exp(i r_*\sin\phi\sinh z)$ and $\rho=e^{-r_*\sin\phi}$.
Appendix~\ref{sec:tba-strip-proof} verifies Assumption~\ref{ass:H} for the
physical TBA correction, with $\rho=e^{-r_*\sin\phi}$, including the
full complex identity for $\mathcal A_H$. For a general $H$, part~(ii)
remains an explicit hypothesis; the regularity and decay in part~(i)
alone do not imply it.

For $\alpha\in\R$, let $v=|\mathcal A_H|$ and
$u_\alpha=e^{-i\pi\alpha/Q}\mathcal A_H/v$.
For the Fourier convention $\widehat f(k)=\int e^{-ikt}f(t)\dd t$,
let $Q_\psi$ be convolution with multiplier
$e^{\psi k}/(1+e^{\pi k})$. Define the trace-class operator
\begin{equation}
 \mathcal T_\alpha=M_{\sqrt v}Q_\phi M_{\sqrt v}M_{u_\alpha}
 +M_{\sqrt v}Q_{\pi-\phi}M_{\sqrt v}M_{\overline{u_\alpha}},
 \qquad \tau_\varepsilon(\alpha)=\det(I+\varepsilon\mathcal T_\alpha).
 \label{eq:newtau}
\end{equation}
All these operators act on $\mathscr H_0=L^2(\mathbb R,dt)$, and
$M_f$ denotes multiplication by $f$. The determinant is the ordinary
Fredholm determinant of $I+\varepsilon\mathcal T_\alpha$ on this
space. For real $\alpha$ it is strictly positive by the reality,
contraction, and continuity argument in
Section~\ref{sec:inverse-source-Hardy}; its logarithm is the real
logarithm. Only real $\alpha$ is used in the theorem and its proof.
Indeed, if $A_+$ and $A_-$ denote the two trace-class summands at
$\alpha=0$, then
\[
 \mathcal T_\alpha=e^{-i\pi\alpha/Q}A_+
                     +e^{i\pi\alpha/Q}A_-.
\]
This gives differentiability in trace norm with respect to real
$\alpha$, which suffices for the determinant differential identity
used below. No complex-$\alpha$ continuation is required.

\paragraph{Scalar normalization.}
Let $\delta_N$ be the unique positive solution of the equilibrium
mass equation \eqref{eq:gb-mass-exact}, with $r=r_*$, and define
\begin{equation}\label{eq:chi-definition}
 \chi_N=\frac14\log\frac{1+\delta_N}{\delta_N}.
\end{equation}
The equilibrium calculation in
Appendix~\ref{sec:exact-general-equilibrium} gives
\begin{equation}
 \chi_N=\log\frac{NC_b}{r_*}+O(N^{-2}),\qquad
 C_b=\frac{\Gamma(1/(2A))\Gamma(1/(2B))}{\sqrt\pi}.
 \label{eq:newchi}
\end{equation}

\begin{theorem}[Ratio asymptotics]
\label{thm:newmain}
Fix real $b\geq1$ and $r_*>0$, and suppose that $H$ satisfies
Assumption~\ref{ass:H}. For either fixed parity
$\varepsilon\in\{-1,1\}$, as $N\to\infty$ along $(-1)^N=\varepsilon$,
\begin{equation}
 \boxed{F_N(\alpha)=\frac{\alpha^2}{2}\chi_N
 +\log\frac{\tau_\varepsilon(\alpha)}{\tau_\varepsilon(0)}+o(1),}
 \label{eq:newconclusion}
\end{equation}
locally uniformly for $\alpha\in\R$.
\end{theorem}

\begin{corollary}[Physical TBA specialization]
\label{cor:physical-TBA-specialization}
Fix real $b\geq1$ and $r_*>0$. Let $H$ be the correction constructed
from the real physical TBA in Appendix~\ref{sec:tba-strip-proof}, with
driving coefficient $R=r_*\sin\phi$. Then $H$ satisfies
Assumption~\ref{ass:H}, and
\eqref{eq:newconclusion} holds.
\end{corollary}
\begin{proof}
The existence, regularity, amplitude representation, and boundary
bounds are established in Appendix~\ref{sec:tba-strip-proof}.
Apply Theorem~\ref{thm:newmain}.
\end{proof}

\section{Muttalib--Borodin ensemble and biorthogonal polynomials}
\label{sec:biorthogonal-ensemble}
\subsection{Change of variables and the ensemble}
In \eqref{eq:newZ}, set $x_j=e^{2B\lambda_j}$, so that
$\lambda_j=(\log x_j)/(2B)$ and $x_j>0$.
Since $A=\theta B$ and $Q=bB$, we have
\[
 d\lambda_j=\frac{dx_j}{2Bx_j},\qquad
 \cosh\lambda_j=\frac{x_j^\eta+x_j^{-\eta}}2,\qquad
 e^{\alpha Q\lambda_j}=x_j^{\alpha b/2}.
\]
The two-body interaction transforms as
\[
 \sinh\bigl(A(\lambda_i-\lambda_j)\bigr)
 \sinh\bigl(B(\lambda_i-\lambda_j)\bigr)
 =\frac{(x_i^\theta-x_j^\theta)(x_i-x_j)}
 {4(x_ix_j)^{A/2}}.
\]
For a vector $x=(x_1,\ldots,x_m)$, write
$\Delta_m(x)=\prod_{1\le i<j\le m}(x_j-x_i)$ and
$x^\theta=(x_1^\theta,\ldots,x_m^\theta)$.
Each variable appears in $N-1$ pair denominators, and the signs
from reversing both Vandermonde factors cancel. Including the
Jacobian therefore gives the exact identity
\begin{equation}\label{eq:MB-partition-transformation}
 Z_{N,b}(\alpha;H)
 =\frac{1}{2^{N(N-1)}(2B)^N}
 \int_{(0,\infty)^N}\Delta_N(x)\Delta_N(x^\theta)
 \prod_{j=1}^N w_{N,\alpha}(x_j)\,dx_j,
\end{equation}
where
\begin{equation}
 w_{N,\alpha}(x)=x^{-1-A(N-1)/2+\alpha b/2}
 e^{-\frac{r_*}{2}(x^\eta+x^{-\eta})+H(\log x/(2B))}.
 \label{eq:newweight}
\end{equation}
After normalization, the integrand in
\eqref{eq:MB-partition-transformation} defines a
Muttalib--Borodin ensemble with parameter $\theta=b^2$ and an
$N$-dependent weight; see \cite{Muttalib95,Borodin99,Claeys-Romano14}.
The prefactor depends only on $N$ and $b$, so it cancels from the
ratio at parameters $\alpha$ and $0$.
Complex continuations of the weight use the logarithm branch
specified in the RH problem. The notation $r_*$ distinguishes the
fixed potential parameter from a distance to the negative insertion
point used later.

\subsection{Bimoments and monic biorthogonal polynomials}
Fix $N$ and the parameters of the weight. Let $W$ denote either
$w_{N,\alpha}$ or $w_{N,\alpha}/(x+c)$, with $c>0$, real $\alpha$, and $H$
satisfying Assumption~\ref{ass:H}. Both weights are positive on
$(0,\infty)$, and all their moments are finite because of their
decay at zero and infinity. We use the standard biorthogonal
construction of Claeys--Romano~\cite{Claeys-Romano14}, with monic normalization.
Define the bimoments and their leading determinants by
\begin{equation}\label{eq:bimoment-definition}
 \mu_{jk}[W]=\int_0^\infty x^{j+\theta k}W(x)\dd x,\qquad
 \mathcal D_m[W]=\det[\mu_{jk}[W]]_{j,k=0}^{m-1},\qquad
 \mathcal D_0[W]=1.
\end{equation}
Here $j,k\ge0$, and the weight, including its parameter $N$, is held
fixed when the determinant size or polynomial degree varies.
With the Vandermonde notation introduced above, Andr\'eief's
identity gives, for $m\ge1$,
\begin{equation}\label{eq:bimoment-positivity}
 \mathcal D_m[W]=\frac1{m!}\int_{(0,\infty)^m}
 \Delta_m(x)\Delta_m(x^\theta)
 \prod_{\ell=1}^m W(x_\ell)\dd x_\ell>0,
\end{equation}
where $x^\theta=(x_1^\theta,\ldots,x_m^\theta)$.
Indeed $x\mapsto x^\theta$ is strictly increasing, so the product
of the two Vandermonde determinants is positive whenever the
coordinates are distinct. This also proves nonsingularity of every
leading bimoment matrix; see the determinantal and integral
construction in~\cite{Claeys-Romano14}.

For each $j\ge0$, let $p_j[W](z)$ and $q_j[W](z)$ be the unique
monic polynomials of degree $j$ characterized by
\begin{equation}\label{eq:monic-biorthogonal-definition}
 \int_0^\infty p_j[W](x)x^{\theta k}W(x)\dd x=0,\qquad
 \int_0^\infty x^kq_j[W](x^\theta)W(x)\dd x=0,
 \quad 0\le k<j.
\end{equation}
For $j=0$ both polynomials equal one. For $j\ge1$, the coefficients
below the leading term solve linear systems with coefficient
matrices $[\mu_{\ell k}]_{\ell,k=0}^{j-1}$ or its transpose;
\eqref{eq:bimoment-positivity} therefore gives existence and uniqueness.
These conditions imply
\begin{equation}\label{eq:monic-biorthogonal-normalization}
 \int_0^\infty p_j[W](x)q_k[W](x^\theta)W(x)\dd x
 =h_j[W]\delta_{jk},\qquad
 h_j[W]=\frac{\mathcal D_{j+1}[W]}{\mathcal D_j[W]}>0,
\end{equation}
where $\delta_{jk}$ is the Kronecker delta. To see the determinant
ratio, change from the monomial bases to the two monic polynomial
bases: the triangular change-of-basis matrices have determinant one,
and the resulting bimoment matrix is diagonal with entries
$h_0,\ldots,h_{m-1}$. Thus $\mathcal D_m=\prod_{j=0}^{m-1}h_j$.

The argument of $q_j$ is its own polynomial variable: $q_j(z)$ is
monic of degree $j$, while $q_j(x^\theta)$ need not be a polynomial
in $x$ when $\theta$ is noninteger. We call $h_j$ the positive
\emph{biorthogonality constant}; for $\theta=1$, $p_j=q_j$ and it is
the usual squared orthogonal-polynomial norm.
The biorthonormal convention of~\cite{Claeys-Romano14} is recovered by replacing
$p_j,q_j$ by $p_j/\sqrt{h_j},q_j/\sqrt{h_j}$.
We suppress $[W]$ when the weight is clear. In particular,
$p_{N,\alpha,c}$ below denotes $p_N[w_{N,\alpha}/(x+c)]$.

Combining \eqref{eq:MB-partition-transformation} with
\eqref{eq:bimoment-positivity} and
\eqref{eq:monic-biorthogonal-normalization}, we obtain
\begin{align}
 Z_{N,b}(\alpha;H)
 &=\frac{N!}{2^{N(N-1)}(2B)^N}\mathcal D_N[w_{N,\alpha}]
 \notag\\
 &=\frac{N!}{2^{N(N-1)}(2B)^N}
   \prod_{j=0}^{N-1}h_j[w_{N,\alpha}].
 \label{eq:MB-bimoment-normalization}
\end{align}
Every factor in this product uses the same fixed weight
$w_{N,\alpha}$. In particular,
$F_N(\alpha)=\log\bigl(\mathcal D_N[w_{N,\alpha}]/
\mathcal D_N[w_{N,0}]\bigr)$.

\section{The vector RH problem and the retained central operator}
\label{sec:newsetup}

\paragraph{The Claeys--Romano map and its inverse branches.}
We use the one-cut map without a hard edge of
Claeys--Romano~\cite[Section~4.3, Lemma~4.3]{Claeys-Romano14}. In their notation
it is
\[
 \widetilde J_{c_0,c_1}(s)
 =(c_1s+c_0)\left(\frac{s+1}{s}\right)^{1/\theta},
 \qquad c_0>c_1>0.
\]
Our principal map is this same map with $c_0=c_{0,N}$ and
$c_1=c_{1,N}$. These parameters are determined by the principal
equilibrium calculation in
Appendix~\ref{sec:exact-general-equilibrium}. Specifically,
$\delta_N>0$ is the unique positive solution of the mass equation
\eqref{eq:gb-mass-exact}, with $r=r_*$, and
\[
 c_{1,N}=\delta_N^{-B/2}(1+\delta_N)^{-(1-1/\theta)/2},
 \qquad c_{0,N}=c_{1,N}(1+\delta_N).
\]
See \eqref{eq:gb-c1exact} and
Proposition~\ref{prop:gb-exact-equilibrium} for the derivation and
uniqueness. With these parameters, we set
\[
 J_N(s)=c_{1,N}(s+1+\delta_N)
 \left(\frac{s+1}{s}\right)^{1/\theta},\qquad
 \mathfrak{s}_N=c_{1,N}^{-1/B}.
\]
The power is analytic off $[-1,0]$ and tends to one at infinity.
The endpoints of the equilibrium support are the critical values
of $J_N$. Explicitly, the two real solutions of $J_N'(s)=0$ are
\[
 s_{\pm,N}=
 \frac{-(\theta-1)\pm
 \sqrt{(\theta-1)^2+4\theta c_{0,N}/c_{1,N}}}{2\theta},
 \qquad s_{-,N}<-1<0<s_{+,N},
\]
and
\[
 x_{\pm,N}=J_N(s_{\pm,N})
 =(c_{1,N}s_{\pm,N}+c_{0,N})
 \left(\frac{s_{\pm,N}+1}{s_{\pm,N}}\right)^{1/\theta}.
\]
The powers in this endpoint formula take their positive real values.
The normalization of $c_{1,N}$ gives $x_{-,N}x_{+,N}=1$.
We call the support interval $[x_{-,N},x_{+,N}]$ the
\emph{equilibrium band}, or simply the \emph{band}.
Let $D_N$ be the bounded domain enclosed by the two conjugate arcs
that $J_N$ maps to this interval.
We denote the exterior and interior inverses from the cited lemma by
\[
\begin{aligned}
 I_{1,N}:\C\setminus[x_{-,N},x_{+,N}]
 &\longrightarrow\C\setminus\overline{D_N},\\
 I_{2,N}:\mathbb H_\theta\setminus[x_{-,N},x_{+,N}]
 &\longrightarrow D_N\setminus[-1,0],
 \qquad \mathbb H_\theta=\{z\ne0:|\arg z|<\pi/\theta\}.
\end{aligned}
\]
We write $I_1,I_2$ when $N$ is understood. Thus the conformal map
and branch geometry are those of \cite{Claeys-Romano14}, with the parameters
fixed by the equilibrium calculation above.
The small geometric scale $\mathfrak{s}_N$ is unrelated to the
field-normalization constant $\kappa_a$.
Appendix~\ref{sec:exact-general-equilibrium} gives
$\delta_N\asymp N^{-4}$ and $\mathfrak{s}_N\asymp N^{-2}$.

\subsection{The vector problem and its normalization}
Recall that the band $[x_{-,N},x_{+,N}]\subset(0,\infty)$ is the support of the
principal equilibrium measure for $V_N$, constructed in
Appendix~\ref{sec:exact-general-equilibrium}. Write
$a_N=x_{-,N}=e^{-2BL_N}$ and $d_N=x_{+,N}=e^{2BL_N}$, where $L_N>0$.
RH jump relations on the band are imposed on its interior
$(a_N,d_N)=(x_{-,N},x_{+,N})$; endpoint behavior is specified separately.

\paragraph{Auxiliary weight modifications.}
For $c>0$, multiplication of a weight $w(x)$ by $x+c$ inserts
$\prod_{j=1}^N(x_j+c)$ into the joint integrand. Division by $x+c$
inserts its reciprocal. These are characteristic-polynomial and
inverse-characteristic-polynomial insertions, respectively, evaluated
at $-c$ up to the conventional sign of the characteristic polynomial.
The informal names ``source'' and ``inverse source'' refer to these
two modifications; ``inverse'' refers to the reciprocal factor.
In this paper we use the explicit modified weight
\[
 w_c(x)=\frac{w_{N,\alpha}(x)}{x+c},\qquad c>0,
\]
and call $c$ the insertion parameter. The factor $(x+c)^{-1}$ has a
pole at the insertion point $-c$, outside the particle domain, so the modified weight remains
positive and nonsingular on $(0,\infty)$. In the auxiliary $s$-plane,
related to the physical coordinate by $z=J_N(s)$, the same point is
$a=I_{1,N}(-c)$. Here $J_N$ and its outer inverse branch $I_{1,N}$
were defined at the beginning of this section; this $s$-plane is
what we call the uniformizing plane. In the resolvent identities, spatial
derivatives hold $c$ fixed and are evaluated at the insertion point
only after differentiation. The
modification is auxiliary: Section~\ref{sec:resolvent-identity} uses
its polynomial to recover the mean resolvent of the original ensemble.

For the analytic domains, recall
\[
 \mathbb H_\theta=\{z\ne0:|\arg z|<\pi/\theta\}.
\]
Powers in $\mathbb H_\theta$ use $-\pi/\theta<\arg z<\pi/\theta$.
The relevant unknown is the row vector
\begin{equation}\label{eq:explicit-Y}
 Y_c(z)=\left(p_{N,\alpha,c}(z),\,
 \frac1{2\pi i}\int_0^\infty
 \frac{p_{N,\alpha,c}(x)w_c(x)}{x^\theta-z^\theta}\,dx\right).
\end{equation}
\paragraph{RH problem for $Y_c$.}
The row vector $Y_c=(Y_{c,1},Y_{c,2})$ solves the following
Riemann--Hilbert problem, in the vector formulation of
\cite[Theorem~1.3]{Claeys-Romano14}, with the condition at zero adapted to $w_c$.
\begin{enumerate}
\item $Y_{c,1}$ is entire and $Y_{c,2}$ is analytic in
$\mathbb H_\theta\setminus(0,\infty)$.
\item With the positive axis oriented from left to right,
\begin{equation}\label{eq:explicit-Yjump}
 Y_{c,+}(x)=Y_{c,-}(x)
 \begin{pmatrix}1&\dfrac{w_c(x)}{\theta x^{\theta-1}}\\0&1\end{pmatrix},
 \qquad x>0.
\end{equation}
\item At infinity,
\[
 Y_{c,1}(z)=z^N(1+O(z^{-1})),\qquad
 Y_{c,2}(z)=O(z^{-(N+1)\theta}).
\]
The second expansion is taken in closed subsectors away from the
positive jump ray and follows from the
orthogonality against $1,x^\theta,\ldots,x^{(N-1)\theta}$.
\item Both components are bounded at zero in their domains. Here all
negative moments exist because of the factor $e^{-r_*x^{-\eta}/2}$.
\item The second component satisfies the sewing condition
\begin{equation}\label{eq:explicit-Ysewing}
 Y_{c,2}(te^{i\pi/\theta})
 =Y_{c,2}(te^{-i\pi/\theta}),\qquad t>0.
\end{equation}
\end{enumerate}
The matrices below act on this row vector. Their size does not turn
the problem into a globally normalized $2\times2$ matrix RH problem:
the two components have different analytic domains, and the sector
sewing condition is part of the problem.

The exact decomposition of the scalar jump weight is
\begin{equation}\label{eq:explicit-weight-split}
 \frac{w_c(z)}{\theta z^{\theta-1}}
 =\omega_{\alpha,c}(z)e^{-NV_N(z)},
\end{equation}
where
\begin{align*}
 V_N(z)&=\frac A2\Log z+\frac{r_*}{2N}(z^\eta+z^{-\eta}),\\
 \omega_\alpha(z)&=\theta^{-1}z^{(1-\theta)/2+\alpha b/2}
 e^{H(\Log z/(2B))},\qquad
 \omega_{\alpha,c}(z)=\frac{\omega_\alpha(z)}{z+c}.
\end{align*}
These functions are continued from the band $(a_N,d_N)$ into the
upper and lower lens regions $\Omega_{N,+}$ and $\Omega_{N,-}$,
defined in Section~\ref{sec:lens-opening}, using the indicated
logarithm. Each region is bounded by the band and the corresponding
lip $\Gamma_{N,\pm}$; see \eqref{eq:explicit-lips} and
Figure~\ref{fig:S-contours}. For each fixed $N$, their closures avoid
$0$ and $-c$.
The unmodified ensemble is obtained by using $w_{N,\alpha}$ and
$\omega_\alpha$ throughout.

\subsection{The equilibrium functions and the transformation \texorpdfstring{$Y\mapsto T$}{Y to T}}
Let $\rho^{\mathrm{eq}}_N(x)$ be the equilibrium probability density
with respect to $dx$, constructed in
Appendix~\ref{sec:exact-general-equilibrium}. Recall the logarithmic
(rapidity) coordinate $\lambda$, related to $x$ by
$x=e^{2B\lambda}$, or $\lambda=(\log x)/(2B)$, where $B=1+b^{-2}$.
The \emph{rapidity density} means the density of this same measure
with respect to $d\lambda$: namely,
$\rho^{\mathrm{eq}}_{\lambda,N}(\lambda)\,d\lambda
=\rho^{\mathrm{eq}}_N(x)\,dx$. Thus
\[
 \rho^{\mathrm{eq}}_{\lambda,N}(\lambda)=2Be^{2B\lambda}\rho^{\mathrm{eq}}_N(e^{2B\lambda}),
 \qquad \int_{-L_N}^{L_N}\rho^{\mathrm{eq}}_{\lambda,N}(\lambda)\,d\lambda=1.
\]
We present and plot the equilibrium measure in the $\lambda$-coordinate
because its symmetry is then explicit. The principal equilibrium
energy is invariant under $\lambda\mapsto-\lambda$, as shown in
Appendix~\ref{sec:exact-general-equilibrium}; uniqueness of the
equilibrium measure therefore gives
$\rho^{\mathrm{eq}}_{\lambda,N}(-\lambda)
=\rho^{\mathrm{eq}}_{\lambda,N}(\lambda)$ on the symmetric support
$[-L_N,L_N]$. In the $x$-coordinate the same symmetry reads
$\rho^{\mathrm{eq}}_N(1/x)=x^2\rho^{\mathrm{eq}}_N(x)$,
expressing invariance of the measure under $x\mapsto x^{-1}$.

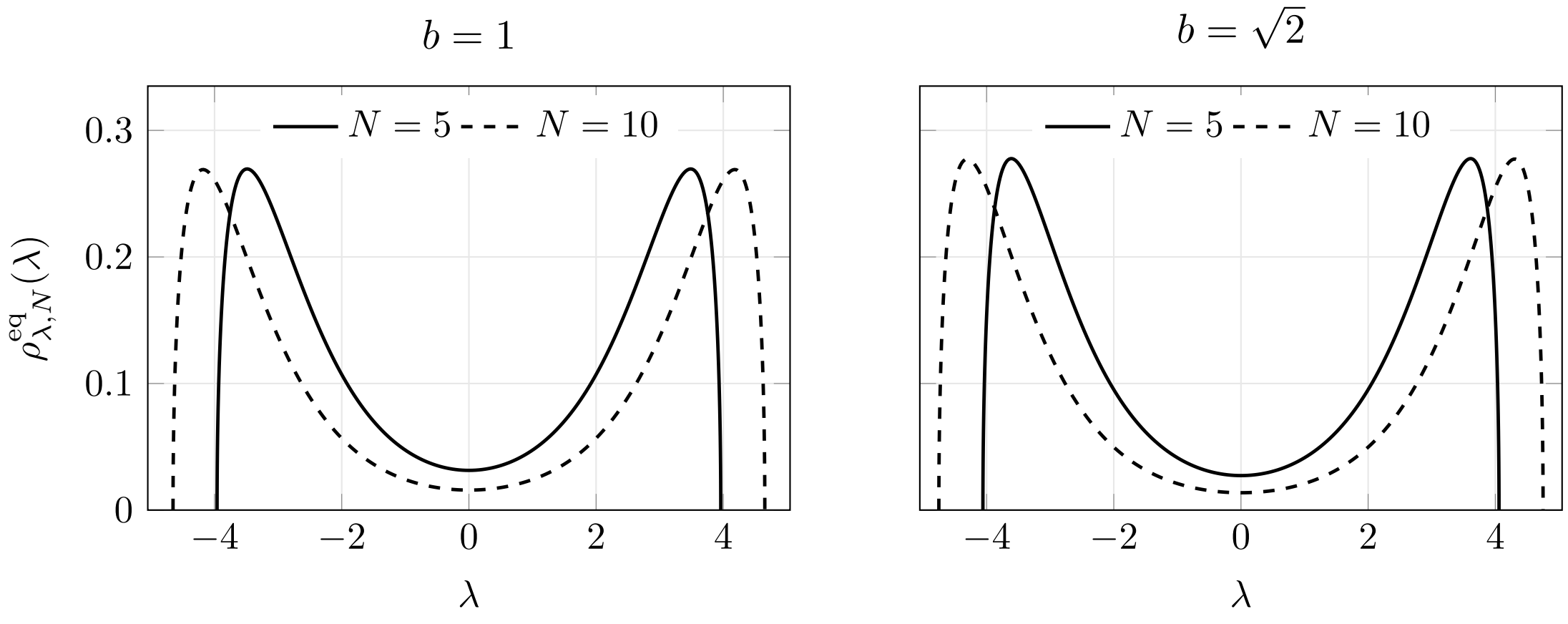
\begin{figure}[htbp]
\centering
\begin{tikzpicture}
\begin{groupplot}[group style={group size=2 by 1,horizontal sep=12mm},
width=.47\textwidth,height=55mm,xmin=-5.05,xmax=5.05,ymin=0,ymax=.335,
xlabel={$\lambda$},grid=major,grid style={gray!18},
tick label style={font=\small},label style={font=\small},
legend style={font=\small,draw=none,at={(.5,.98)},anchor=north,legend columns=2}]
\nextgroupplot[title={$b=1$},ylabel={$\rho^{\mathrm{eq}}_{\lambda,N}(\lambda)$}]
\addplot[color=black,solid,line width=.9pt] coordinates {
(-3.96048630396,0)
(-3.96014699949,0.0117413393359)
(-3.95912914422,0.0234583090109)
(-3.95743291255,0.0351266197877)
(-3.95505859512,0.0467221429367)
(-3.95200659877,0.058220989643)
(-3.94827744643,0.0695995894038)
(-3.94387177707,0.0808347670875)
(-3.93879034559,0.0919038183364)
(-3.93303402266,0.102784583003)
(-3.9266037946,0.113455516319)
(-3.91950076319,0.123895757517)
(-3.9117261455,0.134085195626)
(-3.90328127368,0.144004532194)
(-3.8941675947,0.153635340681)
(-3.88438667016,0.162960122328)
(-3.87394017595,0.17196235826)
(-3.86282990204,0.180626557672)
(-3.85105775211,0.188938301898)
(-3.83862574326,0.196884284235)
(-3.82553600566,0.204452345375)
(-3.81179078215,0.211631504334)
(-3.79739242791,0.218411984778)
(-3.78234341003,0.224785236671)
(-3.76664630708,0.230743953166)
(-3.75030380867,0.236282082703)
(-3.73331871501,0.241394836277)
(-3.71569393641,0.246078689845)
(-3.69743249278,0.250331381882)
(-3.67853751313,0.254151906093)
(-3.65901223502,0.257540499299)
(-3.63886000399,0.26049862454)
(-3.61808427304,0.263028949458)
(-3.59668860197,0.265135320027)
(-3.57467665682,0.266822729723)
(-3.55205220923,0.26809728424)
(-3.52881913578,0.268966161897)
(-3.50498141734,0.269437569876)
(-3.48054313838,0.269520696475)
(-3.45550848627,0.269225659588)
(-3.42988175057,0.26856345163)
(-3.40366732228,0.267545881172)
(-3.37686969313,0.266185511584)
(-3.34949345473,0.26449559699)
(-3.32154329788,0.262490015878)
(-3.29302401168,0.260183202761)
(-3.26394048277,0.25759007825)
(-3.23429769446,0.254725977997)
(-3.20410072589,0.251606580917)
(-3.17335475116,0.248247837151)
(-3.14206503842,0.244665896241)
(-3.11023694902,0.240877035973)
(-3.07787593653,0.236897592347)
(-3.04498754584,0.232743891149)
(-3.01157741221,0.228432181548)
(-2.97765126029,0.223978572145)
(-2.94321490316,0.219398969874)
(-2.90827424131,0.214709022113)
(-2.87283526164,0.209924062327)
(-2.83690403643,0.205059059539)
(-2.80048672232,0.200128571865)
(-2.76358955923,0.195146704301)
(-2.7262188693,0.190127070927)
(-2.6883810558,0.185082761609)
(-2.65008260204,0.180026313263)
(-2.61133007027,0.174969685675)
(-2.57213010054,0.16992424185)
(-2.53248940956,0.164900732797)
(-2.49241478955,0.159909286628)
(-2.4519131071,0.154959401832)
(-2.41099130197,0.150059944519)
(-2.36965638588,0.145219149431)
(-2.32791544137,0.140444624483)
(-2.28577562054,0.135743358593)
(-2.24324414382,0.131121732508)
(-2.20032829876,0.126585532393)
(-2.15703543878,0.122139965862)
(-2.11337298188,0.117789680202)
(-2.06934840939,0.113538782502)
(-2.02496926472,0.109390861402)
(-1.98024315198,0.105349010223)
(-1.93517773477,0.101415851193)
(-1.88978073482,0.0975935605557)
(-1.84405993067,0.0938838943032)
(-1.79802315634,0.090288214344)
(-1.75167830001,0.0868075148888)
(-1.70503330261,0.0834424488793)
(-1.65809615653,0.0801933542908)
(-1.61087490422,0.0770602801601)
(-1.56337763678,0.074043012202)
(-1.51561249263,0.0711410978998)
(-1.46758765609,0.0683538709646)
(-1.41931135595,0.0656804750777)
(-1.37079186412,0.0631198868412)
(-1.32203749416,0.0606709378783)
(-1.27305659988,0.0583323360337)
(-1.22385757391,0.0561026856391)
(-1.17444884624,0.0539805068169)
(-1.12483888281,0.0519642538071)
(-1.07503618403,0.0500523323074)
(-1.02504928333,0.0482431158288)
(-0.974886745715,0.0465349610707)
(-0.924557166273,0.0449262223292)
(-0.874069168721,0.043415264955)
(-0.823431403916,0.0420004778816)
(-0.772652548379,0.0406802852485)
(-0.721741302806,0.0394531571458)
(-0.670706390576,0.0383176195088)
(-0.619556556259,0.0372722631931)
(-0.568300564115,0.0363157522585)
(-0.516947196595,0.035446831495)
(-0.465505252834,0.0346643332177)
(-0.413983547145,0.0339671833628)
(-0.362390907506,0.0333544069097)
(-0.310736174051,0.0328251326572)
(-0.259028197552,0.0323785973773)
(-0.207275837905,0.0320141493685)
(-0.15548796261,0.0317312514285)
(-0.103673445253,0.0315294832626)
(-0.0518411639846,0.0314085433423)
(0,0.031368250225)
(0.0518411639846,0.0314085433423)
(0.103673445253,0.0315294832626)
(0.15548796261,0.0317312514285)
(0.207275837905,0.0320141493685)
(0.259028197552,0.0323785973773)
(0.310736174051,0.0328251326572)
(0.362390907506,0.0333544069097)
(0.413983547145,0.0339671833628)
(0.465505252834,0.0346643332177)
(0.516947196595,0.035446831495)
(0.568300564115,0.0363157522585)
(0.619556556259,0.0372722631931)
(0.670706390576,0.0383176195088)
(0.721741302806,0.0394531571458)
(0.772652548379,0.0406802852485)
(0.823431403916,0.0420004778816)
(0.874069168721,0.043415264955)
(0.924557166273,0.0449262223292)
(0.974886745715,0.0465349610707)
(1.02504928333,0.0482431158288)
(1.07503618403,0.0500523323074)
(1.12483888281,0.0519642538071)
(1.17444884624,0.0539805068169)
(1.22385757391,0.0561026856391)
(1.27305659988,0.0583323360337)
(1.32203749416,0.0606709378783)
(1.37079186412,0.0631198868412)
(1.41931135595,0.0656804750777)
(1.46758765609,0.0683538709646)
(1.51561249263,0.0711410978998)
(1.56337763678,0.074043012202)
(1.61087490422,0.0770602801601)
(1.65809615653,0.0801933542908)
(1.70503330261,0.0834424488793)
(1.75167830001,0.0868075148888)
(1.79802315634,0.090288214344)
(1.84405993067,0.0938838943032)
(1.88978073482,0.0975935605557)
(1.93517773477,0.101415851193)
(1.98024315198,0.105349010223)
(2.02496926472,0.109390861402)
(2.06934840939,0.113538782502)
(2.11337298188,0.117789680202)
(2.15703543878,0.122139965862)
(2.20032829876,0.126585532393)
(2.24324414382,0.131121732508)
(2.28577562054,0.135743358593)
(2.32791544137,0.140444624483)
(2.36965638588,0.145219149431)
(2.41099130197,0.150059944519)
(2.4519131071,0.154959401832)
(2.49241478955,0.159909286628)
(2.53248940956,0.164900732797)
(2.57213010054,0.16992424185)
(2.61133007027,0.174969685675)
(2.65008260204,0.180026313263)
(2.6883810558,0.185082761609)
(2.7262188693,0.190127070927)
(2.76358955923,0.195146704301)
(2.80048672232,0.200128571865)
(2.83690403643,0.205059059539)
(2.87283526164,0.209924062327)
(2.90827424131,0.214709022113)
(2.94321490316,0.219398969874)
(2.97765126029,0.223978572145)
(3.01157741221,0.228432181548)
(3.04498754584,0.232743891149)
(3.07787593653,0.236897592347)
(3.11023694902,0.240877035973)
(3.14206503842,0.244665896241)
(3.17335475116,0.248247837151)
(3.20410072589,0.251606580917)
(3.23429769446,0.254725977997)
(3.26394048277,0.25759007825)
(3.29302401168,0.260183202761)
(3.32154329788,0.262490015878)
(3.34949345473,0.26449559699)
(3.37686969313,0.266185511584)
(3.40366732228,0.267545881172)
(3.42988175057,0.26856345163)
(3.45550848627,0.269225659588)
(3.48054313838,0.269520696475)
(3.50498141734,0.269437569876)
(3.52881913578,0.268966161897)
(3.55205220923,0.26809728424)
(3.57467665682,0.266822729723)
(3.59668860197,0.265135320027)
(3.61808427304,0.263028949458)
(3.63886000399,0.26049862454)
(3.65901223502,0.257540499299)
(3.67853751313,0.254151906093)
(3.69743249278,0.250331381882)
(3.71569393641,0.246078689845)
(3.73331871501,0.241394836277)
(3.75030380867,0.236282082703)
(3.76664630708,0.230743953166)
(3.78234341003,0.224785236671)
(3.79739242791,0.218411984778)
(3.81179078215,0.211631504334)
(3.82553600566,0.204452345375)
(3.83862574326,0.196884284235)
(3.85105775211,0.188938301898)
(3.86282990204,0.180626557672)
(3.87394017595,0.17196235826)
(3.88438667016,0.162960122328)
(3.8941675947,0.153635340681)
(3.90328127368,0.144004532194)
(3.9117261455,0.134085195626)
(3.91950076319,0.123895757517)
(3.9266037946,0.113455516319)
(3.93303402266,0.102784583003)
(3.93879034559,0.0919038183364)
(3.94387177707,0.0808347670875)
(3.94827744643,0.0695995894038)
(3.95200659877,0.058220989643)
(3.95505859512,0.0467221429367)
(3.95743291255,0.0351266197877)
(3.95912914422,0.0234583090109)
(3.96014699949,0.0117413393359)
(3.96048630396,0)
};
\addlegendentry{$N=5$}
\addplot[color=black,dashed,line width=.9pt] coordinates {
(-4.65303766583,0)
(-4.6526390288,0.0127106738712)
(-4.65144318603,0.0253904261714)
(-4.64945034242,0.0380084532548)
(-4.64666083942,0.0505341867593)
(-4.64307515501,0.0629374098352)
(-4.63869390358,0.0751883716886)
(-4.63351783582,0.0872578998964)
(-4.62754783864,0.0991175099627)
(-4.62078493496,0.110739511608)
(-4.61323028356,0.122097111297)
(-4.60488517891,0.13316451055)
(-4.59575105089,0.14391699958)
(-4.58582946458,0.154331045866)
(-4.57512212,0.164384377263)
(-4.5636308518,0.174056059306)
(-4.55135762895,0.183326566386)
(-4.5383045544,0.19217784651)
(-4.52447386473,0.200593379385)
(-4.50986792975,0.208558227606)
(-4.49448925212,0.216059080744)
(-4.47834046689,0.223084292171)
(-4.46142434108,0.229623908486)
(-4.44374377317,0.235669691425)
(-4.42530179265,0.241215132172)
(-4.40610155945,0.246255458018)
(-4.38614636342,0.25078763132)
(-4.3654396238,0.254810340762)
(-4.34398488857,0.258323984932)
(-4.32178583389,0.261330648254)
(-4.29884626346,0.263834069354)
(-4.27517010785,0.265839601954)
(-4.25076142386,0.267354168436)
(-4.22562439377,0.268386206248)
(-4.1997633247,0.268945607364)
(-4.1731826478,0.269043651054)
(-4.14588691755,0.268692930282)
(-4.11788081091,0.267907272075)
(-4.08916912659,0.266701652285)
(-4.05975678419,0.265092105221)
(-4.02964882337,0.263095628669)
(-3.99885040297,0.26073008489)
(-3.96736680014,0.258014098244)
(-3.93520340943,0.25496695013)
(-3.90236574187,0.251608471982)
(-3.86885942403,0.2479589371)
(-3.83469019705,0.24403895213)
(-3.79986391565,0.239869349007)
(-3.76438654712,0.235471078209)
(-3.72826417034,0.230865104139)
(-3.69150297468,0.226072303446)
(-3.654109259,0.221113367057)
(-3.6160894305,0.216008706644)
(-3.5774500037,0.210778366211)
(-3.53819759926,0.205441939375)
(-3.49833894287,0.20001849289)
(-3.45788086411,0.194526496838)
(-3.41683029527,0.188983761833)
(-3.37519427015,0.183407383499)
(-3.33297992286,0.177813694376)
(-3.29019448662,0.172218223324)
(-3.2468452925,0.166635662398)
(-3.20293976815,0.161079841092)
(-3.15848543656,0.155563707766)
(-3.11348991474,0.150099317985)
(-3.06796091247,0.144697829479)
(-3.02190623088,0.139369503316)
(-2.97533376121,0.134123710907)
(-2.92825148343,0.128968946359)
(-2.88066746482,0.123912843728)
(-2.83258985868,0.118962198655)
(-2.78402690285,0.1141229939)
(-2.73498691833,0.109400428267)
(-2.68547830789,0.104798948414)
(-2.63550955457,0.100322283086)
(-2.58508922025,0.095973479281)
(-2.5342259442,0.0917549399231)
(-2.48292844159,0.0876684626138)
(-2.43120550196,0.0837152790747)
(-2.37906598779,0.0798960949207)
(-2.32651883291,0.0762111294345)
(-2.27357304101,0.0726601550459)
(-2.22023768407,0.0692425362517)
(-2.16652190083,0.0659572677431)
(-2.11243489522,0.0628030115406)
(-2.05798593476,0.0597781329645)
(-2.00318434901,0.0568807352988)
(-1.94803952793,0.0541086930335)
(-1.8925609203,0.0514596835958)
(-1.83675803211,0.0489312175031)
(-1.78064042488,0.0465206668941)
(-1.72421771409,0.0442252924114)
(-1.66749956746,0.0420422684293)
(-1.61049570337,0.0399687066333)
(-1.55321588913,0.0380016779744)
(-1.49566993934,0.0361382330299)
(-1.43786771421,0.0343754208154)
(-1.37981911784,0.0327103060998)
(-1.32153409658,0.0311399852813)
(-1.26302263725,0.0296616008899)
(-1.2042947655,0.0282723547841)
(-1.14536054403,0.0269695201135)
(-1.08623007093,0.0257504521207)
(-1.02691347791,0.0246125978566)
(-0.967420928541,0.023553504883)
(-0.907762616579,0.0225708290373)
(-0.847948764167,0.0216623413291)
(-0.787989620099,0.0208259340408)
(-0.727895458066,0.0200596260989)
(-0.667676574893,0.0193615677798)
(-0.607343288777,0.0187300448124)
(-0.546905937515,0.0181634819338)
(-0.486374876735,0.0176604459533)
(-0.425760478125,0.017219648371)
(-0.365073127648,0.0168399476003)
(-0.304323223771,0.0165203508308)
(-0.243521175675,0.0162600155698)
(-0.18267740148,0.0160582508939)
(-0.121802326453,0.0159145184361)
(-0.0609063812239,0.0158284331309)
(0,0.0157997637342)
(0.0609063812239,0.0158284331309)
(0.121802326453,0.0159145184361)
(0.18267740148,0.0160582508939)
(0.243521175675,0.0162600155698)
(0.304323223771,0.0165203508308)
(0.365073127648,0.0168399476003)
(0.425760478125,0.017219648371)
(0.486374876735,0.0176604459533)
(0.546905937515,0.0181634819338)
(0.607343288777,0.0187300448124)
(0.667676574893,0.0193615677798)
(0.727895458066,0.0200596260989)
(0.787989620099,0.0208259340408)
(0.847948764167,0.0216623413291)
(0.907762616579,0.0225708290373)
(0.967420928541,0.023553504883)
(1.02691347791,0.0246125978566)
(1.08623007093,0.0257504521207)
(1.14536054403,0.0269695201135)
(1.2042947655,0.0282723547841)
(1.26302263725,0.0296616008899)
(1.32153409658,0.0311399852813)
(1.37981911784,0.0327103060998)
(1.43786771421,0.0343754208154)
(1.49566993934,0.0361382330299)
(1.55321588913,0.0380016779744)
(1.61049570337,0.0399687066333)
(1.66749956746,0.0420422684293)
(1.72421771409,0.0442252924114)
(1.78064042488,0.0465206668941)
(1.83675803211,0.0489312175031)
(1.8925609203,0.0514596835958)
(1.94803952793,0.0541086930335)
(2.00318434901,0.0568807352988)
(2.05798593476,0.0597781329645)
(2.11243489522,0.0628030115406)
(2.16652190083,0.0659572677431)
(2.22023768407,0.0692425362517)
(2.27357304101,0.0726601550459)
(2.32651883291,0.0762111294345)
(2.37906598779,0.0798960949207)
(2.43120550196,0.0837152790747)
(2.48292844159,0.0876684626138)
(2.5342259442,0.0917549399231)
(2.58508922025,0.095973479281)
(2.63550955457,0.100322283086)
(2.68547830789,0.104798948414)
(2.73498691833,0.109400428267)
(2.78402690285,0.1141229939)
(2.83258985868,0.118962198655)
(2.88066746482,0.123912843728)
(2.92825148343,0.128968946359)
(2.97533376121,0.134123710907)
(3.02190623088,0.139369503316)
(3.06796091247,0.144697829479)
(3.11348991474,0.150099317985)
(3.15848543656,0.155563707766)
(3.20293976815,0.161079841092)
(3.2468452925,0.166635662398)
(3.29019448662,0.172218223324)
(3.33297992286,0.177813694376)
(3.37519427015,0.183407383499)
(3.41683029527,0.188983761833)
(3.45788086411,0.194526496838)
(3.49833894287,0.20001849289)
(3.53819759926,0.205441939375)
(3.5774500037,0.210778366211)
(3.6160894305,0.216008706644)
(3.654109259,0.221113367057)
(3.69150297468,0.226072303446)
(3.72826417034,0.230865104139)
(3.76438654712,0.235471078209)
(3.79986391565,0.239869349007)
(3.83469019705,0.24403895213)
(3.86885942403,0.2479589371)
(3.90236574187,0.251608471982)
(3.93520340943,0.25496695013)
(3.96736680014,0.258014098244)
(3.99885040297,0.26073008489)
(4.02964882337,0.263095628669)
(4.05975678419,0.265092105221)
(4.08916912659,0.266701652285)
(4.11788081091,0.267907272075)
(4.14588691755,0.268692930282)
(4.1731826478,0.269043651054)
(4.1997633247,0.268945607364)
(4.22562439377,0.268386206248)
(4.25076142386,0.267354168436)
(4.27517010785,0.265839601954)
(4.29884626346,0.263834069354)
(4.32178583389,0.261330648254)
(4.34398488857,0.258323984932)
(4.3654396238,0.254810340762)
(4.38614636342,0.25078763132)
(4.40610155945,0.246255458018)
(4.42530179265,0.241215132172)
(4.44374377317,0.235669691425)
(4.46142434108,0.229623908486)
(4.47834046689,0.223084292171)
(4.49448925212,0.216059080744)
(4.50986792975,0.208558227606)
(4.52447386473,0.200593379385)
(4.5383045544,0.19217784651)
(4.55135762895,0.183326566386)
(4.5636308518,0.174056059306)
(4.57512212,0.164384377263)
(4.58582946458,0.154331045866)
(4.59575105089,0.14391699958)
(4.60488517891,0.13316451055)
(4.61323028356,0.122097111297)
(4.62078493496,0.110739511608)
(4.62754783864,0.0991175099627)
(4.63351783582,0.0872578998964)
(4.63869390358,0.0751883716886)
(4.64307515501,0.0629374098352)
(4.64666083942,0.0505341867593)
(4.64945034242,0.0380084532548)
(4.65144318603,0.0253904261714)
(4.6526390288,0.0127106738712)
(4.65303766583,0)
};
\addlegendentry{$N=10$}
\nextgroupplot[title={$b=\sqrt{2}$},yticklabels={}]
\addplot[color=black,solid,line width=.9pt] coordinates {
(-4.05809609042,0)
(-4.05774842349,0.012601149498)
(-4.05670548225,0.0251740476416)
(-4.05496744541,0.0376905403238)
(-4.05253461077,0.0501226675698)
(-4.04940739519,0.0624427596997)
(-4.0455863345,0.0746235324102)
(-4.04107208342,0.086638180425)
(-4.03586541544,0.0984604693691)
(-4.02996722269,0.110064825527)
(-4.02337851581,0.121426423161)
(-4.01610042374,0.132521269061)
(-4.00813419353,0.143326284032)
(-3.99948119016,0.153819381003)
(-3.99014289628,0.163979539469)
(-3.98012091195,0.173786875999)
(-3.9694169544,0.183222710511)
(-3.95803285768,0.192269628063)
(-3.94597057241,0.200911535881)
(-3.9332321654,0.209133715377)
(-3.9198198193,0.216922868879)
(-3.90573583226,0.22426716083)
(-3.89098261749,0.23115625319)
(-3.87556270288,0.23758133478)
(-3.85947873055,0.243535144325)
(-3.84273345641,0.24901198693)
(-3.82532974967,0.254007743765)
(-3.80727059238,0.258519874717)
(-3.78855907887,0.262547413807)
(-3.76919841526,0.266090957194)
(-3.74919191891,0.269152643618)
(-3.72854301782,0.271736127197)
(-3.70725525008,0.273846542519)
(-3.68533226325,0.275490462082)
(-3.6627778137,0.276675846172)
(-3.63959576602,0.277411985364)
(-3.61579009235,0.277709435958)
(-3.59136487166,0.277579948693)
(-3.56632428909,0.277036391249)
(-3.54067263521,0.276092665083)
(-3.51441430531,0.274763617291)
(-3.4875537986,0.273064948209)
(-3.4600957175,0.271013115603)
(-3.43204476682,0.268625236275)
(-3.40340575293,0.265918986037)
(-3.37418358299,0.262912498927)
(-3.34438326406,0.259624266636)
(-3.31400990228,0.256073039017)
(-3.28306870196,0.252277726553)
(-3.25156496472,0.248257305601)
(-3.21950408856,0.244030727149)
(-3.18689156695,0.239616829748)
(-3.15373298788,0.23503425721)
(-3.12003403289,0.230301381543)
(-3.08580047614,0.225436231528)
(-3.05103818336,0.220456427214)
(-3.01575311089,0.215379120566)
(-2.97995130465,0.21022094235)
(-2.9436388991,0.204997955322)
(-2.90682211618,0.199725613662)
(-2.86950726425,0.194418728564)
(-2.83170073702,0.189091439824)
(-2.79340901245,0.183757193208)
(-2.75463865162,0.178428723365)
(-2.71539629765,0.173118041999)
(-2.6756886745,0.167836430999)
(-2.63552258586,0.162594440199)
(-2.594904914,0.157401889439)
(-2.55384261854,0.152267874577)
(-2.5123427353,0.147200777116)
(-2.47041237506,0.142208277093)
(-2.42805872238,0.137297368899)
(-2.38528903434,0.13247437969)
(-2.3421106393,0.127744990072)
(-2.29853093566,0.123114256756)
(-2.25455739057,0.118586636868)
(-2.21019753869,0.114166013653)
(-2.16545898083,0.109855723287)
(-2.12034938273,0.105658582566)
(-2.07487647367,0.101576917211)
(-2.02904804521,0.097612590597)
(-1.98287194982,0.0937670326898)
(-1.93635609952,0.0900412690077)
(-1.88950846458,0.0864359494452)
(-1.84233707208,0.0829513768035)
(-1.79485000462,0.0795875348927)
(-1.74705539884,0.0763441160866)
(-1.69896144412,0.0732205482245)
(-1.6505763811,0.0702160207693)
(-1.60190850031,0.067329510145)
(-1.55296614075,0.0645598041884)
(-1.50375768844,0.0619055256652)
(-1.45429157498,0.0593651548073)
(-1.40457627616,0.0569370508431)
(-1.35462031041,0.0546194724999)
(-1.30443223745,0.0524105974664)
(-1.25402065675,0.0503085408131)
(-1.20339420606,0.0483113723732)
(-1.15256155996,0.0464171330956)
(-1.10153142838,0.0446238503849)
(-1.05031255506,0.0429295524493)
(-0.998913716086,0.0413322816823)
(-0.947343718391,0.0398301071047)
(-0.895611398228,0.0384211359006)
(-0.843725619661,0.0371035240773)
(-0.791695273053,0.0358754862871)
(-0.739529273536,0.0347353048437)
(-0.687236559485,0.033681337972)
(-0.634826090987,0.0327120273247)
(-0.582306848307,0.0318259048035)
(-0.529687830346,0.0310215987182)
(-0.476978053102,0.0302978393173)
(-0.424186548125,0.0296534637231)
(-0.371322360965,0.029087420299)
(-0.318394549631,0.0285987724795)
(-0.26541218303,0.0281867020864)
(-0.212384339418,0.0278505121553)
(-0.159320104843,0.027589629292)
(-0.106228571587,0.0274036055764)
(-0.0531188366133,0.0272921200284)
(0,0.0272549796465)
(0.0531188366133,0.0272921200284)
(0.106228571587,0.0274036055764)
(0.159320104843,0.027589629292)
(0.212384339418,0.0278505121553)
(0.26541218303,0.0281867020864)
(0.318394549631,0.0285987724795)
(0.371322360965,0.029087420299)
(0.424186548125,0.0296534637231)
(0.476978053102,0.0302978393173)
(0.529687830346,0.0310215987182)
(0.582306848307,0.0318259048035)
(0.634826090987,0.0327120273247)
(0.687236559485,0.033681337972)
(0.739529273536,0.0347353048437)
(0.791695273053,0.0358754862871)
(0.843725619661,0.0371035240773)
(0.895611398228,0.0384211359006)
(0.947343718391,0.0398301071047)
(0.998913716086,0.0413322816823)
(1.05031255506,0.0429295524493)
(1.10153142838,0.0446238503849)
(1.15256155996,0.0464171330956)
(1.20339420606,0.0483113723732)
(1.25402065675,0.0503085408131)
(1.30443223745,0.0524105974664)
(1.35462031041,0.0546194724999)
(1.40457627616,0.0569370508431)
(1.45429157498,0.0593651548073)
(1.50375768844,0.0619055256652)
(1.55296614075,0.0645598041884)
(1.60190850031,0.067329510145)
(1.6505763811,0.0702160207693)
(1.69896144412,0.0732205482245)
(1.74705539884,0.0763441160866)
(1.79485000462,0.0795875348927)
(1.84233707208,0.0829513768035)
(1.88950846458,0.0864359494452)
(1.93635609952,0.0900412690077)
(1.98287194982,0.0937670326898)
(2.02904804521,0.097612590597)
(2.07487647367,0.101576917211)
(2.12034938273,0.105658582566)
(2.16545898083,0.109855723287)
(2.21019753869,0.114166013653)
(2.25455739057,0.118586636868)
(2.29853093566,0.123114256756)
(2.3421106393,0.127744990072)
(2.38528903434,0.13247437969)
(2.42805872238,0.137297368899)
(2.47041237506,0.142208277093)
(2.5123427353,0.147200777116)
(2.55384261854,0.152267874577)
(2.594904914,0.157401889439)
(2.63552258586,0.162594440199)
(2.6756886745,0.167836430999)
(2.71539629765,0.173118041999)
(2.75463865162,0.178428723365)
(2.79340901245,0.183757193208)
(2.83170073702,0.189091439824)
(2.86950726425,0.194418728564)
(2.90682211618,0.199725613662)
(2.9436388991,0.204997955322)
(2.97995130465,0.21022094235)
(3.01575311089,0.215379120566)
(3.05103818336,0.220456427214)
(3.08580047614,0.225436231528)
(3.12003403289,0.230301381543)
(3.15373298788,0.23503425721)
(3.18689156695,0.239616829748)
(3.21950408856,0.244030727149)
(3.25156496472,0.248257305601)
(3.28306870196,0.252277726553)
(3.31400990228,0.256073039017)
(3.34438326406,0.259624266636)
(3.37418358299,0.262912498927)
(3.40340575293,0.265918986037)
(3.43204476682,0.268625236275)
(3.4600957175,0.271013115603)
(3.4875537986,0.273064948209)
(3.51441430531,0.274763617291)
(3.54067263521,0.276092665083)
(3.56632428909,0.277036391249)
(3.59136487166,0.277579948693)
(3.61579009235,0.277709435958)
(3.63959576602,0.277411985364)
(3.6627778137,0.276675846172)
(3.68533226325,0.275490462082)
(3.70725525008,0.273846542519)
(3.72854301782,0.271736127197)
(3.74919191891,0.269152643618)
(3.76919841526,0.266090957194)
(3.78855907887,0.262547413807)
(3.80727059238,0.258519874717)
(3.82532974967,0.254007743765)
(3.84273345641,0.24901198693)
(3.85947873055,0.243535144325)
(3.87556270288,0.23758133478)
(3.89098261749,0.23115625319)
(3.90573583226,0.22426716083)
(3.9198198193,0.216922868879)
(3.9332321654,0.209133715377)
(3.94597057241,0.200911535881)
(3.95803285768,0.192269628063)
(3.9694169544,0.183222710511)
(3.98012091195,0.173786875999)
(3.99014289628,0.163979539469)
(3.99948119016,0.153819381003)
(4.00813419353,0.143326284032)
(4.01610042374,0.132521269061)
(4.02337851581,0.121426423161)
(4.02996722269,0.110064825527)
(4.03586541544,0.0984604693691)
(4.04107208342,0.086638180425)
(4.0455863345,0.0746235324102)
(4.04940739519,0.0624427596997)
(4.05253461077,0.0501226675698)
(4.05496744541,0.0376905403238)
(4.05670548225,0.0251740476416)
(4.05774842349,0.012601149498)
(4.05809609042,0)
};
\addlegendentry{$N=5$}
\addplot[color=black,dashed,line width=.9pt] coordinates {
(-4.75078442354,0)
(-4.75037741231,0.0136209061512)
(-4.74915644837,0.0272061438172)
(-4.74712174092,0.0407201860395)
(-4.7442736386,0.0541277883195)
(-4.74061262941,0.0673941283653)
(-4.73613934065,0.08048494407)
(-4.7308545388,0.0933666691449)
(-4.72475912938,0.106006565845)
(-4.71785415681,0.118372854239)
(-4.71014080421,0.130434837486)
(-4.70162039323,0.142163022605)
(-4.69229438381,0.153529236231)
(-4.6821643739,0.164506734876)
(-4.67123209922,0.175070309205)
(-4.65949943297,0.185196381881)
(-4.64696838549,0.19486309851)
(-4.63364110389,0.204050411235)
(-4.61951987174,0.212740154533)
(-4.60460710865,0.220916112754)
(-4.58890536983,0.228564078966)
(-4.5724173457,0.235671904614)
(-4.55514586139,0.242229539578)
(-4.53709387629,0.248229062125)
(-4.51826448352,0.25366469834)
(-4.49866090938,0.25853283058)
(-4.47828651286,0.262831994557)
(-4.457144785,0.26656286469)
(-4.43523934832,0.269728227424)
(-4.41257395622,0.272332942315)
(-4.38915249228,0.274383890772)
(-4.36497896966,0.275889912473)
(-4.34005753036,0.276861729644)
(-4.31439244455,0.27731185952)
(-4.2879881098,0.277254515519)
(-4.26084905037,0.276705497831)
(-4.23297991638,0.27568207431)
(-4.20438548308,0.274202852747)
(-4.17507064997,0.272287645758)
(-4.14504044,0.269957329665)
(-4.11429999869,0.267233698868)
(-4.08285459325,0.26413931727)
(-4.0507096117,0.260697368375)
(-4.0178705619,0.256931505663)
(-3.98434307067,0.252865704825)
(-3.95013288277,0.248524119342)
(-3.91524585993,0.243930940819)
(-3.87968797987,0.239110265311)
(-3.84346533525,0.234085966762)
(-3.80658413263,0.228881578467)
(-3.76905069142,0.223520183336)
(-3.73087144277,0.218024313519)
(-3.6920529285,0.212415859808)
(-3.65260179998,0.206715991039)
(-3.61252481694,0.200945083591)
(-3.57182884637,0.195122660929)
(-3.53052086131,0.189267342994)
(-3.48860793968,0.183396805191)
(-3.44609726304,0.177527746579)
(-3.40299611537,0.171675866836)
(-3.35931188184,0.165855851496)
(-3.31505204751,0.16008136493)
(-3.27022419608,0.154365050489)
(-3.22483600856,0.148718537236)
(-3.178895262,0.14315245266)
(-3.1324098281,0.13767644079)
(-3.0853876719,0.132299185114)
(-3.03783685041,0.127028435724)
(-2.98976551121,0.121871040135)
(-2.94118189109,0.116832977238)
(-2.89209431458,0.111919393867)
(-2.84251119261,0.107134643506)
(-2.79244102098,0.102482326656)
(-2.74189237895,0.0979653324546)
(-2.69087392778,0.093585881133)
(-2.63939440921,0.0893455669438)
(-2.587462644,0.0852454012286)
(-2.53508753038,0.0812858553122)
(-2.48227804257,0.0774669029503)
(-2.4290432292,0.0737880620839)
(-2.37539221177,0.0702484356847)
(-2.32133418311,0.0668467515032)
(-2.2668784058,0.0635814005581)
(-2.21203421053,0.060450474233)
(-2.1568109946,0.0574517998708)
(-2.1012182202,0.0545829747763)
(-2.04526541288,0.0518413985654)
(-1.98896215985,0.0492243038127)
(-1.93231810841,0.0467287849712)
(-1.8753429642,0.0443518255553)
(-1.81804648963,0.0420903235901)
(-1.76043850215,0.0399411153464)
(-1.70252887259,0.0379009973925)
(-1.64432752348,0.0359667470006)
(-1.58584442732,0.034135140961)
(-1.52708960487,0.0324029728581)
(-1.46807312348,0.0307670688727)
(-1.40880509532,0.029224302178)
(-1.34929567566,0.0277716060023)
(-1.28955506114,0.0264059854302)
(-1.22959348799,0.0251245280218)
(-1.16942123032,0.0239244133239)
(-1.10904859835,0.022802921352)
(-1.04848593661,0.0217574401184)
(-0.987743622208,0.0207854722825)
(-0.926832063017,0.0198846409946)
(-0.865761695923,0.0190526950066)
(-0.80454298502,0.0182875131147)
(-0.743186419817,0.0175871080016)
(-0.681702513448,0.0169496295382)
(-0.620101800863,0.0163733676036)
(-0.558394837026,0.0158567544771)
(-0.496592195108,0.0153983668522)
(-0.434704464674,0.014996927519)
(-0.372742249869,0.0146513067568)
(-0.310716167597,0.014360523474)
(-0.24863684571,0.0141237461289)
(-0.186514921179,0.01394029346)
(-0.124361038277,0.0138096350499)
(-0.0621858467507,0.0137313917441)
(0,0.0137053359377)
(0.0621858467507,0.0137313917441)
(0.124361038277,0.0138096350499)
(0.186514921179,0.01394029346)
(0.24863684571,0.0141237461289)
(0.310716167597,0.014360523474)
(0.372742249869,0.0146513067568)
(0.434704464674,0.014996927519)
(0.496592195108,0.0153983668522)
(0.558394837026,0.0158567544771)
(0.620101800863,0.0163733676036)
(0.681702513448,0.0169496295382)
(0.743186419817,0.0175871080016)
(0.80454298502,0.0182875131147)
(0.865761695923,0.0190526950066)
(0.926832063017,0.0198846409946)
(0.987743622208,0.0207854722825)
(1.04848593661,0.0217574401184)
(1.10904859835,0.022802921352)
(1.16942123032,0.0239244133239)
(1.22959348799,0.0251245280218)
(1.28955506114,0.0264059854302)
(1.34929567566,0.0277716060023)
(1.40880509532,0.029224302178)
(1.46807312348,0.0307670688727)
(1.52708960487,0.0324029728581)
(1.58584442732,0.034135140961)
(1.64432752348,0.0359667470006)
(1.70252887259,0.0379009973925)
(1.76043850215,0.0399411153464)
(1.81804648963,0.0420903235901)
(1.8753429642,0.0443518255553)
(1.93231810841,0.0467287849712)
(1.98896215985,0.0492243038127)
(2.04526541288,0.0518413985654)
(2.1012182202,0.0545829747763)
(2.1568109946,0.0574517998708)
(2.21203421053,0.060450474233)
(2.2668784058,0.0635814005581)
(2.32133418311,0.0668467515032)
(2.37539221177,0.0702484356847)
(2.4290432292,0.0737880620839)
(2.48227804257,0.0774669029503)
(2.53508753038,0.0812858553122)
(2.587462644,0.0852454012286)
(2.63939440921,0.0893455669438)
(2.69087392778,0.093585881133)
(2.74189237895,0.0979653324546)
(2.79244102098,0.102482326656)
(2.84251119261,0.107134643506)
(2.89209431458,0.111919393867)
(2.94118189109,0.116832977238)
(2.98976551121,0.121871040135)
(3.03783685041,0.127028435724)
(3.0853876719,0.132299185114)
(3.1324098281,0.13767644079)
(3.178895262,0.14315245266)
(3.22483600856,0.148718537236)
(3.27022419608,0.154365050489)
(3.31505204751,0.16008136493)
(3.35931188184,0.165855851496)
(3.40299611537,0.171675866836)
(3.44609726304,0.177527746579)
(3.48860793968,0.183396805191)
(3.53052086131,0.189267342994)
(3.57182884637,0.195122660929)
(3.61252481694,0.200945083591)
(3.65260179998,0.206715991039)
(3.6920529285,0.212415859808)
(3.73087144277,0.218024313519)
(3.76905069142,0.223520183336)
(3.80658413263,0.228881578467)
(3.84346533525,0.234085966762)
(3.87968797987,0.239110265311)
(3.91524585993,0.243930940819)
(3.95013288277,0.248524119342)
(3.98434307067,0.252865704825)
(4.0178705619,0.256931505663)
(4.0507096117,0.260697368375)
(4.08285459325,0.26413931727)
(4.11429999869,0.267233698868)
(4.14504044,0.269957329665)
(4.17507064997,0.272287645758)
(4.20438548308,0.274202852747)
(4.23297991638,0.27568207431)
(4.26084905037,0.276705497831)
(4.2879881098,0.277254515519)
(4.31439244455,0.27731185952)
(4.34005753036,0.276861729644)
(4.36497896966,0.275889912473)
(4.38915249228,0.274383890772)
(4.41257395622,0.272332942315)
(4.43523934832,0.269728227424)
(4.457144785,0.26656286469)
(4.47828651286,0.262831994557)
(4.49866090938,0.25853283058)
(4.51826448352,0.25366469834)
(4.53709387629,0.248229062125)
(4.55514586139,0.242229539578)
(4.5724173457,0.235671904614)
(4.58890536983,0.228564078966)
(4.60460710865,0.220916112754)
(4.61951987174,0.212740154533)
(4.63364110389,0.204050411235)
(4.64696838549,0.19486309851)
(4.65949943297,0.185196381881)
(4.67123209922,0.175070309205)
(4.6821643739,0.164506734876)
(4.69229438381,0.153529236231)
(4.70162039323,0.142163022605)
(4.71014080421,0.130434837486)
(4.71785415681,0.118372854239)
(4.72475912938,0.106006565845)
(4.7308545388,0.0933666691449)
(4.73613934065,0.08048494407)
(4.74061262941,0.0673941283653)
(4.7442736386,0.0541277883195)
(4.74712174092,0.0407201860395)
(4.74915644837,0.0272061438172)
(4.75037741231,0.0136209061512)
(4.75078442354,0)
};
\addlegendentry{$N=10$}
\end{groupplot}
\end{tikzpicture}

\caption{Principal equilibrium probability densities in the logarithmic
(rapidity) coordinate $\lambda$ for
$r_*=1$, $N=5$ (solid), and $N=10$ (dashed), with $b=1$ and $b=\sqrt2$.
Each support is a single interval with a strictly positive interior
density, despite the two pronounced maxima.}
\label{fig:equilibrium-densities}
\end{figure}

Figure~\ref{fig:equilibrium-densities} illustrates this principal equilibrium
problem for $r_*=1$. The endpoint equation
\eqref{eq:gb-mass-exact} is solved before evaluating the positive-density
formula \eqref{eq:gb-positive-density}; no large-$N$ replacement of
these formulas is made in the plot. The resulting endpoints are
\[
\begin{array}{c|cc}
 &N=5&N=10\\\hline
 b=1&L_N=3.96048630&L_N=4.65303767\\
 b=\sqrt2&L_N=4.05809609&L_N=4.75078442
\end{array}
\]
The plotted measures have unit mass. Numerical quadrature gives mass
errors below $2\times10^{-14}$ and residuals below $5\times10^{-14}$
in the differentiated Euler--Lagrange equation at
$\lambda/L_N=0.2,0.55,0.85$. These are numerical checks, not rigorous
error enclosures.
The principal measure does not include the subleading factors containing
$H$ or $\alpha$, and differs from the finite-$N$ mean empirical density.

The two equilibrium logarithmic functions are defined by
\begin{align}
 g_N(z)&=\int_{a_N}^{d_N}\log(z-x)\rho^{\mathrm{eq}}_N(x)\,dx,
 &&z\in\C\setminus(-\infty,d_N],\label{eq:explicit-g}\\
 \widetilde g_N(z)&=\int_{a_N}^{d_N}\log(z^\theta-x^\theta)
 \rho^{\mathrm{eq}}_N(x)\,dx,
 &&z\in\mathbb H_\theta\setminus(0,d_N].\label{eq:explicit-gtilde}
\end{align}
Both logarithms have arguments in $(-\pi,\pi)$ before analytic
continuation. In particular,
\[
 g_N(z)=\log z+O(z^{-1}),\qquad
 \widetilde g_N(z)=\theta\log z+O(z^{-\theta}).
\]
The Euler--Lagrange constant $\ell_N$ is fixed by
\begin{equation}\label{eq:explicit-EL}
 g_{N,+}(x)+\widetilde g_{N,-}(x)
 =g_{N,-}(x)+\widetilde g_{N,+}(x)
 =V_N(x)+\ell_N,\qquad a_N<x<d_N.
\end{equation}
Put
\begin{equation}\label{eq:explicit-phase}
 \varphi_N(z)=g_N(z)+\widetilde g_N(z)-V_N(z)-\ell_N.
\end{equation}
On the interior of the band $[a_N,d_N]=[x_{-,N},x_{+,N}]$,
\begin{equation}\label{eq:explicit-band-phase}
 \varphi_{N,+}(x)=-\varphi_{N,-}(x)
 =g_{N,+}(x)-g_{N,-}(x)
 =2\pi i\int_x^{d_N}\rho^{\mathrm{eq}}_N(y)\,dy.
\end{equation}
On either exterior interval define the real function
\begin{equation}\label{eq:explicit-gap-phase}
 \Psi_N(x)=\int_{a_N}^{d_N}
 \log\bigl|(x-y)(x^\theta-y^\theta)\bigr|\rho^{\mathrm{eq}}_N(y)\,dy
 -V_N(x)-\ell_N<0.
\end{equation}
For $0<x<a_N$ one has $\varphi_{N,\pm}(x)=\Psi_N(x)\pm2\pi i$;
for $x>d_N$ one has $\varphi_N(x)=\Psi_N(x)$.

Define
\begin{equation}\label{eq:explicit-T}
 T_c(z)=Y_c(z)
 \begin{pmatrix}e^{-Ng_N(z)}&0\\0&e^{N(\widetilde g_N(z)-\ell_N)}\end{pmatrix}.
\end{equation}
\paragraph{RH problem for $T_c$.}
The transformed row vector solves the following Riemann--Hilbert
problem.
\begin{enumerate}
\item $T_{c,1}$ is analytic in $\C\setminus[a_N,d_N]$, and
$T_{c,2}$ is analytic in $\mathbb H_\theta\setminus(0,\infty)$.
\item The positive axis is oriented to the right. On the band interior
$(a_N,d_N)=(x_{-,N},x_{+,N})$,
\begin{equation}\label{eq:explicit-Tband}
 T_{c,+}=T_{c,-}
 \begin{pmatrix}
 e^{-N\varphi_{N,+}}&\omega_{\alpha,c}\\
 0&e^{N\varphi_{N,+}}
 \end{pmatrix}.
\end{equation}
On the gaps $(0,a_N)\cup(d_N,\infty)$ it is
\begin{equation}\label{eq:explicit-Tgap}
 T_{c,+}=T_{c,-}
 \begin{pmatrix}1&\omega_{\alpha,c}e^{N\Psi_N}\\0&1\end{pmatrix}.
\end{equation}
\item At infinity in the respective domains,
\[
 T_{c,1}(z)=1+O(z^{-1}),\qquad T_{c,2}(z)=O(z^{-\theta}).
\]
Both components are bounded at $0,a_N,d_N$ within their domains.
\item On the sector edges,
\[
 T_{c,2}(te^{i\pi/\theta})=T_{c,2}(te^{-i\pi/\theta}),\qquad t>0.
\]
\end{enumerate}
There is no additional jump for $T_{c,1}$ on the negative axis:
$g_{N,+}-g_{N,-}=2\pi i$ there and $N$ is an integer. Likewise,
$\widetilde g_N(te^{i\pi/\theta})-
\widetilde g_N(te^{-i\pi/\theta})=2\pi i$, so the second
component retains the sewing condition \eqref{eq:explicit-Ysewing}.

\subsection{Opening the lens: the transformation \texorpdfstring{$T\mapsto S$}{T to S}}
\label{sec:lens-opening}
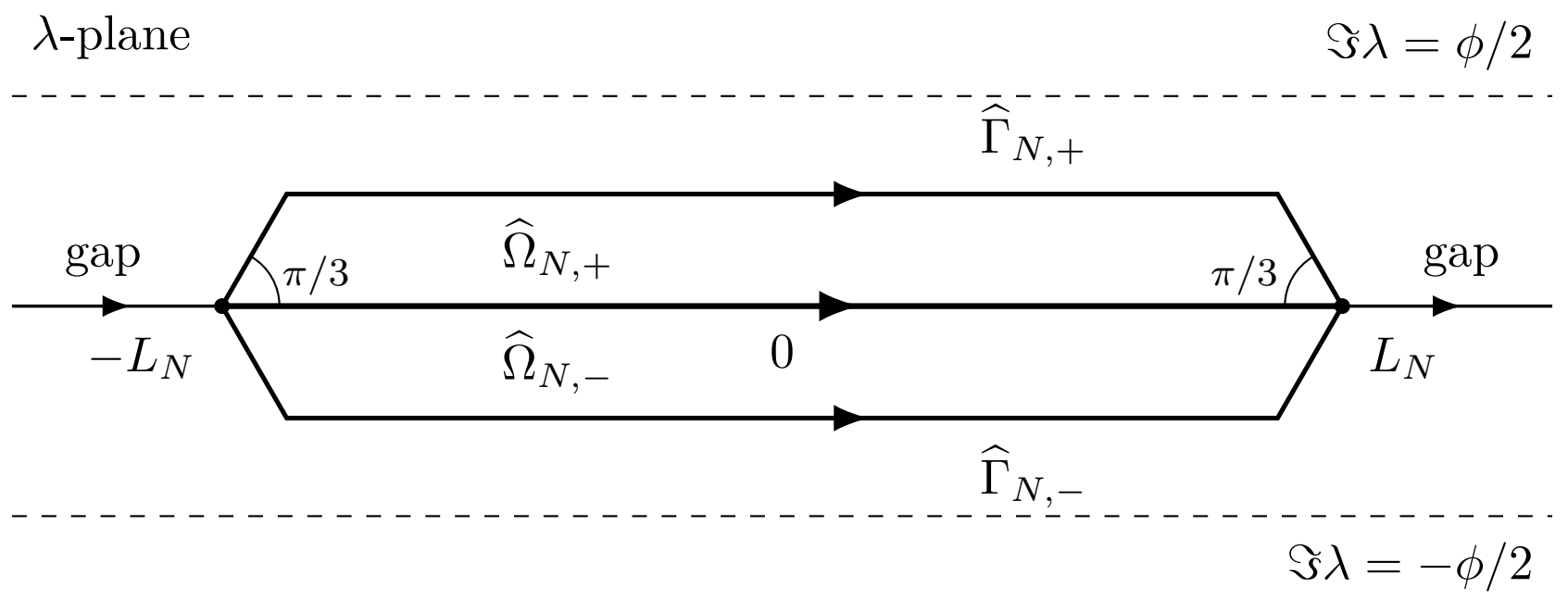
\begin{figure}[htbp]
\centering
\begin{tikzpicture}[x=1cm,y=1cm,>=Latex,
 every node/.style={font=\small},
 oriented/.style={postaction={decorate},decoration={markings,
 mark=at position .57 with {\arrow{Latex}}}}]
\draw[black,dashed] (-5.5,1.5)--(5.5,1.5);
\draw[black,dashed] (-5.5,-1.5)--(5.5,-1.5);
\node[anchor=south east,text=black] at (5.5,1.55)
 {$\Im\lambda=\phi/2$};
\node[anchor=north east,text=black] at (5.5,-1.55)
 {$\Im\lambda=-\phi/2$};
\draw[oriented,line width=.65pt] (-5.5,0)--(-4,0);
\draw[oriented,line width=1.2pt] (-4,0)--(4,0);
\draw[oriented,line width=.65pt] (4,0)--(5.5,0);
\draw[oriented,black,line width=.95pt]
 (-4,0)--(-3.53812,.8)--(3.53812,.8)--(4,0);
\draw[oriented,black,line width=.95pt]
 (-4,0)--(-3.53812,-.8)--(3.53812,-.8)--(4,0);
\fill (-4,0) circle (1.6pt);
\fill (4,0) circle (1.6pt);
\node[anchor=north east] at (-4.05,-.09) {$-L_N$};
\node[anchor=north west] at (4.05,-.09) {$L_N$};
\node[anchor=north] at (0,-.08) {$0$};
\node[anchor=south,text=black] at (1.8,.86)
 {$\widehat\Gamma_{N,+}$};
\node[anchor=north,text=black] at (1.8,-.86)
 {$\widehat\Gamma_{N,-}$};
\node at (-1.6,.40) {$\widehat\Omega_{N,+}$};
\node at (-1.6,-.40) {$\widehat\Omega_{N,-}$};
\node[anchor=south] at (-4.85,.08) {gap};
\node[anchor=south] at (4.85,.08) {gap};
\draw[thin,black] (-3.59,0) arc[start angle=0,end angle=60,radius=.41];
\node[font=\scriptsize] at (-3.33,.22) {$\pi/3$};
\draw[thin,black] (3.59,0) arc[start angle=180,end angle=120,radius=.41];
\node[font=\scriptsize] at (3.30,.22) {$\pi/3$};
\node[anchor=south west] at (-5.5,1.63) {$\lambda$-plane};
\end{tikzpicture}

\caption{Jump contours of $S_c$ in the logarithmic (rapidity) coordinate $\lambda$. All solid
contours are oriented to the right; the thick segment is the band.
The hatted lips and lens regions are mapped to $\Gamma_{N,\pm}$ and
$\Omega_{N,\pm}$ by $z=e^{2B\lambda}$. The dashed lines are the
boundaries of the strip $|\Im\lambda|<\phi/2$, identified for the
second component; they are not additional matrix-jump contours.
The endpoint angle is $\pi/3$ relative to the adjacent band, and
corners may be rounded. The vertical scale is schematic.}
\label{fig:S-contours}
\end{figure}

Choose conjugate upper and lower lips $\Gamma_{N,+}$ and
$\Gamma_{N,-}$ joining $a_N$ to $d_N$ inside $\mathbb H_\theta$.
Orient both lips from $a_N$ to $d_N$, and orient the real axis to the
right. The $+$ side of an oriented contour is its left side. Write
$\Omega_{N,+}$ and $\Omega_{N,-}$ for the upper and lower lens interiors.
A concrete shape is most easily specified in the complex logarithmic coordinate $\lambda$,
where $z=e^{2B\lambda}$ and the band is $[-L_N,L_N]$.
For $0<h_0<\phi/2$ and $L_N>h_0/\sqrt3$, the piecewise smooth arcs
\begin{equation}\label{eq:explicit-lips}
 \lambda=t\pm i h_N(t),\qquad
 h_N(t)=\min\{h_0,\sqrt3(L_N-|t|)\},\qquad -L_N\leq t\leq L_N,
\end{equation}
give horizontal central portions and transverse endpoint departures.
Figure~\ref{fig:S-contours} displays these contours in rapidity coordinates.
The upper departures have angles $\pi/3$ at $-L_N$ and $2\pi/3$
at $L_N$; the lower arcs are their reflections. Corners can be rounded
by an equivalent analytic contour deformation. Their exponential images lie strictly
inside $\mathbb H_\theta$, since $2Bh_0<\pi/\theta$.
For steepest descent the height is chosen small enough, with local
deformations if needed, that $\Re\varphi_N>0$ on the open lips.
The following jump identities are exact independently of the subsequent
uniform estimates on this choice of contours.

Let $E_{ij}$ be the elementary matrix and set
\[
 L_{N,c}^{\pm}(z)=I+
 \omega_{\alpha,c}(z)^{-1}e^{-N\varphi_N(z)}E_{21},
 \qquad z\in\Omega_{N,\pm},
\]
where $\varphi_N$ is the continuation from the corresponding half-plane.
The factorization
\begin{equation}\label{eq:explicit-factorization}
 \begin{pmatrix}e^{-N\varphi_{N,+}}&\omega_{\alpha,c}\\
 0&e^{N\varphi_{N,+}}\end{pmatrix}
 =L_{N,c}^{-}
 \begin{pmatrix}0&\omega_{\alpha,c}\\-\omega_{\alpha,c}^{-1}&0\end{pmatrix}
 L_{N,c}^{+}
\end{equation}
holds on $(a_N,d_N)$, with the boundary values understood. Define
\begin{equation}\label{eq:explicit-S}
 S_c(z)=
 \begin{cases}
 T_c(z)(L_{N,c}^{+}(z))^{-1},&z\in\Omega_{N,+},\\
 T_c(z)L_{N,c}^{-}(z),&z\in\Omega_{N,-},\\
 T_c(z),&z\text{ outside the lenses}.
 \end{cases}
\end{equation}
In particular, $S_{c,2}=T_{c,2}$ everywhere in its domain.

\paragraph{RH problem for $S_c$.}
The row vector $S_c$ solves the following Riemann--Hilbert problem.
\begin{enumerate}
\item The first component is analytic off the band
$[a_N,d_N]=[x_{-,N},x_{+,N}]$ and the two lips in
$\C$, and the second is analytic in
$\mathbb H_\theta\setminus(0,\infty)$.
\item On the oriented contour
\[
 \Sigma_{S,N}=(0,a_N)\cup[a_N,d_N]\cup(d_N,\infty)
 \cup\Gamma_{N,+}\cup\Gamma_{N,-},
\]
the row-vector jump relation is $S_{c,+}=S_{c,-}J_{S,c}$, where
\begin{equation}\label{eq:explicit-Sjumps}
 J_{S,c}(z)=
 \begin{cases}
 \begin{pmatrix}0&\omega_{\alpha,c}(z)\\
 -\omega_{\alpha,c}(z)^{-1}&0\end{pmatrix},&z\in(a_N,d_N),\\[4mm]
 I+\omega_{\alpha,c}(z)^{-1}e^{-N\varphi_N(z)}E_{21},
 &z\in\Gamma_{N,+}\cup\Gamma_{N,-},\\[2mm]
 I+\omega_{\alpha,c}(z)e^{N\Psi_N(z)}E_{12},
 &z\in(0,a_N)\cup(d_N,\infty).
 \end{cases}
\end{equation}
\item The asymptotic and local conditions are
\begin{align*}
 S_{c,1}(z)&=1+O(z^{-1}),&S_{c,2}(z)&=O(z^{-\theta})
 &&(z\to\infty),\\
 S_{c,j}(z)&=O(1)&&&&(z\to0,a_N,d_N)
\end{align*}
within the corresponding domains and sectors.
\item The second component obeys the sewing condition
\[
 S_{c,2}(te^{i\pi/\theta})=S_{c,2}(te^{-i\pi/\theta}),\qquad t>0.
\]
\end{enumerate}
The sector edges are thus identified boundaries for the second
component, not extra ordinary matrix jumps. Reversing the two changes
of unknown recovers \eqref{eq:explicit-Y}, so these conditions
determine the vector uniquely.

The endpoint neighborhoods still have square-root equilibrium edges
and Airy local coordinates. In contrast, the central portions of the
lip jumps need not tend to the identity. Indeed, on a fixed upper
rapidity line, Appendix~\ref{sec:exact-general-equilibrium} gives
\[
 e^{-N\varphi_N(e^{2B(t+ih_0)})}
 =(-1)^N\exp\{ir_*\sin\phi\sinh(t+ih_0)+o(1)\}
\]
locally uniformly in real $t$. Consequently the central operator below
is retained in the model; it is not discarded by a small-norm argument.

\subsection{The outer factor and the retained central operator}
\paragraph{RH problem for the outer row $P_c$.}
The outer factor solves the following Riemann--Hilbert problem on
the band $[a_N,d_N]=[x_{-,N},x_{+,N}]$.
\begin{enumerate}
\item $P_{c,1}$ is analytic in $\C\setminus[a_N,d_N]$, and
$P_{c,2}$ is analytic in $\mathbb H_\theta\setminus[a_N,d_N]$.
\item The band is oriented to the right, and
\[
 P_{c,+}(x)=P_{c,-}(x)
 \begin{pmatrix}0&\omega_{\alpha,c}(x)\\
 -\omega_{\alpha,c}(x)^{-1}&0\end{pmatrix},\qquad a_N<x<d_N.
\]
\item At infinity, $P_{c,1}(z)=1+O(z^{-1})$ and
$P_{c,2}(z)=O(z^{-\theta})$ in their respective domains.
Both components are bounded at zero, and
$P_{c,j}(z)=O((z-e)^{-1/4})$ as $z\to e\in\{a_N,d_N\}$.
\item The second component satisfies
\[
 P_{c,2}(te^{i\pi/\theta})=P_{c,2}(te^{-i\pi/\theta}),\qquad t>0.
\]
\end{enumerate}
We now construct this row explicitly using the Claeys--Romano map.
Let $D_N$ be the bounded Jordan domain enclosed by the two preimage
arcs, with counterclockwise boundary $\gamma_N$ (also denoted $\gamma$).
Let $\gamma_{N,+}$ be its upper boundary arc. The branches $I_{1,N}$
and $I_{2,N}$ take values in $\mathbb C\setminus\overline{D_N}$ and
$D_N\setminus[-1,0]$, respectively.
Put
\[
 \mathcal C_N(s)=\frac1{2\pi i}\oint_{\gamma_N}
 \frac{\log\omega_\alpha(J_N(\xi))}{\xi-s}\,d\xi,
 \qquad R_N(s)=\sqrt{(s-s_-)(s-s_+)},
\]
where the cut of $R_N$ is $\gamma_{N,+}$ and $R_N(s)/s\to1$.
The logarithm on the contour is the real logarithm of the positive
band weight. Write $a_c=I_{1,N}(-c)$ and define
\[
 \mathscr P_{c}^{\mathrm{out}}(s)
 =\frac{s e^{\mathcal C_N(s)}}{R_N(s)}
   \frac{J_N(s)+c}{c_{1,N}(s-a_c)},\qquad s\notin\overline{D_N},
\]
\[
 \mathscr P_{c}^{\mathrm{in}}(s)
 =\frac{s e^{\mathcal C_N(s)}}{R_N(s)}
   \frac1{c_{1,N}(s-a_c)},\qquad s\in D_N.
\]
The first expression is continued at $a_c$ by replacing its last factor
there by $J_N'(a_c)/c_{1,N}$. The band row is
\[
 P_c(z)=\bigl(\mathscr P_c^{\mathrm{out}}(I_{1,N}(z)),
             \mathscr P_c^{\mathrm{in}}(I_{2,N}(z))\bigr).
\]
Plemelj's formula gives $\mathcal C_{N,\mathrm{in}}-
\mathcal C_{N,\mathrm{out}}=\log\omega_\alpha$.
The square root changes sign on the upper arc and not on the lower arc.
Together with the ratio $(J_N+c)^{-1}$ of the insertion factors,
these relations yield
\[
 P_{c,1,+}=-\omega_{\alpha,c}^{-1}P_{c,2,-},\qquad
 P_{c,2,+}=\omega_{\alpha,c}P_{c,1,-}.
\]
Thus $P_{c,+}=P_{c,-}
\left(\begin{smallmatrix}0&\omega_{\alpha,c}\\
-\omega_{\alpha,c}^{-1}&0\end{smallmatrix}\right)$.
The normalization is $P_{c,1}=1+O(z^{-1})$ and
$P_{c,2}=O(z^{-\theta})$ at infinity. The interior scalar factor is
analytic across the slit $[-1,0]$, which gives the sector sewing for
$P_{c,2}$.

Every component is nonzero at every finite regular point of its
domain. Indeed the exponential never vanishes, the only zero of the
factor $s$ is $s=0$ (the interior preimage of infinity), and the
exterior conformal map has $J_N(s)+c=0$ only at $s=a_c$, where the
zero is canceled by $s-a_c$ with nonzero derivative. At a band endpoint
$e$, the components have order $O((z-e)^{-1/4})$ with nonzero leading
coefficients. This proves pointwise legitimacy of componentwise division.
It does not give a uniform positive lower bound on the unscaled
components: at fixed $N$ and a finite interior lift,
$\mathscr P_c^{\mathrm{in}}(s)=O(c^{-1})$ as $c\to\infty$.
Uniform matching estimates must instead control the appropriate
weight-compensated component ratios.

For the unmodified weight write $\mathscr P_\alpha(s)
=s e^{\mathcal C_N(s)}/R_N(s)$, with its two domain restrictions
understood. The insertion factors above are denoted
$D_c^{\rm out}$ and $D_c^{\rm in}$.

The exterior model is
\begin{equation}
 p^{\rm mod}_{N,\alpha,c}(z)
 =e^{Ng_N(z)}\mathscr P_\alpha(I_1(z))
 D_c^{\rm out}(I_1(z))\,C_{N,\alpha,c}(I_1(z)).
 \label{eq:newpolymodel}
\end{equation}
The negative axis is outside the lenses and matching disks.

Choose a fixed height $0<h_0<\phi/2$ and a smooth cutoff equal to one
for $|t|\leq L_N/3$ and zero for $|t|\geq L_N/3+1$.
The outer preimages of the two rapidity lips are $z_\pm(t)$ and their
partners are $p_\pm(t)$. For the unmodified weight the scalar retained
jump coefficient is
\[
 a_\pm(t)=\chi_N^{\rm cut}(t)^2
 \omega_\alpha(e^{2B(t\pm ih_0)})^{-1}
 e^{-N\varphi_N(e^{2B(t\pm ih_0)})}
 \frac{\mathscr P_\alpha(p_\pm(t))}{\mathscr P_\alpha(z_\pm(t))}.
\]
Orientation signs are included in the oriented differentials $z_\pm'(t)\dd t$.

\paragraph{Scalar RH problem for $C_{N,\alpha,c}$.}
First write $C=C_{N,\alpha}$ for the unmodified weight, and let
$\Sigma_{C,N}$ be the retained integration contours parametrized by
$z_\pm(t)$. The scalar Riemann--Hilbert problem is:
\begin{enumerate}
\item $C$ is analytic in $\C\setminus\Sigma_{C,N}$.
\item On each oriented contour,
\[
 C_+(z_\tau(t))-C_-(z_\tau(t))
 =a_\tau(t)C(p_\tau(t)),\qquad \tau\in\{+,-\}.
\]
This is a nonlocal jump: $p_\tau(t)$ is a partner point off the
integration contours, rather than the same point approached from
the other side.
\item $C(s)=1+O(s^{-1})$ at infinity, and $C$ is bounded at the
cutoff endpoints.
\end{enumerate}
For the modified weight, replace $a_\tau(t)$ by
\[
 a_{\tau,c}(t)=a_\tau(t)
 \frac{z_\tau(t)-a_c}{p_\tau(t)-a_c},\qquad a_c=I_{1,N}(-c),
\]
and denote the solution by $C_{N,\alpha,c}$.
The unmodified problem has the Cauchy representation and partner operator
\begin{align}
 C(s)&=1+\sum_\tau\int
 \frac{a_\tau(v)z_\tau'(v)C(p_\tau(v))}
 {2\pi i(z_\tau(v)-s)}\dd v,\label{eq:newcoreC}\\
 K_{\sigma\tau}(t,v)&=
 \frac{a_\tau(v)z_\tau'(v)}{2\pi i(z_\tau(v)-p_\sigma(t))}.
 \label{eq:newcoreK}
\end{align}
The smooth cutoff makes this a compact-contour trace-class problem;
double-exponential balancing will make its estimates uniform in $N$.
Both the integration and partner-evaluation contours stay off the negative
observation axis. The boundary deformation and component identification
described in Section~\ref{sec:newcorelimit} concern the limiting
Fredholm determinant.

\section{A negative-axis resolvent identity and the central trace identity}
\label{sec:resolvent-identity}

Let $w$ be a positive weight on $(0,\infty)$ with all positive and
negative moments.  For $c>0$, let $p_{N,c}$ be the monic degree-$N$
biorthogonal polynomial for $w(x)/(x+c)$, with the dual variable
$x^\theta$.  The derivative below always holds $c$ fixed.

\begin{proposition}[Negative-axis resolvent]
Let $\rho^{\mathrm{mean}}_N(x)=w(x)S_N(x,x)$ be the mean counting density for the
unmodified weight $w$, normalized by $\int_0^\infty\rho^{\mathrm{mean}}_N(x)\,dx=N$.
Thus $N^{-1}\rho^{\mathrm{mean}}_N$ is the mean empirical density; it
is distinct from the equilibrium probability density $\rho^{\mathrm{eq}}_N$. Then
\begin{equation}\label{eq:weak-resolvent-polynomial}
 R_N(-c):=\int_0^\infty\frac{\rho^{\mathrm{mean}}_N(x)}{-c-x}\,dx
 =\frac{p_{N,c}'(-c)}{p_{N,c}(-c)}.
\end{equation}
For the logarithmic insertion
$\ell_b(x)=\partial_\alpha\log w_{N,\alpha}(x)=\tfrac b2\log x$,
this gives
\begin{equation}\label{eq:weak-derivative-resolvent}
 F_N'(\alpha)=\frac b2\int_0^\infty
 \left\{\frac N{1+c}+R_{N,\alpha}(-c)\right\}\,dc.
\end{equation}
The integral converges absolutely for each fixed $N$.
\end{proposition}
\begin{proof}
Write $Z[w]$ and $Z[w_c]$ for the partition functions with the
same normalization. Heine's integral representation at $z=-c$ gives
\[
 (-1)^N p_{N,c}(-c)=\frac{Z[w]}{Z[w_c]}>0.
\]
Both integrals are strictly positive: on ordered positive variables,
$\Delta(X)\Delta(X^\theta)>0$. Thus the denominator below is nonzero
without requiring a theorem about all the polynomial zeros.
Cancellation of the modified normalization in Heine's formula yields
\[
 \frac{p_{N,c}(z)}{p_{N,c}(-c)}
 =\mathbb E_w\prod_{j=1}^N\frac{X_j-z}{X_j+c}.
\]
At $z=-c$ the product is one. Its spatial derivative, holding $c$
fixed, is $-\sum_j(X_j+c)^{-1}$, proving
\eqref{eq:weak-resolvent-polynomial}.
Finally,
\[
 \log x=\int_0^\infty
 \left\{\frac1{1+c}-\frac1{x+c}\right\}\,dc.
\]
For fixed $x$ the integrand has the constant sign of $x-1$, and its
absolute integral is $|\log x|$.  Integrability of the logarithmic
moment justifies Fubini and proves \eqref{eq:weak-derivative-resolvent}.
\end{proof}

In rapidity variables $c=e^{2Bd}$, define the smoothed counting function
\begin{equation}\label{eq:weak-softcount}
 A_N(d)=-cR_N(-c)
 =\mathbb E\sum_{j=1}^N
 \frac1{1+e^{2B(\lambda_j-d)}}.
\end{equation}
The logistic density has mean zero, so
\begin{equation}\label{eq:weak-smoothed-derivative}
 F_N'(\alpha)=Q\int_{\mathbb R}
 \{N\mathbf1_{d>0}-A_N(d)\}\,dd.
\end{equation}
This is also obtained from \eqref{eq:weak-derivative-resolvent}: the
difference between $Nc/(1+c)$ and $N\mathbf1_{d>0}$ has zero integral
in $d$.  The identity needs no asymptotic density calculation.

\begin{proposition}[A sufficient off-real RH error]
Suppose that, for $|d|\le R_N$ and $c=e^{2Bd}$,
\[
 p_{N,c}(z)=p^{\rm app}_{N,c}(z)(1+E_{N,c}(z))
\]
holds on the disk $|z+c|\le\delta c$, where $0<\delta<1$ is fixed,
the model has no zero there, $E_{N,c}$ is holomorphic, and
$\sup|E_{N,c}|\le\varepsilon_N(d)\le\tfrac12$.
Then
\[
 |c(R_N(-c)-R_N^{\rm app}(-c))|
 \le\frac2\delta\varepsilon_N(d),\qquad
 R_N^{\rm app}(-c)=\left.\partial_z\log p^{\rm app}_{N,c}(z)
 \right|_{z=-c}.
\]
Consequently the error in the truncated integral representation of $F_N'(\alpha)$
\eqref{eq:weak-smoothed-derivative} is at most
$2Q\delta^{-1}\int_{-R_N}^{R_N}\varepsilon_N(d)\,dd$.
In particular, an $O(N^{-\gamma})$ bound uniform on an interval of
length $O(\log N)$ gives $O(N^{-\gamma}\log N)$.
\end{proposition}
\begin{proof}
The resolvent difference is $E_{N,c}'(-c)/(1+E_{N,c}(-c))$.
Cauchy's estimate bounds the numerator by
$\varepsilon_N(d)/(\delta c)$ and the denominator below by $1/2$.
\end{proof}

\begin{lemma}[Exponential-moment bounds and tails of the derivative integral]
Assume $H'$ is bounded, $r_*>0$, and $\alpha$ ranges over a compact
interval.  For every fixed $t>0$ there is a constant $C_t$, independent
of $N$, such that
\[
 \mathbb E\sum_{j=1}^N e^{t|\lambda_j|}\le C_tN^{t+1}.
\]
In particular,
\begin{equation}\label{eq:weak-actual-tail}
 \int_{|d|>R}|A_N(d)-N\mathbf1_{d>0}|\,dd
 \le C N^{2B+1}e^{-2BR}.
\end{equation}
Thus $R=C_0\log N$ gives an $o(1)$ tail whenever
$C_0>1+(2B)^{-1}$.
\end{lemma}
\begin{proof}
Write $K(u)=A\coth(Au)+B\coth(Bu)$ and
$M_k=\mathbb E\sum_j\cosh(k\lambda_j)$.  The integration-by-parts (Schwinger--Dyson) identity
for a smooth test $g$ reads
\[
 r_*\mathbb E\sum_jg(\lambda_j)\sinh\lambda_j
 =\mathbb E\sum_j\{g'(\lambda_j)+(H'(\lambda_j)+\alpha Q)g(\lambda_j)\}
 +\mathbb E\sum_{i<j}(g(\lambda_i)-g(\lambda_j))K(\lambda_i-\lambda_j).
\]
Use $g=\tanh\lambda$.  Since
$|g(x)-g(y)|\le\min(2,|x-y|)$ and
$|K(u)|\le A+B+2/|u|$, each pair term is bounded by
$2(A+B)+2$.  Also $\sinh\lambda\tanh\lambda
=\cosh\lambda-\operatorname{sech}\lambda$.
It follows that $M_1\le C N^2$.

For an integer $k\ge1$, use $g=\sinh(k\lambda)$.  The elementary
identity
\[
 \frac{|\sinh(kx)-\sinh(ky)|}{\cosh(kx)+\cosh(ky)}
 =\tanh\left(\frac{k|x-y|}{2}\right)
\]
and $a\coth(a|u|)\le a+|u|^{-1}$ give
\[
 (g(x)-g(y))K(x-y)
 \le(A+B+k)\{\cosh(kx)+\cosh(ky)\}.
\]
Using $2\sinh\lambda\sinh(k\lambda)
=\cosh((k+1)\lambda)-\cosh((k-1)\lambda)$, the same integration-by-parts identity implies
$M_{k+1}\le M_{k-1}+C_kN M_k$.  Since $M_0=N$, induction gives
$M_k\le C_kN^{k+1}$.  Interpolation with respect to the expected
counting measure divided by $N$ proves the assertion for every real
$t>0$.  All integrations by parts can first be performed with smooth
cutoffs; the confining weight supplies finite moments at fixed $N$.

For $d>R$, $N-A_N(d)\le e^{-2Bd}\mathbb E\sum_j e^{2B\lambda_j}$;
for $d<-R$, $A_N(d)\le e^{2Bd}\mathbb E\sum_j e^{-2B\lambda_j}$.
Integrating and using the moment bound gives
\eqref{eq:weak-actual-tail}.
\end{proof}

\subsection{A differential identity for the central Fredholm determinant}
We work at finite $N$ with the central contour truncated by the smooth
cutoff, keeping its two components at a fixed height $0<h_0<\phi/2$.
The operator acts on a compact contour; it is not assumed to have
finite rank. Here ``central'' refers to the part of the lens where
the jump has a nonzero limiting amplitude. The integration and
partner-evaluation contours stay separated from the negative
observation axis.

For a trace-norm differentiable family $K_\alpha$, with
$I-K_\alpha$ invertible, the standard Fredholm determinant identity is
\[
 \partial_\alpha\log\det(I-K_\alpha)
 =-\Tr\bigl((I-K_\alpha)^{-1}\partial_\alpha K_\alpha\bigr).
\]
The proposition below represents this trace as an integral over the
negative observation parameter. That representation is specific to
the central operator considered here.

Let $z(v)$ be an integration-contour point and $p(v)$ its partner
evaluation point. These are distinct from the lifted insertion point
$a=I_1(-c)$. The notation includes a sum over the two contour
components. Write
\[
 K(t,v)=\frac{k(v)z'(v)}{2\pi i\{z(v)-p(t)\}},\qquad
 \mathcal R=(I-K)^{-1}.
\]
Assume $K$ is trace class and $I-K$ is invertible.  Put
$a_0=-1-\delta_N$, and suppose all integration and partner-evaluation points avoid
$(-\infty,a_0]$.  Logarithms below have that cut and agree with the
branches of the exact $\alpha$-dependent scalar factor. For $a<a_0$, set
\[
 r_a(t)=\frac1{p(t)-a},\qquad
 \ell_a(f)=\int\frac{k(v)z'(v)}{2\pi i\{z(v)-a\}}f(v)\,dv,
 \qquad E(a)=\ell_a\mathcal R r_a.
\]
The calculation for the modified weight in
Section~\ref{sec:source-dressing} identifies $E(a)$ with the spatial
logarithmic derivative of its central factor, evaluated at the lifted
insertion point $a$.  That identification is a separate algebraic dressing lemma;
the trace identity here needs only the displayed definitions.

\begin{proposition}[Exact central logarithmic derivative]
Suppose $K=K_\alpha$ is differentiable in trace norm and the dependence
of its jump amplitude on $\alpha$ is
\begin{equation}\label{eq:weak-alpha-amplitude}
 \partial_\alpha k(v)
 =\frac b2\{\log(p(v)-a_0)-\log(z(v)-a_0)\}k(v).
\end{equation}
Then
\begin{equation}\label{eq:weak-core-trace}
 \frac b2\int_{-\infty}^{a_0}E(a)\,da
 =\partial_\alpha\log\det(I-K).
\end{equation}
\end{proposition}
\begin{proof}
The compact contours have positive distance from the evaluation ray
near its finite endpoint.  At $a=-\infty$,
$\|r_a\|=O(|a|^{-1})$ and $\|\ell_a\|=O(|a|^{-1})$.
Hence the integral of the rank-one operators $r_a\otimes\ell_a$
converges in trace norm.  Its kernel is obtained from
\[
 \int_{-\infty}^{a_0}\frac{da}{(z-a)(p-a)}
 =\frac{\log(z-a_0)-\log(p-a_0)}{z-p}.
\]
This scalar identity holds first for positive real $z-a_0,p-a_0$
and then by analytic continuation on the cut plane.  Thus, writing
$L=M_{\log(p-a_0)}$, the integrated operator $W$ has kernel
\[
 W(t,v)=K(t,v)\{\log(z(v)-a_0)-\log(p(t)-a_0)\}.
\]
Equation \eqref{eq:weak-alpha-amplitude} gives the operator identity
\[
 \frac b2W=-\partial_\alpha K+\frac b2[K,L].
\]
The multiplier $L$ is bounded on the compact contour.  Since
$\mathcal R$ commutes with $K$, cyclicity for a trace-class factor
gives $\Tr(\mathcal R[K,L])=0$.  Trace-norm Fubini therefore yields
\[
 \frac b2\int E(a)\,da
 =-\Tr(\mathcal R\partial_\alpha K)
 =\partial_\alpha\log\det(I-K),
\]
as asserted.
\end{proof}

For the exact uniformization
\[
 J_N(s)=c_{1,N}(s+1+\delta_N)
 \left(\frac{s+1}{s}\right)^{1/\theta},
\]
the negative real inverse point $a=I_{1,N}(-c)$ runs from $a_0$ to
$-\infty$ as $c$ runs from zero to infinity, and
$I_{1,N}'(-c)\,dc=-da$.  Thus the core contribution to the
resolvent integral for $F_N'(\alpha)$ is precisely the left side of
\eqref{eq:weak-core-trace}. The exact dependence on $\alpha$ of the
band-factor ratio is
\[
 \left(\frac{p(v)+1+\delta_N}{z(v)+1+\delta_N}\right)^{\alpha b/2},
\]
which proves \eqref{eq:weak-alpha-amplitude} with the stated branches.
There is no rationality assumption on $\theta$ in this calculation.

\subsection{The outer contribution to the derivative}
Compare the resolvent models at $\alpha$ and $0$; the
equilibrium exponential and the scalar factor associated with the insertion cancel.
The remaining outer $\alpha$-dependent factor is exactly
\[
 \frac{P_{\alpha}^{\rm out}(a)}{P_0^{\rm out}(a)}
 =\left(\frac a{a+1}\right)^{\alpha b/(2\theta)}.
\]
Consequently its contribution to the model expression for
$F_N'(\alpha)-F_N'(0)$ is
\begin{align*}
 \frac b2\int_0^\infty
 \left.\partial_z\log
 \frac{P_{\alpha}^{\rm out}(I_1(z))}
 {P_0^{\rm out}(I_1(z))}\right|_{z=-c}\,dc
 &=\frac{\alpha b^2}{4\theta}
 \int_{-\infty}^{a_0}\left(\frac1a-\frac1{a+1}\right)\,da\\
 &=\frac\alpha4\log\frac{1+\delta_N}{\delta_N}
 =\alpha\chi_N.
\end{align*}
Here $b^2=\theta$. Therefore the RH model for the modified weight gives the
following expression for $F_N'(\alpha)-F_N'(0)$:
\[
 \alpha\chi_N+
 \partial_\alpha\log\det(I-K_{N,\alpha})
 -\left.\partial_\alpha\log\det(I-K_{N,\alpha})\right|_{\alpha=0}.
\]
The $\alpha$-derivative of the logarithm of the finite-$N$ central determinant,
evaluated at $\alpha=0$, is retained at finite $N$; it tends to zero
by convergence to the even limiting determinant. The full comparison in Section~\ref{sec:global-source-comparison}
supplies the integrated error $O(N^{-1}\log N)$.

\section{The modified central problem and an integrable resolvent bound}
\label{sec:source-dressing}

Write the exact central scalar problem in the form
\[
 C(s)=1+\sum_\tau\int
 \frac{a_\tau(v)z_\tau'(v)C(p_\tau(v))}
 {2\pi i\{z_\tau(v)-s\}}\,dv,
\]
with the cutoff included in $a_\tau$. Its partner-value operator is
\[
 K_{\sigma\tau}(t,v)=
 \frac{a_\tau(v)z_\tau'(v)}
 {2\pi i\{z_\tau(v)-p_\sigma(t)\}}.
\]
All algebra below can first be performed on compact, separated contours,
where the multiplication gauges are bounded and boundedly invertible.
The limiting identities follow either by trace-norm approximation or by
cancellation of the gauge factors in Fredholm integrands.

For the modified weight $w(x)/(x+c)$, let $a=I_1(-c)$ be the outer
preimage of the negative observation point. The exact scalar Szego
factors are
\[
 D^{\rm out}_{c,-}(s)=\frac{J(s)+c}{c_1(s-a)},\qquad
 D^{\rm in}_{c,-}(s)=\frac1{c_1(s-a)}.
\]
The apparent outer singularity at $s=a$ is removable. Their ratio on
the band is $(J(s)+c)^{-1}$, and the outer factor tends to one at
infinity. The corresponding central coefficient is multiplied by
$(z-a)/(p-a)$.

\begin{proposition}[Logarithmic derivative at the lifted insertion point]
Let $R_a(t)=(p_\sigma(t)-a)^{-1}$ and
\[
 \ell_\infty f=\sum_\tau\int
 \frac{a_\tau(v)z_\tau'(v)}{2\pi i}f_\tau(v)\,dv,
 \qquad D_a=(I-K)^{-1}R_a.
\]
Suppose the base and modified central problems are solvable.
Define the global meromorphic continuation
\[
 D_a(s)=\frac1{s-a}+\sum_\tau\int
 \frac{a_\tau(v)z_\tau'(v)D_{a,\tau}(v)}
 {2\pi i\{z_\tau(v)-s\}}\,dv.
\]
Then, with $\delta_a=1-\ell_\infty D_a$,
\[
 C_{c,-}(s)=\frac{(s-a)D_a(s)}{\delta_a},\qquad
 C_{c,-}(a)=\frac1{\delta_a},
\]
and the spatial logarithmic derivative at the lifted insertion point is
\begin{equation}\label{eq:source-logderivative-cancellation}
 \frac{C_{c,-}'(a)}{C_{c,-}(a)}
 =\sum_\tau\int
 \frac{a_\tau(v)z_\tau'(v)D_{a,\tau}(v)}
 {2\pi i\{z_\tau(v)-a\}}\,dv
 =:\Psi(a).
\end{equation}
In particular the normalization denominator $\delta_a$ cancels from
this logarithmic derivative.
\end{proposition}
\begin{proof}
Set $G(s)=C_{c,-}(s)/(s-a)$. Multiplying the modified integral equation
by $(p_\sigma(t)-a)^{-1}$ and using
\[
 \frac{z-a}{(p-a)(z-p)}=\frac1{z-p}+\frac1{p-a}
\]
gives $(I-K)G=R_a(1+\ell_\infty G)$. Therefore
$G=D_a/\delta_a$. The displayed continuation of $D_a$ has residue one
at $a$, and its coefficient of $s^{-1}$ at infinity is $\delta_a$.
This proves the formula for the modified central factor and its normalization. Write
$D_a(s)=(s-a)^{-1}+d_a(s)$ near $a$, with $d_a$ analytic.
Then $C_{c,-}(s)=[1+(s-a)d_a(s)]/\delta_a$, so its logarithmic
derivative at $a$ equals $d_a(a)$, which is precisely
\eqref{eq:source-logderivative-cancellation}.
\end{proof}

The determinant lemma also gives
\[
 \delta_a=\frac{\det(I-K_{c,-})}{\det(I-K)},
\]
since the gauged modified kernel is $K+R_a\otimes\ell_\infty$.
This identity is useful for nonvanishing, but a lower bound on
$C_{c,-}(a)$ is not required to estimate
\eqref{eq:source-logderivative-cancellation}.

We now prove an integrated estimate for the exact finite central model.
This concerns the central model only; comparison with the actual
polynomial RH solution, including Airy disks and omitted wings, is
proved in Section~\ref{sec:global-source-comparison}.

Suppose $\mathfrak{s}_N\to0$ and write
\[
 z_{\tau,N}(v)=-1+\mathfrak{s}_N Z_{\tau,N}(v),\qquad
 p_{\sigma,N}(t)=-1+\mathfrak{s}_N P_{\sigma,N}(t).
\]
Fix a lens height $0<h_0<\phi/2$. The limiting
integration and partner-evaluation rays then stay a positive angle from the negative
real axis. Suppose the map estimates imply
\begin{equation}\label{eq:negative-source-separation}
 |Z_{\tau,N}(v)+\rho|\ge c(e^{2v}+\rho),\qquad
 |P_{\sigma,N}(t)+\rho|\ge c(e^{2t}+\rho),\qquad
 |Z'_{\tau,N}(v)|\le Ce^{2v}
\end{equation}
on the retained core, uniformly for $\rho\ge0$, together with local
convergence of these maps and derivatives. Let $\chi_N^{\rm cut}$ be the core
cutoff and suppose the uncut jump amplitudes obey
\[
 |a_{\tau,N}(v)|\le C e^{C|v|}e^{-c_1\cosh v},
\]
and converge locally uniformly. Choose
$q_0(t)=e^{-c_0\cosh t}$, $0<c_0<c_1$. Redistributing the two cutoff
factors gives the balanced operator $\widetilde K_N$ and vectors
\begin{align*}
 f_{N,\rho,\sigma}(t)
 &=\frac{q_0(t)\chi_N^{\rm cut}(t)}{P_{\sigma,N}(t)+\rho},\\
 \ell_{N,\rho}u
 &=\sum_\tau\int
 \frac{\chi_N^{\rm cut}(v)a_{\tau,N}(v)Z'_{\tau,N}(v)}
 {2\pi i q_0(v)\{Z_{\tau,N}(v)+\rho\}}u_\tau(v)\,dv.
\end{align*}
The scaled regular part is exactly
\begin{equation}\label{eq:scaled-source-resolvent}
 \mathfrak{s}_N\Psi_N(-1-\mathfrak{s}_N\rho)
 =\ell_{N,\rho}(I-\widetilde K_N)^{-1}f_{N,\rho}.
\end{equation}

\begin{proposition}[Uniform integrable convergence in the insertion parameter]
Assume the preceding bounds, $\widetilde K_N\to\widetilde K_\infty$
in operator norm, and invertibility of $I-\widetilde K_\infty$.
Then there exists $\epsilon_N\to0$ such that
\begin{equation}\label{eq:source-uniform-integrable}
 \left|\mathfrak{s}_N\Psi_N(-1-\mathfrak{s}_N\rho)
       -\Psi_\infty(-\rho)\right|
 \le\frac{\epsilon_N}{(1+\rho)^2},\qquad\rho\ge0.
\end{equation}
Also each scaled regular part is bounded by $C(1+\rho)^{-2}$.
The conclusions are uniform in additional parameters whenever all
hypotheses and inverse bounds are uniform in those parameters.
\end{proposition}
\begin{proof}
The elementary inequality
\[
 \frac{1+\rho}{e^{2t}+\rho}\le\max\{1,e^{-2t}\}
\]
and \eqref{eq:negative-source-separation} give
\[
 \|f_{N,\rho}\|_2\le\frac C{1+\rho},\qquad
 \|\ell_{N,\rho}\|\le\frac C{1+\rho}.
\]
Indeed the respective dominating functions are constants times
$q_0(t)\max(1,e^{-2t})$ and
$e^{C|t|}e^{-(c_1-c_0)\cosh t}\max(1,e^{2t})$.
They belong to $L^2$ and do not depend on $N$ or $\rho$.

On every bounded rapidity interval, the rescaled rational factors
converge uniformly for the entire range $\rho\ge0$. For example,
\[
 (1+\rho)\left|\frac1{P_N(t)+\rho}-\frac1{P_\infty(t)+\rho}\right|
 \le C_M|P_N(t)-P_\infty(t)|,\qquad |t|\le M,
\]
with $C_M$ independent of $\rho$. The same applies to
$Z_N'(t)/(Z_N(t)+\rho)$. Splitting at $|t|=M$, using the displayed
$L^2$ envelopes on the tails and local convergence on the compact
interval, proves
\[
 \sup_{\rho\ge0}(1+\rho)\|f_{N,\rho}-f_{\infty,\rho}\|_2\to0,
 \qquad
 \sup_{\rho\ge0}(1+\rho)\|\ell_{N,\rho}-\ell_{\infty,\rho}\|\to0.
\]
The resolvent identity gives uniform inverse bounds and convergence of
the inverses. Expanding the difference in
\eqref{eq:scaled-source-resolvent} into its three factors now proves
\eqref{eq:source-uniform-integrable}.
\end{proof}

For the Lukyanov outer map, $a=I_1(-c)$ runs exactly through
$(-\infty,-1-\delta_N)$ as $c$ ranges through $(0,\infty)$.
Thus $\rho$ runs through $(\delta_N/\mathfrak{s}_N,\infty)$; if
$\delta_N=O(\mathfrak{s}_N^2)$, the omitted interval at the limiting lower
endpoint contributes $O(\mathfrak{s}_N)$. Integrating
\eqref{eq:source-uniform-integrable} therefore proves
\[
 \int_{-\infty}^{-1-\delta_N}\Psi_N(a)\,da
 -\int_{-\infty}^0\Psi_\infty(a)\,da=o(1).
\]
The interior-height restriction is important: at $b=1$ and boundary
height $h_0=\phi/2$, the limiting integration rays lie on the negative
real axis, where the insertion points are evaluated. The integral over
the insertion parameter is therefore taken before the limiting
Fredholm contour deformation.

\section{The central limit and uniform inversion at finite \texorpdfstring{$N$}{N}}
\label{sec:newcorelimit}
This section records the identification and the uniform estimates used
by the comparisons for the modified weights. Write
$a_\alpha=(1-\theta)/2+\alpha b/2$ in this section. The power part of
the scalar band factor has the exact branches
\[
 D^{\rm out}(s)=\left(\frac{s}{s+1}\right)^{a_\alpha/\theta},\qquad
 D^{\rm in}(s)=\theta^{-1}[c_{1,N}(s+1+\delta_N)]^{a_\alpha}.
\]
Thus in the upper or lower lip coefficient its ratio is
\begin{equation}
 \omega_\alpha(x)^{-1}\frac{D^{\rm in}(p)}{D^{\rm out}(z)}
 =\left(\frac{p+1+\delta_N}{z+1+\delta_N}\right)^{a_\alpha}
 e^{\mathcal C_{H,N}(p)-\mathcal C_{H,N}(z)-H(\lambda)}.
 \label{eq:newpowercancel}
\end{equation}
Here $\mathcal C_{H,N}$ is the scalar Cauchy transform on the band
preimage contour. For $|t|\leq L_N/3+1$ and a fixed height $h_0$,
the two inverse-map estimates are
\begin{align}
 z_{\pm,N}(t)+1&=\mathfrak{s}_Ne^{2t\pm i(2h_0+\phi)}
 [1+O(\mathfrak{s}_N\cosh 2t)],\notag\\
 p_{\pm,N}(t)+1&=\mathfrak{s}_Ne^{2t\pm i(2h_0-\phi)}
 [1+O(\mathfrak{s}_N\cosh 2t)].\label{eq:newmapcore}
\end{align}
They hold through fixed holomorphic derivatives, by the implicit
equation for $J_N$. The square-root ratio is
$-e^{i\phi}(1+o(1))$ on the upper lip and
$e^{-i\phi}(1+o(1))$ on the lower. Combining this with
\eqref{eq:newpowercancel} and the central phase fixes the two oriented
coefficients as
\[
 \varepsilon e^{-i\phi}a_{+,h_0}(t),\qquad
 -\varepsilon e^{i\phi}a_{-,h_0}(t),\qquad \varepsilon=(-1)^N,
\]
where $a_{-,h_0}=\overline{a_{+,h_0}}$ for real data and
\[
 a_{+,h_0}(t)=e^{-i\pi\alpha/Q}
 \exp\{i r_*\sin\phi\sinh(t+ih_0)
 +\mathcal C_H(e^{2t+i(2h_0-\phi)})
 -\mathcal C_H(e^{2t+i(2h_0+\phi)})-H(t+ih_0)\}.
\]
After multiplication of the partner unknown by $e^t$ and constant
component signs, the limiting two-contour kernel is
\begin{equation}
 K_{\infty,h_0}(t,v)=\frac\varepsilon{2\pi i}
 \begin{pmatrix}
 -a_{+,h_0}(v)/\sinh(t-v-i\phi)&a_{-,h_0}(v)/\sinh(t-v+2ih_0)\\
 -a_{+,h_0}(v)/\sinh(t-v-2ih_0)&a_{-,h_0}(v)/\sinh(t-v+i\phi)
 \end{pmatrix}.
 \label{eq:newunfolded}
\end{equation}
To describe the boundary-height representation, introduce the limiting
coordinates
\[
 Z_\pm(t)=e^{2t\pm i(2h_0+\phi)},\qquad
 P_\pm(t)=e^{2t\pm i(2h_0-\phi)}.
\]
They are the limits of $(z_{\pm,N}+1)/\mathfrak{s}_N$ and
$(p_{\pm,N}+1)/\mathfrak{s}_N$, respectively. At an interior height
$0<h_0<\phi/2$, the partner-evaluation contours are distinct. At
the boundary height $h_0=\phi/2$,
\[
 P_+(t)=P_-(t)=e^{2t},\qquad Z_\pm(t)=e^{2t\pm2i\phi}.
\]
Thus the partner components can be identified. The integration rays
coincide only when $b=1$, for which $\phi=\pi/2$ and
$Z_+(t)=Z_-(t)=-e^{2t}$. The original lines
$\lambda=t\pm i\phi/2$ remain distinct in the $\lambda$-plane.

Correspondingly, the two rows of \eqref{eq:newunfolded} agree at the
boundary height. Writing the block operator as
\[
 K_{\infty,\phi/2}=\begin{pmatrix}K_+&K_-\\K_+&K_-\end{pmatrix},
\]
define $\mathsf A f=(f,f)$ and
$\mathsf B(f_+,f_-)=K_+f_++K_-f_-$. Then
$\mathsf A\mathsf B=K_{\infty,\phi/2}$ and
$\mathsf B\mathsf A=K_++K_-$. In the trace-class realization,
Sylvester's identity gives
\[
 \det(I-K_{\infty,\phi/2})
 =\det(I-K_+-K_-)=\tau_\varepsilon(\alpha).
\]
The descriptive terms ``unfolded'' and ``folded'' mean, respectively,
the two-component representation at interior height and the scalar
representation after boundary deformation and identification of the
partner components. We use these explicit descriptions below.
The block reduction is algebraic. Relating it to a determinant
initially defined at $h_0<\phi/2$ also requires an admissible analytic
contour deformation; the deformation and operator estimates are
treated separately below.

\paragraph{Contour deformation of the limiting determinant.}
The following argument concerns the limiting operator and is independent
of the finite-$N$ comparison. Fix $0<h_*<\phi/2$, put
$\delta_h=\phi/2-h$ for $h\in[h_*,\phi/2]$, and set
\[
 \Gamma_+(h)=\mathbb R-i\delta_h,\qquad
 \Gamma_-(h)=\mathbb R+i\delta_h.
\]
For real $\alpha$ define
\[
 A_+(z)=e^{-i\pi\alpha/Q}g(z+i\phi/2),\qquad
 A_-(z)=\overline{A_+(\bar z)},\qquad
 m_{\tau,d}(z)=\frac{e^{2z+2\tau i\phi}+e^{2d}}
                         {e^{2z}+e^{2d}},\quad \tau=\pm1.
\]
The base problem is obtained by setting $m_{\tau,d}=1$. The assumed
identity for $\mathcal A_H$, applied by the identity theorem to the
holomorphic Cauchy-factor expression, gives
$a_{+,h}(t)=e^{-i\pi\alpha/Q}g(t+ih)$ and its reflected counterpart.
On $h\in[h_*,\phi/2]$, the arguments of the two Cauchy transforms
stay at positive angular separation from their integration rays.
Their integrands are uniformly bounded by a constant times $|H(v)|$,
so the explicit amplitude formula gives
\[
 |a_{+,h}(t)|\leq C\exp\{-r_*\sin\phi\sin(h_*)\cosh t\}.
\]
The same bound with slightly smaller decay constants holds through
fixed derivatives, by Cauchy estimates on slightly larger closed
strips. This verifies the decay condition in the lemma below directly
from Assumption~\ref{ass:H}.
Consequently the substitutions $u=t-\sigma i\delta_h$ and
$v=s-\tau i\delta_h$ express the two-contour kernel, including the
insertion multiplier, as
\[
 \mathfrak K_{\sigma\tau}(u,v)
 =-\frac{\varepsilon\tau}{2\pi i}
   \frac{A_\tau(v)m_{\tau,d}(v)}{\sinh(u-v-\tau i\phi)},
 \qquad u\in\Gamma_\sigma(h),\quad v\in\Gamma_\tau(h).
\]
We use a common balancing function $b(t)=e^{-\gamma\cosh t}$,
with $\gamma>0$ smaller than the uniform amplitude-decay constant,
and denote the resulting trace-class realization on
$L^2(\mathbb R,dt)\oplus L^2(\mathbb R,dt)$ by
$\widetilde K_{h,d}$. Explicitly its kernel is the displayed kernel
in real parameters multiplied by $b(t)/b(s)$.

\begin{lemma}[Limiting determinant deformation and uniform inversion]
\label{lem:limiting-deformation}
Suppose the amplitudes and their fixed derivatives have a common
bound $C_j e^{C_j|t|}e^{-c\cosh t}$ on the closed strips swept out by
$\Gamma_\pm(h)$, and the analytic conditions above hold. Then
$\det(I-z\widetilde K_{h,d})$ is independent of
$h\in[h_*,\phi/2]$ for every $z\in\mathbb C$ and real $d$.
If $I-\widetilde K_{\phi/2,d}$ is invertible for every real $d$ and
for the two limits $d=\pm\infty$, then $I-\widetilde K_{h_*,d}$ is
invertible and its inverse is uniformly bounded in
$d\in[-\infty,+\infty]$. The same conclusion holds uniformly for
$\alpha$ in a compact real interval when the boundary invertibility
and amplitude estimates hold there.
\end{lemma}
\begin{proof}
The balanced kernels and their fixed derivatives have integrable
exponential envelopes in both real variables. Applying
$1+t^2-\partial_t^2$ to the kernel produces a Hilbert--Schmidt
operator. Factoring through the Hilbert--Schmidt inverse of this
harmonic oscillator proves trace class. The same bounds give
trace-norm continuity by dominated convergence.
During independent contour deformations, the imaginary parts of all
integration variables lie in $[-\delta_{h_*},\delta_{h_*}]$.
Thus the shifted imaginary differences in the Cauchy denominators
have absolute values between $2h_*$ and $2\phi-2h_*$, an interval
strictly inside $(0,\pi)$. No denominator zero is crossed, and
$|\sinh(u-v-\tau i\phi)|^{-1}$ is bounded by
$C/\cosh(\Re u-\Re v)$. The insertion multiplier has poles only on
$\Im z=\pi/2\pmod\pi$, outside the swept strips, and
\[
 |m_{\tau,d}(z)|\leq\sec(\delta_{h_*})
\]
there, uniformly in $d$. The same assertion for fixed derivatives
follows from holomorphy on a slightly larger strip.

The balancing factors cancel in every Fredholm minor. After extracting
one amplitude from each column, Hadamard's inequality bounds the
absolute value of an $n\times n$ minor by
$n^{n/2}C^n\prod_j e^{-c'\cosh(\Re v_j)}$. Summation over the
$2^n$ component choices and division by $n!$ give a summable majorant
for the Fredholm series, locally uniformly in $z$ and uniformly in
$h,d$. Cauchy's theorem can therefore be applied successively to
each integration variable; the vertical closing segments vanish by
the amplitude decay. Every term is unchanged when both contours
are moved to the real line. This proves determinant invariance.
The two rows then agree, and the block reduction above identifies
the boundary determinant with the scalar Fredholm determinant.
The common-envelope realization and the realization balanced by
$\sqrt v$ have identical Fredholm minors, since their balancing
factors cancel. Their determinants therefore agree; no bounded
similarity by the inverse balancing multiplier is asserted.

For a trace-class operator, a nonzero Fredholm determinant is
equivalent to invertibility of $I-\widetilde K_{h,d}$. At fixed
$h_*$, the insertion multipliers tend pointwise to $1$ and
$e^{2\tau i\phi}$ as $d\to+\infty$ and $d\to-\infty$,
respectively, and are uniformly bounded. Multiplication of a fixed
trace-class operator by these multipliers is continuous in trace
norm, including the two endpoints: approximate the operator by
finite-rank operators and apply dominated convergence to their
vectors. Continuity of inversion on this compact parameter set
therefore gives the uniform inverse bound. Including a compact real
interval of $\alpha$ gives the last assertion. No equality of
operator norms on different contour realizations is used.
\end{proof}

We give the uniform finite-$N$ estimates, since compact pointwise
convergence alone would be insufficient. For fixed $h_0>0$, the
integration--evaluation angular differences are $\pm2\phi$ and $\pm4h_0$.
Hence
\[
 |z_{\tau,N}(v)-p_{\sigma,N}(t)|
 \geq c\mathfrak{s}_N(e^{2t}+e^{2v}),\qquad
 |z'_{\tau,N}(v)|\leq C\mathfrak{s}_Ne^{2v},
\]
on the retained window, including analogous differentiated bounds.
This follows first for rays from
$|Re^{iu}-Se^{iv}|\geq|\sin((u-v)/2)|(R+S)$ and then from
\eqref{eq:newmapcore}; its maximum relative error is $O(N^{-4/3})$.

For the $H$-Cauchy factor put $w=\xi+1$ on a band arc. Direct
differentiation of the map gives
\[
 \frac{J_N'(\xi)}{J_N(\xi)}
 =\frac{\theta w^2-(\theta+1)w-\delta_N}
 {\theta w(w-1)(w+\delta_N)},\qquad
 \frac{\dd\xi}{\dd v}=\frac{2B}{J_N'(\xi)/J_N(\xi)}.
\]
On these arcs $|w(v)|\asymp\mathfrak{s}_Ne^{2v}$ and
\[
 \frac{|w'(v)|}{|w(v)|}
 \leq C[1+(L_N+v)^{-1/2}+(L_N-v)^{-1/2}].
\]
The latter bound is the simple-critical-point estimate; the former
uses $|w|\geq c\delta_N$ in the exact map. Split the integral into
$|v|\leq2L_N/3$ and the endpoint wings. On the central part the
angular separation bounds the denominator by
$c\mathfrak{s}_N(e^{2v}+e^{2t})$. On each wing $|w(v)|$ is respectively
much larger or much smaller than the observation scale. Multiplication
by the exponentially decreasing $H(v)$ gives an integrable bound,
also for fixed observation derivatives. The endpoint square-root
singularities are integrable. This proves uniform boundedness of the
finite Cauchy differences on the retained contours and their local
convergence to \eqref{eq:newCH}.

The phase estimate gives
\[
 \Re[-N\Phi_+(t+ih_0)]
 \leq-r_*\sin\phi\sin h_0\cosh t+CN^{-1}\cosh2t
 \leq-c\cosh t
\]
on the retained core. Every fixed derivative of a jump amplitude is
therefore bounded by $C_j e^{C_j|t|}e^{-c\cosh t}$.
Redistribute the two cutoff factors and use the balancing
$q_0(t)=e^{-c_0\cosh t}$, $0<c_0<c$. The balanced kernel and its
mixed derivatives are dominated by
\[
 C_m e^{C_m(|t|+|v|)}
 e^{-c_0\cosh t-(c-c_0)\cosh v}.
\]
Local convergence and this common envelope prove trace-norm
convergence by the harmonic-oscillator factorization described in the
completion section. They also prove convergence of the derivatives
with respect to $\alpha$.
For the modification $w(x)/(x+c)$ the exact ratio and effective-center lemma
below give the same conclusions uniformly for all $c>0$.

Lemma~\ref{lem:limiting-deformation} reduces invertibility on an
interior contour to nonvanishing of the scalar boundary determinant.
The Hardy-space contraction and positivity argument in the next
section supplies that input, including both limiting insertion
centers. Its compactness conclusion then gives the uniform interior
inverse. The trace-norm comparison above transfers this inverse bound
to the retained finite-$N$ central operators. This sequence uses no
global comparison with the exact polynomial RH solution.

\section{Uniform central inversion for the modified weight}
\label{sec:inverse-source-Hardy}

The logarithmic-resolvent route needs only the modified weight
$w(x)/(x+c)$, $c>0$.  Its limiting central operator admits a uniform
inverse for every insertion position.  This statement is stronger than
pointwise invertibility and uses the analytic phase of the physical
amplitude, not only its absolute value.

Put
\[
 \phi=\frac\pi{1+b^2},\quad h=\frac\phi2,\quad a_0=\frac\pi2,
 \quad y=a_0-h,\quad d=\frac{\log c}{2B}.
\]
Let $g$ be bounded and holomorphic on $0<\Im z<a_0$, with boundary bounds
\[
 |g(t)|\leq1,\qquad |g(t+ia_0)|\leq\rho<1,
\]
and suppose $a(t)=e^{-i\pi\alpha/Q}g(t+ih)$ has an integrable,
square-integrable envelope.  The bare and physical TBA amplitudes have
these properties, with $\rho=e^{-r_*\sin\phi}$.
Define the strip Blaschke factor
\[
 B_y(z)=\frac{e^{2z}-e^{2iy}}{e^{2z}-e^{-2iy}},\qquad
 g_d(z)=e^{2i\phi}B_y(z-d)g(z).
\]
The function $B_y$ is holomorphic in the strip, has one zero at $iy$,
has modulus one on both horizontal boundaries, and has modulus at most
one inside.  Hence $g_d$ has the same two boundary bounds as $g$,
uniformly for every real $d$.

At the central evaluation line,
\begin{equation}
 a_d(t):=e^{-i\pi\alpha/Q}g_d(t+ih)=a(t)M_d(t),\qquad
 M_d(t)=\frac{1+e^{2(t-d)+2i\phi}}{1+e^{2(t-d)}}.
 \label{eq:inverse-source-M}
\end{equation}
In particular $|M_d(t)|\leq1$.  This is the exact phase appropriate
to division by $x+c$: it tends to $1$ as $t-d\to-\infty$ and to
$e^{2i\phi}$ as $t-d\to+\infty$.

Let $Q_\phi,Q_{\pi-\phi}$ be convolution operators with Fourier
multipliers
\[
 q_\phi(k)=\frac{e^{\phi k}}{1+e^{\pi k}},\qquad
 q_{\pi-\phi}(k)=\frac{e^{(\pi-\phi)k}}{1+e^{\pi k}},
\]
and set
\[
 L_d=Q_\phi M_{a_d}+Q_{\pi-\phi}M_{\overline{a_d}}.
\]

\begin{proposition}[Contraction uniform in the insertion center]
For real $\alpha$ and every real $d$,
\begin{equation}
 \|L_d\|\leq c_{\phi,\rho}<1,\qquad
 c_{\phi,\rho}^2=
 \sup_{k\in\mathbb R}
 \frac{\cosh((\pi-\phi)k)+\rho^2\cosh(\phi k)}
 {1+\cosh(\pi k)}.
 \label{eq:inverse-source-contract}
\end{equation}
Thus both $(I\pm L_d)^{-1}$ exist with norm at most
$(1-c_{\phi,\rho})^{-1}$, uniformly in $d$ and real $\alpha$.
\end{proposition}

\begin{proof}
Let $\mathscr H$ be the Hardy space on $0<\Im z<a_0$ whose squared norm is
\[
 \|F\|_{\mathscr H}^2=\|F(\cdot)\|_{L^2(dt)}^2
                      +\|F(\cdot+ia_0)\|_{L^2(dt)}^2.
\]
Here $g$ is bounded and holomorphic on this strip, as required in the
strip condition in Assumption~\ref{ass:H}(ii). Let $E_u$ be
restriction to height $u$, and put $y=a_0-h$. Plancherel's formula gives
\[
 E_hE_h^*=Q_{\pi-\phi},\qquad E_yE_y^*=Q_\phi.
\]
Define $\mathsf E:\mathscr H\oplus\mathscr H\to\mathscr H_0$ by
$\mathsf E(F_1,F_2)=E_hF_1+E_yF_2$. Its norm is at most one, since
\[
 \mathsf E\mathsf E^*=Q_{\pi-\phi}+Q_\phi,\qquad
 q_{\pi-\phi}(k)+q_\phi(k)
 =\frac{\cosh((\pi/2-\phi)k)}{\cosh(\pi k/2)}\leq1.
\]
Write $\widetilde g=e^{-i\pi\alpha/Q}g_d$ and
$\widetilde g^\sharp(z)=\overline{\widetilde g(\bar z+ia_0)}$.
Restriction commutes with multiplication, so
\[
 L_d^*=E_hM_{\widetilde g}E_h^*
       +E_yM_{\widetilde g^\sharp}E_y^*.
\]
The boundary modulus bounds for these two multipliers are $(1,\rho)$
and $(\rho,1)$, respectively. Hence
\[
 \|L_d^*f\|_2^2
 \leq \|M_{\widetilde g}E_h^*f\|_{\mathscr H}^2
      +\|M_{\widetilde g^\sharp}E_y^*f\|_{\mathscr H}^2
 \leq\frac1{2\pi}\int_{\mathbb R}m_{\phi,\rho}(k)|\widehat f(k)|^2\,dk,
\]
where
\[
 m_{\phi,\rho}(k)
 =\frac{e^{-\phi k}(1+\rho^2e^{-\pi k})
       +e^{-(\pi-\phi)k}(\rho^2+e^{-\pi k})}
       {(1+e^{-\pi k})^2}
 =\frac{\cosh((\pi-\phi)k)+\rho^2\cosh(\phi k)}
       {1+\cosh(\pi k)}.
\]
Indeed the lower and upper boundary transforms of $E_u^*f$ are
$e^{-uk}\widehat f(k)/(1+e^{-\pi k})$ and
$e^{-(u+a_0)k}\widehat f(k)/(1+e^{-\pi k})$. This proves the stated
bound. The identity
\[
 m_{\phi,\rho}(k)=
 \frac{\cosh((\pi/2-\phi)k)}{\cosh(\pi k/2)}
 -(1-\rho^2)\frac{\cosh(\phi k)}{1+\cosh(\pi k)}
\]
shows that it is strictly less than one at finite $k$, and it tends to
zero at infinity. The bound is uniform in $d$ and real $\alpha$.

\end{proof}

For differentiability and determinants it is convenient to use a
\emph{fixed} balancing envelope.  Write $v=|a|$, $u=a/v$, and
$A_v=M_{\sqrt v}$.  At any zeros the choice of $u$ is immaterial.
Set
\begin{equation}
 T_d=A_vQ_\phi A_vM_{uM_d}
     +A_vQ_{\pi-\phi}A_vM_{\bar u\overline{M_d}}.
 \label{eq:inverse-source-fixedbalance}
\end{equation}
Factoring $T_d=A_vB_d$ gives $B_dA_v=L_d$.  Therefore
\begin{align}
 \|T_d\|_1&\leq \frac1{\pi\sin\phi}\int_{\mathbb R}v(t)\,dt=:C_1,
 \label{eq:inverse-source-tracebound}\\
 \|(I\pm T_d)^{-1}\|
 &\leq1+\frac{2\|v\|_\infty}{1-c_{\phi,\rho}}=:C_2.
 \label{eq:inverse-source-balancedinverse}
\end{align}
Indeed $A_vQ_\phi A_v$ and $A_vQ_{\pi-\phi}A_v$ are positive
trace class. The convolution kernels of $Q_\phi,Q_{\pi-\phi}$ have
diagonal value $1/(2\pi\sin\phi)$; the sandwiched kernels have
diagonal $v(t)/(2\pi\sin\phi)$ and trace
$(2\pi\sin\phi)^{-1}\int_{\mathbb R}v(t)\,dt$.
The inverse identity is the elementary $AB/BA$ resolvent formula.
In particular
\begin{equation}
 e^{-C_1C_2}\leq|\det(I\pm T_d)|\leq e^{C_1},
 \label{eq:inverse-source-detbounds}
\end{equation}
uniformly in $d$.  For the lower bound use
$(I\pm T_d)^{-1}=I\mp T_d(I\pm T_d)^{-1}$ and the upper determinant
bound applied to its trace-class difference from $I$.

\paragraph{Reality, positivity, and the logarithm.}
Let $\mathsf J f=\bar f$ be complex conjugation on $\mathscr H_0$.
The real Fourier multipliers satisfy
$q_\phi(-k)=q_{\pi-\phi}(k)$, so
$\mathsf JQ_\phi\mathsf J=Q_{\pi-\phi}$.
For real parameters the two summands defining $T_d$ are therefore
exchanged by $\mathsf J$, and $\mathsf JT_d\mathsf J=T_d$.
Consequently $\det(I+\varepsilon sT_d)$ is real for every real $s$.
For $0\leq s\leq1$, the factorization $T_d=A_vB_d$, $B_dA_v=L_d$
and $\|sL_d\|<1$ show that $I+\varepsilon sT_d$ is invertible.
Its determinant varies continuously from $1$ at $s=0$ without vanishing;
hence $\det(I+\varepsilon T_d)>0$. The same argument applies to
$\tau_\varepsilon(\alpha)$, using the unmodified amplitude.
The logarithms at real $\alpha$ are the real logarithms.

All fixed-order $d$ derivatives of $M_d$ are uniformly bounded.
For instance $|\partial_dM_d|\leq\sin\phi$.
Consequently $T_d$ and its inverse have uniformly bounded derivatives
of every fixed order in trace and operator norm, respectively.
This avoids differentiating $\sqrt{|a_d|}$, which is inconvenient
when the modified jump amplitude vanishes.  Uniform holomorphic continuation
to a sufficiently small complex neighborhood of real $d$ follows
from the same resolvent bound and the poles of the logistic denominator,
which stay at imaginary distance $\pi/2$ from the real $d$ axis.

\paragraph{Returning to a fixed interior lens height.}
The Hardy-space contraction and positivity argument supplies the
boundary invertibility required by Lemma~\ref{lem:limiting-deformation}.
For $d=+\infty$ the insertion multiplier is $1$; for $d=-\infty$ it
is a constant phase, absorbed into the real parameter $\alpha$.
The lemma therefore proves invertibility at every fixed
$0<h_0<\phi/2$, uniformly in $d$ and on compact real intervals of
$\alpha$. This uses determinant invariance and compactness, not an
equality of operator norms on different contours. The limiting
argument is independent of the finite-$N$ global comparison in
Section~\ref{sec:global-source-comparison}.

\subsection{Exact finite-\texorpdfstring{$N$}{N} insertion factor and effective center}
For
\[
 J_N(s)=c_{1,N}(s+1+\delta_N)
                 \left(\frac{s+1}{s}\right)^{1/\theta},
\]
let $a_c=I_{1,N}(-c)<-1-\delta_N$.  The scalar band factor for $w(x)/(x+c)$ is exactly
\[
 D_c^{\rm out}(s)=\frac{J_N(s)+c}{c_{1,N}(s-a_c)},\qquad
 D_c^{\rm in}(s)=\frac1{c_{1,N}(s-a_c)}.
\]
The apparent singularity of the outer expression at $a_c$ is removable;
its value there is $J_N'(a_c)/c_{1,N}\ne0$.  The outer normalization
at infinity is one.  The resulting central jump coefficient is the
base coefficient multiplied by
\begin{equation}
 m_{N,c}(t)=\frac{z_N(t)-a_c}{p_N(t)-a_c}.
 \label{eq:inverse-source-exactratio}
\end{equation}
Here $z_N,p_N$ are the outer point and its partner; lower-lip formulas
are the corresponding reflected ones.

Write $a_c=-1-r_c$, $r_c>\delta_N$, and
$\mathfrak{s}_N=c_{1,N}^{-1/B}$.  The exact relation is
\[
 c=c_{1,N}(r_c-\delta_N)
              \left(\frac{r_c}{1+r_c}\right)^{1/\theta}.
\]
An effective insertion center that works even for $c$ outside the central
scaling range is
\begin{equation}
 d_N(c)=\frac12\log\frac{r_c}{\mathfrak{s}_N}.
 \label{eq:inverse-source-effectivecenter}
\end{equation}
Using $d_N(c)$ avoids any need for a relative approximation of $r_c$
by $\mathfrak{s}_Nc^{1/B}$ when $c$ is extremely small or large.

\begin{lemma}[Insertion-ratio error uniform over all $c>0$]
Suppose on a retained central window one has
\[
 z_N(t)+1=\mathfrak{s}_Ne^{2t+2i\phi}(1+e_z(t)),\qquad
 p_N(t)+1=\mathfrak{s}_Ne^{2t}(1+e_p(t)),
\]
with $|e_z|+|e_p|\leq\epsilon_N<1/2$.  Then
\[
 \sup_{c>0,t}|m_{N,c}(t)-M_{d_N(c)}(t)|\leq C\epsilon_N,
 \qquad \sup_{c>0,t}|m_{N,c}(t)|\leq1+C\epsilon_N.
\]
The same assertion applies on any fixed-height contour with the
corresponding phase factors and a uniform denominator separation.
\end{lemma}
\begin{proof}
Set $R=r_c/\mathfrak{s}_N>0$ and $u=e^{2t}>0$.  The exact ratio is
\[
 \frac{R+ue^{2i\phi}(1+e_z)}{R+u(1+e_p)},
\]
whereas its model is $(R+ue^{2i\phi})/(R+u)$.
The denominator has modulus at least $(R+u)/2$, and subtraction
therefore bounds the difference by $C\epsilon_Nu/(R+u)$.
This is uniform for every $R>0$, proving both claims.
\end{proof}

Thus, once the base finite-$N$ central kernel is compared with its
limiting kernel in the required operator ideal, multiplication by the
insertion ratio preserves that comparison uniformly for all $c>0$.
For example, in a common realization,
\[
 K_{N,c}=K_NM_{m_{N,c}},\qquad K_{d_N}=KM_{M_{d_N}},
\]
gives
\[
 \|K_{N,c}-K_{d_N}\|_{\mathfrak S}
 \leq (1+C\epsilon_N)\|K_N-K\|_{\mathfrak S}
       +C\epsilon_N\|K\|_{\mathfrak S},
\]
for the operator norm or trace norm, with the two reflected channels
understood.  Cutoff tails are harmless for this comparison whenever
the base tail estimate is available, since $|M_d|\leq1$ uniformly.

\paragraph{Scope.}
The uniform inverse theorem for the limiting family of modified operators is proved
above.  The insertion-ratio lemma proves that division by $x+c$
does not introduce an additional loss of uniformity into an already
established finite-$N$ central-kernel comparison. The endpoint and full-contour estimates are proved in
Section~\ref{sec:global-source-comparison}, using the unmodified
full operator and its exact scalar dressing. The central-kernel
convergence used above is independent of that global comparison.

\subsection{Central observation bounds}
Write $a=I_{1,N}(-c)=-1-r$ and abbreviate
$\mathfrak{s}=\mathfrak{s}_N$ in this subsection.
\begin{proposition}\label{prop:observation}
Under the finite central trace-norm and inverse estimates of Sections~\ref{sec:source-dressing}--\ref{sec:inverse-source-Hardy}, the central model $C_{N,c}$ satisfies
\[
 m\le |C_{N,c}(a)|\le M,
 \qquad |C'_{N,c}(a)|\le\frac{C\mathfrak{s}}{(r+\mathfrak{s})^2},
\]
uniformly in $c>0$ and compact real $\alpha$ intervals.
\end{proposition}
\begin{proof}
Let $K_{N,c}$ be the modified central operator. The exact dressing
identity of Section~\ref{sec:source-dressing} is
\[
 C_{N,c}(a)=\frac{\det(I-K_N)}{\det(I-K_{N,c})}.
\]
Suppose the trace norms are at most $M_0$ and inverse norms at most
$B_0$. For either operator $K$,
\[
 |\det(I-K)|\le e^{M_0},\qquad
 |\det(I-K)|^{-1}
 =|\det(I+K(I-K)^{-1})|\le e^{M_0B_0}.
\]
The determinant ratio is therefore bounded above and below uniformly.
The gauge proving the determinant identity is performed on the compact
retained supports; balancing factors cancel from Fredholm minors, so
no uniform bound on the inverse gauge is required.
Section~\ref{sec:source-dressing} also gives
\[
 \frac{C'_{N,c}(a)}{C_{N,c}(a)}=\Psi_N(a),\qquad
 |\Psi_N(a)|\le\frac{C\mathfrak{s}}{(r+\mathfrak{s})^2}.
\]
Multiplying by the upper bound for $|C_{N,c}(a)|$ proves the claim.
\end{proof}

\section{Full scalar dressing and the global comparison}
\label{sec:global-source-comparison}

\paragraph{Role of the global comparison.}
Assumption~\ref{ass:H} concerns the weight and its central
amplitude. It does not contain an assumption about a finite-$N$ RH
error. This section proves the integrated
comparison \eqref{eq:newmainerror} using full scalar dressing and
weighted contour estimates. The logarithmic loss is sufficient for
the finite part; no sharper integrated $O(N^{-1})$ rate is asserted.

Fix $\theta\ge1$, the physical parameter $r_*>0$, a compact set of
real values of $\alpha$, and $H$ satisfying Assumption~\ref{ass:H}.
Constants are independent of $N$
and of the insertion parameter $c>0$. Abbreviate
$c_1=c_{1,N}$, $\delta=\delta_N$, and $\mathfrak{s}=\mathfrak{s}_N$.
Write
\[
 J(s)=c_1(s+1+\delta)((s+1)/s)^{1/\theta},\quad
 \delta\asymp N^{-4},\quad\mathfrak{s}\asymp N^{-2},\quad
 a=I_1(-c)=-1-r,\quad r>\delta.
\]
The inputs are the principal equilibrium construction and phase
estimates, ordinary relative soft-edge coordinates, and the uniform
inverse bound for the retained central operator of the unmodified weight.
We use the complete scalar dressing identity to reduce the comparison
to an estimate for the unmodified operator. The resulting integrated
bound is $O(N^{-1}\log N)$, which is sufficient for the constant term.
A phase estimate needed for the
omitted lips, through fixed derivatives, is
\[
 N\Phi_+(t+ih)=i\pi N-i r_*\sin\phi\sinh(t+ih)
       +O(N^{-1}\cosh2t),\quad |t|\le L_N-C_*,\quad e^{L_N}\asymp N.
\]
Choose $0<h<\phi/2$ small and fixed and then $C_*$ sufficiently large.
All uses of the map involve only its two prescribed inverse branches.

\subsection{Exact insertion factors and their uniform bounds}
The scalar band multiplier for $w_N/(x+c)$ is
\[
 D_c^{\rm out}(s)=\frac{J(s)+c}{c_1(s-a)},\qquad
 D_c^{\rm in}(s)=\frac1{c_1(s-a)}.
\]
The outer factor is normalized to one at infinity. Its singularity
at $a$ is removable, with value $J'(a)/c_1\ne0$.
An active outer--inner pair $(z,p)$ acquires the multiplier
$M_a(z,p)=(z-a)/(p-a)$. The two off-diagonal matching coefficients
acquire this factor or its reciprocal; diagonal ones are unchanged.

On fixed relative left matching contours, $s+1=\delta\zeta$, with
$\zeta$ in compact contours near $-1/(1+\theta)$ separated from $-1$.
Since $r\ge\delta$, $|s-a|\asymp r+\delta$. On right matching
contours near $1/\theta>0$, $|s-a|\asymp1+r$. Thus $M_a$, its
inverse, and fixed relative derivatives are uniformly bounded.
On central lips, $z+1,p+1$ have moduli comparable to
$\mathfrak{s} e^{2t}$ and arguments separated from the negative ray. Hence
\[
 |z-a|\asymp|p-a|\asymp r+\mathfrak{s} e^{2t}.
\]
The insertion multiplier and its fixed rapidity derivatives are uniformly
bounded there as well. No new endpoint or local singularity occurs.

\subsection{Exact row-vector Airy removal}
\paragraph{Local RH problem for $P_c^{(e)}$.}
For $e\in\{a_N,d_N\}$, let $U_e$ be the endpoint disk specified
below. The local row parametrix solves the following
Riemann--Hilbert problem with a matching condition:
\begin{enumerate}
\item $P_c^{(e)}$ is analytic in $U_e\setminus\Sigma_{S,N}$.
\item On the oriented local arcs,
\[
 P_{c,+}^{(e)}(z)=P_{c,-}^{(e)}(z)J_{S,c}(z),
 \qquad z\in U_e\cap\Sigma_{S,N},
\]
where the jump matrices are those of \eqref{eq:explicit-Sjumps}.
\item The row is bounded at $e$ within every local sector.
\item On $\partial U_e$, away from the contour intersections,
\[
 \frac{P_{c,j}^{(e)}(z)}{P_{c,j}(z)}=1+O(N^{-1}),\qquad j=1,2,
\]
uniformly in $c>0$ and on compact real intervals of $\alpha$.
\end{enumerate}
The construction below gives $P_c^{(e)}=E_eA_0$. Its uniform
matching follows from the compensated component ratios estimated
below; no inverse of a row vector is used.

Uniform relative Airy coordinates follow directly for real $\theta$.
Put $\eta=1/(2B)$ and $u=1/(2A)$. The exterior spectral function is
\[
 \mathcal N_{\rm out}(s)=1+
 \frac{r_*\eta\sin(\pi u)}{2\pi N}\int_0^1
 \frac{W_\delta(v)}{v+s}\,dv,
\]
where
\[
 W_\delta(v)=c_1^\eta(1+\delta-v)^\eta((1-v)/v)^u
 +c_1^{-\eta}(1+\delta-v)^{-\eta}((1-v)/v)^{-u}.
\]
Near $s_b=1/\theta+O(\delta)$, this integral and all fixed derivatives
converge, since $c_1^\eta/N$ has a positive finite limit,
$c_1^{-\eta}/N=O(N^{-2})$, and the endpoint powers are integrable.
In particular
\[
 \mathcal N'_{\rm out}(s_b)\longrightarrow
 -C\int_0^1\frac{(1-v)^{1/2}v^{-u}}{(v+1/\theta)^2}\,dv<0.
\]
If $s_o=s_b+h$ and $s_i=s_b-h+O(h^2)$ are paired, then
\[
 \varphi(J(s_b+h))=
 \frac23\mathcal N'_{\rm out}(s_b)\frac{J''(s_b)}{J(s_b)}h^3
 +O(h^4),\qquad \frac{J''(s_b)}{J(s_b)}\to\frac{\theta^2}{\theta+1}.
\]
The cubic coefficient is uniformly nonzero. The exact involution
$s\mapsto-(1+\delta)(s+1)/(s+1+\delta)$, satisfying
$J(\mathfrak r(s))=J(s)^{-1}$, transfers this to the left relative
coordinate $(s-s_a)/\delta$. This identifies nondegenerate limiting Airy coordinates. The endpoint
disks, sector conventions, and compensated matching bounds used below
are specified in Section~\ref{sec:full-contour-estimates}.

At the right endpoint put $\psi=-\varphi_N/2$,
$f=(3\psi/2)^{2/3}$, and $\zeta=N^{2/3}f$; the positive branch is
chosen on the exterior gap. At the left endpoint first subtract the
integer upper/lower phase $\pm2\pi i$ from $\varphi_N$, which does not
change its integer-$N$ exponential. Let $U$ be the fixed invertible
connection matrix in the usual Airy normalization
\[
 \Phi_{\rm Ai}(\zeta)=\zeta^{-\sigma_3/4}U
 (I+O(\zeta^{-3/2}))e^{-2\zeta^{3/2}\sigma_3/3}.
\]
Here $\omega_c=\omega_{\alpha,c}$. In an endpoint disk set
\[
 \mathsf B=\zeta^{-\sigma_3/4}U\omega_c^{-\sigma_3/2},\qquad
 A_0=\Phi_{\rm Ai}(\zeta)e^{N\psi\sigma_3}\omega_c^{-\sigma_3/2}.
\]
On an endpoint disk $U_e$, with the sector conventions fixed, define
\[
 E_e(z)=P_c(z)\mathsf B(z)^{-1}.
\]
This is a $1\times2$ local row, not a global exterior matrix; no
inverse of $E_e$ or normalization at infinity is used. At the right
endpoint one can choose
$U=2^{-1/2}\left(\begin{smallmatrix}1&i\\i&1\end{smallmatrix}\right)$,
for which
\[
 U^{-1}\operatorname{diag}(-i,i)U
 =\begin{pmatrix}0&1\\-1&0\end{pmatrix}.
\]
Thus $P_c$ and $\mathsf B$ have the same band jump, and $E_e$ extends
holomorphically across the punctured band. Their endpoint orders give
$E_e(z)=O(|z-e|^{-1/2})$; the isolated singularity is removable.
At the left endpoint the Airy coordinate reverses the upper and lower
sides. Inserting $\sigma_3$ after $U$, with the corresponding reflected
Airy sector convention, restores the same band jump and gives the same
removability argument. This proves fixed-$N$ analyticity of the local
prefactor. Uniform matching bounds follow from the compensated ratios
estimated below.
The bare band row is $P_c=E_e\mathsf B$ and the local Airy row is $E_eA_0$.
Both $A_0,\mathsf B$ are invertible matrices. Therefore
$\widehat S=S A_0^{-1}\mathsf B$ inside the disk, with $\widehat S=S$
outside, is legitimate exact right multiplication of a row.
It replaces the Airy-ray jumps by the bare band jump and a matching
jump. Dividing by the two components of $P_c$ and gluing the sheets
gives a scalar unknown $q$, normalized to one at infinity and bounded
at the lifted endpoints. No inverse of a row, or division by a central
endpoint value, is used.

Here is the explicit matching jump. Orient a physical matching circle
counterclockwise, with the interior as its plus side, and put
$M=A_0^{-1}\mathsf B$. Then $\widehat S_+=\widehat S_-M$. If $s$ belongs
to its sheet-$j$ lift, let $\pi_i(s)=I_i(J(s))$. The scalar jump is
\[
 q_+(s)-q_-(s)=\sum_i a_{ji}(s)q_-(\pi_i(s)),\qquad
 a_{ji}(s)=[M_{ij}(J(s))-\delta_{ij}]
                   \frac{P_{c,i}(J(s))}{P_{c,j}(J(s))}.
\]
\paragraph{The component ratios needed for matching.}
Write $P=(P_1,P_2)$ for the band row without the modification
$(x+c)^{-1}$, and set
\[
 \mathcal Q_c(z)=\omega_{\alpha,c}(z)
                    \frac{P_{c,1}(z)}{P_{c,2}(z)},\qquad
 \mathcal Q_{\mathrm{base}}(z)=\omega_\alpha(z)
                    \frac{P_1(z)}{P_2(z)}.
\]
These definitions are used on the matching circles, where both
components are nonzero. The explicit insertion factors give the exact
identity
\[
 \mathcal Q_c(z)=\mathcal Q_{\mathrm{base}}(z)
 \frac{I_{2,N}(z)-a_c}{I_{1,N}(z)-a_c},\qquad a_c=I_{1,N}(-c).
\]
Indeed $D_c^{\mathrm{out}}(I_1(z))/D_c^{\mathrm{in}}(I_2(z))
=(z+c)(I_2(z)-a_c)/(I_1(z)-a_c)$, which cancels the factor
$(z+c)^{-1}$ in $\omega_{\alpha,c}$.
If the Airy expansion on these circles gives
\[
 A_0^{-1}\mathsf B
 =\omega_{\alpha,c}^{\sigma_3/2}(I+\mathcal E_N)
  \omega_{\alpha,c}^{-\sigma_3/2},\qquad
 \mathcal E_N=O(N^{-1}),
\]
then its off-diagonal scalarized coefficients are
$(\mathcal E_N)_{12}\mathcal Q_c$ and
$(\mathcal E_N)_{21}\mathcal Q_c^{-1}$. Thus the required bound is
\[
 \sup_{c>0}\sup_{z\text{ on matching circles}}
 \bigl(|\mathcal Q_c(z)|+|\mathcal Q_c(z)^{-1}|\bigr)\le C,
\]
with the analogous fixed-order relative derivative bounds where used.
The insertion-dependent ratio and its reciprocal are controlled by the
separation estimates above; the base-ratio bound is established in
Section~\ref{sec:full-contour-estimates}.
Absolute lower bounds for the separate unscaled components, uniform in
$c$, would be inappropriate: for fixed $N$ and finite regular $z$,
$P_{c,2}(z)=c^{-1}P_2(z)(1+O(c^{-1}))$ and
$P_{c,1}(z)=P_1(z)(1+O(c^{-1}))$ as $c\to\infty$.
The compensated ratios remove this normalization dependence.

The complete lift is one small closed contour about the simple
critical point, consisting of two sheet arcs. Its partner involution
is reflection to leading order in the local critical coordinate, so
its arclength Jacobian and inverse Jacobian are uniformly bounded.
The insertion factors in these two off-diagonal entries are exactly
$M_a$ and $M_a^{-1}$. The Cauchy representation of this jump evaluates
$q_-$ at these partner points and requires no matrix inverse of $P_c$.

More explicitly, let $\Sigma_E$ be the matching lifts and $\Sigma_W$
the omitted integration-contour lifts. Let $\widehat a_\sigma$ denote
the lip jump coefficient, with the cutoff omitted. The exact scalar equation is
\[
 q(s)=1+\frac1{2\pi i}\sum_\sigma\int
 \frac{\chi(t)^2 \widehat a_\sigma(t)q(p_\sigma(t))z_\sigma'(t)}
      {z_\sigma(t)-s}\,dt
 +\frac1{2\pi i}\sum_j\int_{\Sigma_{E,j}}
 \frac{\sum_i a_{ji}(t)q_-(\pi_i(t))}{t-s}\,dt
 +\mathcal C_W[q](s).
\]
On a matching contour take the minus Cauchy boundary value. The
partner map preserves the exterior side of the physical matching
circle, so the displayed minus value is the required one. Core
partner curves have no boundary Cauchy singularity. The last term has
the same form for omitted lips and gaps, including $1-\chi^2$ on the
transition portion. Subtracting these Cauchy integrals from $q$ leaves
an entire normalized function; the origin and critical points are
removable once the stated sectorwise local construction and all
boundary-side conventions are fixed. Section~\ref{sec:full-contour-estimates} specifies the full-contour
realization, its boundary sides, and the required operator estimates.

\subsection{The exact full scalar dressing}\label{sec:full-dressing}
\paragraph{Full scalar RH problem for $q$.}
After the vector transforms, exact Airy removal and gluing of the
two components, the scalar unknown solves the following nonlocal
Riemann--Hilbert problem on the complete lifted contour
$\Sigma=\bigcup_j\Sigma_j$:
\begin{enumerate}
\item $q$ is analytic off $\Sigma$ and has the prescribed boundary
values in the weighted spaces of Section~\ref{sec:full-contour-estimates}.
\item On each oriented arc $\Sigma_j$,
\[
 q_+(t)-q_-(t)=\sum_i a_{ji}(t)q_-(\pi_i(t)).
\]
The partner maps and boundary sides are those fixed in the preceding
Airy-removal construction. When a partner lies off the contour, its
value is the ordinary analytic value of $q$.
\item $q(s)=1+O(s^{-1})$ at infinity. Its sectorwise values are
bounded at the lifted band endpoints, the finite gap endpoints,
and the contour attachments, in particular as $s\to0$ or $s\to-1$
within the corresponding domains.
\end{enumerate}
Its Cauchy representation is
\begin{equation}\label{eq:fullscalar}
 q(s)=1+\frac1{2\pi i}\sum_j\int_{\Sigma_j}
 \frac{\sum_i a_{ji}(t)q_-(\pi_i(t))}{t-s}\dd t.
\end{equation}
The sum includes lip, matching, and gap channels. Each channel has an
integration point $t$ and a partner evaluation point $\pi_i(t)$.
Matching boundary values are the exterior values of the counterclockwise
matching contour; incident arcs keep their specified sides.
In this section all base coefficients and operators are
\emph{unmodified}: they contain no insertion parameter $c$.

\begin{lemma}[Full insertion law]\label{lem:dress}
The reciprocal insertion changes every scalar channel coefficient by
\[
 a_{ji}^{(c)}(t)=a_{ji}(t)\frac{t-a}{\pi_i(t)-a}.
\]
If the full Cauchy realization and its inverse exist, then
\begin{equation}\label{eq:fullpsi}
 \left.\partial_s\log q_c(s)\right|_{s=a}
 =\ell_a(I-\mathbb K)^{-1}r_a,
\end{equation}
where $r_a(p)=(p-a)^{-1}$, $\mathbb K$ is the unmodified
partner-value operator, and $\ell_a$ is its Cauchy observation at $a$.
No trace-class property of $\mathbb K$ is required.
\end{lemma}
\begin{proof}
For a lip channel the insertion factors of the scalar band row give
$(t-a)/(\pi_i(t)-a)$ directly. For matching channels use
\[
 M_c=\omega_c^{\sigma_3/2}(I+\mathcal E_N)
             \omega_c^{-\sigma_3/2},\qquad
 a_{ji}^{(c)}=(M_{c,ij}-\delta_{ij})P_{c,i}/P_{c,j}.
\]
The powers of $x+c$ cancel between $M_c$ and the band row, leaving
the stated ratio. A diagonal channel has $\pi_i(t)=t$ and is unchanged.
For a gap the triangular jump has its scalar integration point on the
interior sheet, $t=I_2(x)$, and partner $p=I_1(x)$. Its coefficient
is proportional to $\omega_c(x)P_{c,1}(x)/P_{c,2}(x)$. Since
\[
 \frac{D_c^{\rm out}(p)}{D_c^{\rm in}(t)}
 =(x+c)\frac{t-a}{p-a},
\]
the factor $(x+c)^{-1}$ in $\omega_c$ gives exactly $(t-a)/(p-a)$.

Set $G(s)=q_c(s)/(s-a)$ and let $\ell_\infty$ be the observation
functional with the Cauchy denominator omitted. The identity
\[
 \frac{t-a}{(s-a)(t-s)}=\frac1{t-s}+\frac1{s-a}
\]
gives
\[
 (I-\mathbb K)G=r_a(1+\ell_\infty G).
\]
Let $D_a=(I-\mathbb K)^{-1}r_a$ and
$d_a=1-\ell_\infty D_a$. Then $G=D_a/d_a$.
The meromorphic continuation has the expansion
\[
 D_a(s)=\frac1{s-a}+\ell_sD_a.
\]
Hence $q_c(s)=[1+(s-a)\ell_sD_a]/d_a$, which proves
\eqref{eq:fullpsi}. Nonvanishing at $a$ follows from the existing
modified polynomial problem and its exact negative-axis positivity.
One can first do the algebra on truncated contours. The same identities
pass to a bounded full-contour realization when the displayed source
and observation functionals converge. Diagonal Plemelj terms obey the
same identity: their multiplier ratio at coincident points is one.
\end{proof}

\subsection{An integrable comparison estimate}\label{sec:integrable-full-comparison}
The estimate needed for the constant term is $o(1)$, not necessarily
$O(N^{-1})$. The following comparison permits a logarithmic loss.
Use central and two matching spaces with norm
\[
 \|u\|_{\rm sc}^2=\|u_0\|_2^2+\mathfrak{s}^{-2}\|u_-\|_2^2+
 \|u_+\|_2^2.
\]
Matching norms are arclength norms before this scaling. Write
\[
 \mathbb K^{(0)}=\begin{pmatrix}K_N&0&0\\A_-&0&0\\A_+&0&0\end{pmatrix},
 \qquad \|A_-\|_{\rm sc}\le C,\quad \|A_+\|_{\rm sc}\le C.
\]
Here the block norms use the central input norm and the indicated
scaled output norm. In particular, an unscaled left-block bound
$O(\sqrt\delta)=O(\mathfrak{s})$ becomes $O(1)$ after multiplication
of its output norm by $\mathfrak{s}^{-1}$.
\begin{proposition}\label{prop:weaker}
Suppose the unmodified full-contour problem has this triangular leading
operator, a perturbation of size $O(N^{-1})$ in the scaled norm,
and omitted blocks smaller than every power of $N^{-1}$.
Suppose the matching contours have lengths $O(\delta)$ and $O(1)$,
separation from $a=-1-r$ comparable to $r+\delta$ and $r+1$,
bounded partner Jacobians, and $O(N^{-1})$ scalar coefficients.
Then on the retained and matching contours
\begin{align}\label{eq:majorant}
 |\Psi_{\rm full}(-1-r)-\Psi_N(-1-r)|
 \le\frac C N\left\{
 \frac{\mathfrak{s}}{(r+\mathfrak{s})(r+\delta)}+
 \frac{\delta}{(r+\delta)^2}+
 \frac1{(1+r)(r+\delta)}\right\}.
\end{align}
The omitted source and observation bounds in Section~\ref{sec:omitted-observation} give an
additional error $\epsilon_N/(r+\delta)^2$, uniformly for all $r>\delta$.
\end{proposition}
\begin{proof}
The inverse of $I-\mathbb K^{(0)}$ has diagonal entries
$(I-K_N)^{-1},I,I$, and lower-left entries
$A_-(I-K_N)^{-1}$ and $A_+(I-K_N)^{-1}$. It is uniformly bounded
in the scaled norm. Schur elimination of the omitted variables followed
by a Neumann series gives a uniformly bounded full inverse, differing
from this inverse by $O(N^{-1})$.

The source and observation bounds are
\[
 \|r_a\|_{\rm sc}\le\frac C{r+\delta},\qquad
 \|\ell_{a,0}\|\le\frac{C\mathfrak{s}}{r+\mathfrak{s}},
\]
\[
 \|\ell_{a,-}\|_{{\rm sc},*}\le\frac{C\delta}{N(r+\delta)},
 \qquad \|\ell_{a,+}\|\le\frac C{N(1+r)}.
\]
For the source, the unscaled left norm is at most
$C\sqrt\delta/(r+\delta)$, the right norm at most $C/(1+r)$,
and the balanced central norm at most $C/(r+\mathfrak{s})$.
For the observations apply Cauchy--Schwarz to the matching integrals;
the additional left factor $\mathfrak{s}\asymp\sqrt\delta$ comes from
the dual scaled norm. The central bound follows from its amplitude
envelope and $|z_N(t)+1+r|\ge c(\mathfrak{s} e^{2t}+r)$.
To keep track of the leading matching terms, put
\[
 B_0=(I-\mathbb K^{(0)})^{-1},\qquad
 D_{a,0}=(I-K_N)^{-1}r_{a,0},\qquad
 \|D_{a,0}\|\le\frac C{r+\mathfrak{s}}.
\]
On the retained and matching spaces, let $B$ be the inverse after
Schur elimination; omit for the moment the superpolynomially small
source and observation corrections. Then
\[
 B_0r_a=
 \begin{pmatrix}
 D_{a,0}\\ r_{a,-}+A_-D_{a,0}\\ r_{a,+}+A_+D_{a,0}
 \end{pmatrix}.
\]
By \eqref{eq:fullpsi}, the difference of observations is
\[
 \ell_a(B-B_0)r_a
 +\ell_{a,-}(r_{a,-}+A_-D_{a,0})
 +\ell_{a,+}(r_{a,+}+A_+D_{a,0}).
\]
The bound $\|A_-\|_{\rm sc}\le C$ suffices: for all sufficiently
large $N$, $\mathfrak{s}\ge\delta$, and hence
\begin{equation}\label{eq:left-block-absorption}
 |\ell_{a,-}A_-D_{a,0}|
 \le\frac{C\delta}{N(r+\delta)(r+\mathfrak{s})}
 \le\frac{C\delta}{N(r+\delta)^2}.
\end{equation}
The direct left source term has the same final bound. Both right
terms are bounded by $C/[N(1+r)(r+\delta)]$. Finally,
$\|B-B_0\|_{\rm sc}\le C/N$ gives
$C\mathfrak{s}/[N(r+\mathfrak{s})(r+\delta)]$ from the central
observation; its matching observations are bounded by $O(N^{-2})$
versions of the other two terms. These estimates prove
\eqref{eq:majorant}. The omitted source and observation corrections
are estimated in Section~\ref{sec:omitted-observation}.
\end{proof}

Each term in \eqref{eq:majorant} is integrable over $r\ge\delta$.
The three integrals are respectively
\[
 \frac{\mathfrak{s}}{\mathfrak{s}-\delta}
 \log\frac{\mathfrak{s}+\delta}{2\delta},\qquad
 \frac12,\qquad
 \frac1{1-\delta}\log\frac{1+\delta}{2\delta}.
\]
Their sum is $O(\log N)$. Since $\partial_x=J'(a)^{-1}\partial_s$
and $|dc|=|J'(a)|\dd r$, the Jacobian cancels from the integrated
resolvent error.

\subsection{Uniform insertion bounds on the omitted gaps}\label{sec:omitted-observation}
The only source denominator that could become small as $c\downarrow0$
is on the exterior partner of the left gap. The integration source
itself lies on the interior sheet and stays separated from the
negative observation ray. We retain the square-root jump weight in
estimating this source; discarding it would lose uniformity.

Write $a_0=-1-\delta$, $p=I_1(x)$ and $z=I_2(x)$ on the left gap.
Outside the fixed relative Airy disk, $0<x\le\gamma x_{-,N}$ for
some fixed $\gamma<1$. The rescaled inverse map gives
\[
 p-a_0\asymp\delta x/x_{-,N},\qquad z\in(s_-,-1).
\]
For completeness, near $a_0$ the simple zero of $J$ gives the first
comparison with a nonzero derivative; on the remaining compact
relative gap it follows by continuity of the rescaled inverse,
which is separated from the critical point by the disk. Put
$x=x_{-,N}e^{-2Bu}$. Since
$p-a=(p-a_0)+(r-\delta)$,
\[
 \frac{r+\delta}{|p-a|}\le C(1+e^{2Bu}),\qquad r\ge\delta.
\]
The gap estimate below bounds its half-amplitude by
$N^M e^{-cN(e^u-1-u)}$, with constants adjusted after taking the square
root. It absorbs $e^{2Bu}$ uniformly. The bounded source arclength
therefore gives
\[
 \|r_{a,W}\|\le\frac{\epsilon_N}{r+\delta},\qquad
 \epsilon_N=o(N^{-m})\quad\hbox{for every fixed }m>0.
\]
On the right gap $p\ge s_++c_*>0$ and
$|p-a|\ge c(1+r)$; the same conclusion follows directly. On omitted
lips the angular separation gives $|p-a|\ge c(r+\delta)$.
All omitted integration sources satisfy
$|z-a|\ge c(r+\delta)$: left-gap sources lie to the right of $s_-$,
right-gap sources lie in $(0,s_+)$, and lip sources have fixed angular
separation in their appropriate relative charts. Consequently
\[
 \|\ell_{a,W}\|\le\frac{\epsilon_N}{r+\delta}.
\]
Omitted terms in the full resolvent comparison are thus bounded by
$\epsilon_N/(r+\delta)^2$ after the inverse bounds below. Their total
integral is at most $\epsilon_N/(2\delta)$, still smaller than every
power of $N^{-1}$. These estimates use the unmodified gap amplitudes;
no uniform boundedness of the bare insertion multiplier is asserted.

\subsection{Contours and unmodified operator estimates}\label{sec:full-contour-estimates}
The following construction records the concrete estimates needed to
apply Proposition~\ref{prop:weaker}. All operators in this section are
unmodified; the source parameter enters only through Lemma~\ref{lem:dress}.
\paragraph{Choice of contours and the size of omitted jumps.}
Choose fixed sufficiently small endpoint disks in the relative physical
coordinates $x/x_{+,N}$ and $x/x_{-,N}$, and choose the two lens lips to
be $\lambda=t\pm ih$ until $|t|=L_N-C_*$, with $h>0$ sufficiently
small and $C_*$ fixed sufficiently large. In the remaining fixed relative
endpoint charts continue them into the Airy rays. The intersections with
the disk boundaries are transverse and their angles are bounded away
from zero. One can use the preimages of radial Airy rays inside these
disks and interpolate in a fixed compact relative chart outside them.
The critical points are simple; the corresponding rescaled maps and
phases converge holomorphically on these compact charts, so the
interpolation, disk boundaries and their lifts have uniformly bounded
relative derivatives and nonzero limiting intersection angles.

The gap phase has a quantitative bound requiring no new asymptotics.
Let
\[
 \mathcal U(\lambda)=\frac{r_*}{N}\cosh\lambda-
 \int\log|\sinh A(\lambda-\mu)\sinh B(\lambda-\mu)|
 \,d\nu_N(\mu),
\]
where $\nu_N$ is the equilibrium probability measure in rapidity.
Outside the support its second derivative is $r_*\cosh\lambda/N$
plus the integral of the two positive squared-cosecant-hyperbolic
terms. Writing $\lambda=L_N+u$ on the right gap, it therefore satisfies $\mathcal U'(L_N)=0$ and
$N\mathcal U''(\lambda)\ge r_*\cosh\lambda$. Consequently
\[
 N\{\mathcal U(L_N+u)-\mathcal U(L_N)\}
 \ge r_*\int_0^u(u-v)\cosh(L_N+v)\,dv
 \ge cN(e^u-1-u),\qquad u\ge0.
\]
The left gap obeys the same bound by reflection. Outside the endpoint
disks, $u\ge u_*>0$. Any power of $x$, $N$, or the relative endpoint
coordinate multiplying the gap exponential is absorbed to give a bound
$N^M e^{-cN}$, uniformly over the complete gap, including $x\to0$ and
$x\to\infty$. Indeed such powers are at most $N^M e^{M u}$ and
$\sup_{u\ge u_*}\exp\{Mu-cN(e^u-1-u)\}\le C e^{-c' N}$.

For the middle lips the inverse-map proof and the weighted integral
estimate in Appendix~\ref{sec:exact-general-equilibrium} extend to $|t|\le L_N-C_*$ with constants
independent of $N$: the two small parameters $|s+1|$ and
$\delta/|s+1|$ are at most $Ce^{-2C_*}$. The error/main-term ratio in
$N\Phi$ is at most $C N^{-1}\cosh t\le Ce^{-C_*}$. Taking $C_*$ large
therefore yields $|e^{-N\varphi}|\le C e^{-c\cosh t}$ on the fixed
height lips. On the remaining compact relative endpoint lip pieces,
strict positivity of the limiting band density permits a sufficiently
narrow lens with $\Re\varphi\ge c_*>0$ outside the disks; its jumps
are $O(e^{-c_*N})$. Thus all omitted lip coefficients, whose support
has $|t|\ge L_N/3$, are bounded by
\[
 \varepsilon_N=N^M e^{-cN^{1/3}}.
\]
This includes transition factors $1-(\chi_N^{\rm cut})^2$.

\paragraph{The omitted-contour space.}
Each scalar jump channel has an integration point $z_j$ and a partner
point $p_j$, with scalar coefficient $b_j(z_j)$. Use integration
arclength $d\sigma_j=|dz_j|$ and write $dz_j=\tau_jd\sigma_j$, where
$|\tau_j|=1$. For omitted channels set
\[
 \eta_j=|b_j|^{1/2},\qquad
 \lambda_j=\begin{cases}b_j/|b_j|^{1/2},&b_j\ne0,\\0,&b_j=0,\end{cases}
 \qquad v_j(z_j)=\eta_j(z_j)q(p_j(z_j)).
\]
Let $\mathcal H_W=\bigoplus_jL^2(\Sigma_{W,j},d\sigma_j)$.
The forcing is $(\eta_j)_j$. No division by a zero coefficient is used
when reconstructing $q$: its omitted density is $\lambda_jv_j$.
The omitted-to-omitted operator has kernel
\[
 W_{ij}(z_i,z_j)=\frac{\eta_i(z_i)\lambda_j(z_j)\tau_j(z_j)}
 {2\pi i\{z_j-p_i(z_i)\}}.
\]
For a coincident boundary contour take the exterior boundary value
prescribed by the counterclockwise matching contour. At a junction
values on a single point do not affect the $L^2$ operator; each
incident arc retains its prescribed one-sided value.

Here the integration lifts are all bounded in the $s$-plane. In
particular the left-gap integration lift lies between $s_-$ and $-1$,
and its exterior partner between $-1-\delta$ and $s_-$. The right-gap
integration lift lies between $0$ and $s_+$, and its exterior partner
between $s_+$ and $+\infty$. The endpoint disks remove a fixed
relative neighborhood of $s_\pm$. Hence the two gap lifts are
separated by at least $c\delta$ on the left and $c$ on the right.
The ratio $|dz_j|/|dp_j|$ is bounded by a power of $N$. For the
right-gap tail this follows from $I_2(x)=O(x^{-\theta})$ and
$I_1(x)=x/c_1+O(1)$; the ratio decays as $O_N(x^{-\theta-1})$.
For the left-gap tail, $I_2(x)+1=O_N(x^\theta)$ and
$I_1(x)+1+\delta=O_N(x)$, so the ratio is $O_N(x^{\theta-1})$,
which is bounded since $\theta\ge1$. The constants are powers of
$c_1$ and $\delta^{-1}$, hence powers of $N$.

\begin{lemma}[Polynomial bounds for the contour operators]
\label{lem:polynomial-contour-bounds}
Let $\Sigma_j$ be the integration arcs and $\widehat\Sigma_j=p_j(\Sigma_j)$
the partner arcs of the above contour realization. On ordinary arclength
spaces define
\[
 (\mathcal C_{ij}f)(p)=\frac1{2\pi i}
       \int_{\Sigma_j}\frac{f(z)}{z-p}\,dz,
 \qquad
 (\mathcal P_i g)(z)=g(p_i(z)).
\]
On coincident arcs, $\mathcal C_{ij}$ denotes the prescribed one-sided
Cauchy boundary value, including its Plemelj term. At attachments the
arcs are split and keep their prescribed sides. The geometric conditions used here are the following, with constants
independent of $N$ for fixed $b$ and fixed contour choices.
After splitting at attachments and adding $O(\log N)$ dyadic charts, the
bounded arcs have total length and diameter at most $CN^d$. Their
regular parametrizations, inverse speeds and $C^2$ bounds are at most
$CN^d$. Common smooth pieces are treated as a single chart. Distinct
incident half arcs have angles at least $cN^{-d}$; nonincident chart
pieces are either separated by $cN^{-d}$ or belong to one common smooth
chart. These properties include a polynomial chord--arc bound on each
chart. The maps $p_i$ are one-to-one on the split arcs, with
\[
 \sup_{\widehat\Sigma_i}\frac{|dz_i|}{|dp_i|}\le CN^d.
\]
All integration arcs are bounded. The only unbounded partner arc is
a straight right-gap tail. The constants and the power $d$ may increase
from line to line.

Under these conditions, which hold for the contours above,
for each fixed $b\ge1$
there are $C,d>0$, independent of $N$, such that
\[
 \|\mathcal C_{ij}\|_{L^2(\Sigma_j)\to L^2(\widehat\Sigma_i)}
 +\|\mathcal P_i\|_{L^2(\widehat\Sigma_i)\to L^2(\Sigma_i)}
 +\|\mathcal P_i\mathcal C_{ij}\|_{L^2(\Sigma_j)\to L^2(\Sigma_i)}
 \le C N^d .
\]
The same polynomial bound holds for self, mixed omitted/matching, and
mixed omitted/retained-central blocks, after applying the matching
normalizations and the central balancing used in this note.
\end{lemma}
\begin{proof}
The stated geometric conditions follow for the contours above as follows. On compact
relative endpoint charts the rescaled maps converge in $C^2$, their
limiting derivatives are nonzero, and the chosen attachments are
transverse. On the finite rapidity pieces, use dyadic charts in
$\delta\lesssim |s+1|\lesssim1$. There are $O(\log N)$ such charts.
The angular separation of the two sheets and
\[
 \frac{J'(s)}{J(s)}=
 \frac{\theta(s+1)^2-(\theta+1)(s+1)-\delta}
 {\theta(s+1)s(s+1+\delta)}
\]
give polynomial bounds for the chart derivatives, inverse speeds,
and separations. The regular charts can be split further a polynomial
number of times, if necessary, to obtain the stated chord--arc bounds.
For the straight gap tails use the displayed endpoint expansions for
$I_1,I_2$, rather than an inverse-speed bound at $x=0$.
The left integration and partner tails are separated by $c\delta$;
the right tails are separated by $c$. The same expansions give the
stated bound for $|dz_i|/|dp_i|$, including the two infinite physical
ends. Since $\delta^{-1}$ and $c_1^{\pm1}$ are polynomial in $N$,
all these losses are polynomial.

Here are the operator estimates in these charts. On a common smooth
arc, subtract the straight-line principal-value kernel. If $z=\gamma(u)$,
then the difference between
$\gamma'(v)/(\gamma(v)-\gamma(u))$ and $1/(v-u)$ is bounded by the
$C^2$ and inverse-speed constants; its operator norm is polynomial by
Schur's test. The principal-value term is the Hilbert transform, and
the Plemelj multiplication term is bounded. At an attachment the
kernel between distinct incident half arcs is bounded by
$CN^d/(u+v)$. The Carleman operator with kernel $(u+v)^{-1}$ has
$L^2(0,\infty)$ norm $\pi$: the unitary substitution
$u=e^x$ turns it into convolution with
$[2\cosh((x-y)/2)]^{-1}$, whose $L^1$ norm is $\pi$.
For separated bounded pieces, the Hilbert--Schmidt bound is
\[
 \|\mathcal C_{ij}\|\le
 \frac{|\Sigma_j|^{1/2}|\widehat\Sigma_i|^{1/2}}
      {2\pi\,\operatorname{dist}(\Sigma_j,\widehat\Sigma_i)},
\]
which is polynomial. Finally, on the unbounded right partner tail,
choose $T>2\sup_{z\in\bigcup_j\Sigma_j}|z|$ bounded by a power of
$N$. For $p>T$, $|z-p|^{-1}\le 2/p$, so
\[
 \int_T^\infty\!\int_{\Sigma_j}\frac{|dz|\,dp}{|z-p|^2}
 \le \frac{4|\Sigma_j|}{T}.
\]
This proves the Cauchy bounds also on that tail. Summing the chart
bounds, with their bounded or polynomial overlap, preserves a
polynomial bound.

For pullback, change variables along each partner arc:
\[
 \|\mathcal P_i g\|_{L^2(|dz_i|)}^2
 =\int_{\widehat\Sigma_i}|g(p)|^2\frac{|dz_i|}{|dp|}\,|dp|
 \le CN^d\|g\|_{L^2(|dp|)}^2.
\]
This gives the composition estimate. The left matching normalization
multiplies norms by $\mathfrak{s}^{-1}\asymp\delta^{-1/2}$.
Central retained arcs have finite
rapidity support $|t|=O(\log N)$, and their changes between rapidity
and arclength have polynomial bounds in the same charts. In a mixed
central block the remaining multipliers are $q_0$ or the coefficient
cofactor $a_N/q_0$, both uniformly bounded with fixed exponential
moments. No inverse of the balancing envelope is used as a bounded
operator. These observations prove the final assertion.
\end{proof}

Since $\|\eta_j\|_\infty=\|\lambda_j\|_\infty\le
\varepsilon_N^{1/2}$, all omitted blocks satisfy
\[
 \|U\|+\|V\|+\|W\|+\|g\|_{\mathcal H_W}
 \le N^{M'}e^{-c'N^{1/3}}=o(N^{-m})
\]
for every fixed $m$. Only bounded-operator estimates are used.

\paragraph{The matching blocks.}
Normalize the matching spaces by writing
$\bar u_-=\mathfrak{s}^{-1}u_-$ and $\bar u_+=u_+$.
Since $\mathfrak{s}\asymp\sqrt\delta$, this is uniformly equivalent
to the relative-arclength normalization $\delta^{-1/2}u_-$.
The bare power factors give, at any paired matching points,
\[
 \mathcal Q_{\rm base}
 =G_N(z,p)
 \left(\frac{z+1+\delta}{p+1+\delta}\right)^{a_\alpha}
 e^{H(\lambda)-\mathcal C_H(p)+\mathcal C_H(z)},
\]
where $G_N(z,p)=zR_N(p)/(pR_N(z))$ and
$R_N(s)=\sqrt{(s-s_-)(s-s_+)}$ has the branches of the band row.
On a fixed right relative matching contour each ratio converges to a
finite nonzero function. On the left use $s+1=\delta\zeta$; the
common powers of $\delta$ cancel, and the same assertion holds.
The $H$ Cauchy term and its relative derivatives are bounded: near
the observation endpoint subtract its analytic local value before
applying the Cauchy kernel, and on the complement use
$|d\xi|\le C|\xi+1|d\lambda$ and integrability of the exponentially
decaying $H(\lambda)$. The local rescaled contour is fixed and smooth.
Thus $\mathcal Q_{\rm base}$ and its reciprocal have uniformly bounded
fixed relative derivatives. Ordinary Airy matching consequently gives
$|a_{ji}|\le C/N$ in these charts.

The self matching Cauchy operators are dilation invariant and the
partner involution has bounded relative Jacobian. The two matching
contours are separated by a fixed distance. On the central contour,
$|p_N(t)+1|\asymp\mathfrak{s} e^{2t}$ and the left matching contour has
length $O(\delta)$. Since $q_0(t)e^{-2t}\in L^2$, the left-to-central
block has norm $O(\delta/(N\mathfrak{s}))=O(\mathfrak{s}/N)$ in the normalized
spaces. The right-to-central block is $O(N^{-1})$. The remaining
cross estimates are immediate Cauchy--Schwarz estimates. Collecting
them, in the order central/left/right,
\[
 E=\frac1N
 O\!\begin{pmatrix}0&\mathfrak{s}&1\\0&1&1\\0&\delta&1\end{pmatrix}.
\]
For the retained central contribution, evaluation on the left matching
contour has pointwise operator bound $O(1)$ and hence unnormalized
$L^2$ bound $O(\sqrt\delta)=O(\mathfrak{s})$. Evaluation on the right has
bound $O(\mathfrak{s})$, since its denominator is bounded away from zero.
Thus the distinction between the two norms is
\[
 \|A_-\|_{\rm unscaled}\le C\sqrt\delta,\qquad
 \|A_-\|_{\rm sc}\le C\frac{\sqrt\delta}{\mathfrak{s}}\le C,
 \qquad \|A_+\|_{\rm sc}\le C\mathfrak{s}.
\]
These are precisely the bounds required for the uniformly bounded
triangular leading inverse; no smallness of the scaled left block
is used in Proposition~\ref{prop:weaker}.

\paragraph{Identification with the exact RH solution.}
All matching contours are the full two-sheet lifts of the physical
matching circles, oriented counterclockwise with exterior minus values.
The partner map preserves the exterior side. Lip and gap sources carry
the orientations inherited from their exact scalar jumps. Split at each
attachment and keep the incident boundary sides; point values at the
junctions do not change the operators. These conventions determine the
full equation \eqref{eq:fullscalar}.

To justify that equation without an extra existence theorem, start
with the existing orthogonal/biorthogonal polynomial for the positive
weight. Apply the exact $Y\mapsto T\mapsto S$ transformations and the
invertible Airy right multiplications specified above. Divide
by the nonzero band-row components and glue the two sheets. The
resulting scalar function has precisely the jumps used above; it is
bounded at the endpoints and at the origin and tends to one at
infinity. At an attachment its sectorwise expressions are obtained
from the same analytic vector by the prescribed invertible factors,
so there is no extra pole or additive point contribution. Subtracting
its Cauchy jump integrals leaves an entire function equal to one.
The resulting boundary values belong to the stated balanced spaces:
the polynomial growth at the gap ends is dominated by the gap
exponentials. Thus they solve the operator equation. The block inverse
proved below identifies these exact values uniquely.

\begin{proposition}[Unmodified full operator estimate]\label{prop:blocks}
On the contour realization above, the retained and matching block is
$\mathbb K^{(0)}+E$ with $\|E\|_{\rm sc}\le C/N$. In the omitted
space just specified,
\[
 \|U\|+\|V\|+\|W\|=O(N^{-m})
\]
for every fixed $m>0$. Its constants are independent of the insertion
parameter, because this is the unmodified operator.
\end{proposition}
\begin{proof}
The matching-block table above gives the first bound. The central
block has bounded inverse by the retained central analysis.
The half-amplitude factors on omitted channels each have supremum
$N^M e^{-cN^{1/3}}$, with possibly different constants. Multiplication
on either side of the Cauchy/partner operators bounded in
Lemma~\ref{lem:polynomial-contour-bounds}
therefore gives the displayed estimate for all three omitted blocks.
The norm of the omitted forcing has the same bound since the
integration source curves have bounded total arclength. The
left scaled matching norm costs only another power of $N$.
Schur elimination of $W$ changes the retained/matching block by
$U(I-W)^{-1}V=O(N^{-m})$. The triangular leading inverse is uniformly
bounded, and a Neumann series completes its inversion.
\end{proof}

\subsection{Completion of the integrated comparison}
\label{sec:full-comparison-completion}
\begin{proposition}[Integrated comparison]\label{prop:integrated-comparison}
Fix real $b\geq1$, $r_*>0$, a function $H$ satisfying
Assumption~\ref{ass:H}, and a compact interval $I\subset\R$.
Let $p_{N,\alpha,c}$ be the monic degree-$N$ polynomial for
$w_{N,\alpha}(x)/(x+c)$, and let $p^{\rm mod}_{N,\alpha,c}$ be
the exterior RH model defined in \eqref{eq:newpolymodel}.
Then, for all sufficiently large $N$, with $c$ held fixed in each
spatial derivative,
\begin{equation}
 \sup_{\alpha\in I}\int_0^\infty
 \left|\left.\partial_z\log
 \frac{p_{N,\alpha,c}(z)}{p^{\rm mod}_{N,\alpha,c}(z)}
 \right|_{z=-c}\right|\dd c\leq C_I\frac{\log N}{N}.
 \label{eq:newmainerror}
\end{equation}
Here $C_I$ is independent of $N$, $\alpha\in I$, and $c>0$.
\end{proposition}
\begin{proof}
Apply the full rank-one identity of Lemma~\ref{lem:dress}, the
operator bounds of Proposition~\ref{prop:blocks}, and the comparison
of Proposition~\ref{prop:weaker}. The omitted contribution is bounded
by $\epsilon_N/(r+\delta)^2$, uniformly over the entire insertion
range, with $\epsilon_N$ smaller than every power of $N^{-1}$.
Its integral is at most $\epsilon_N/(2\delta)$.
The other three terms of \eqref{eq:majorant} have total integral
$O(N^{-1}\log N)$. The change from $r$ to $c$ cancels the spatial
Jacobian. To make the polynomial identification explicit, write
$s=I_{1,N}(z)$. On the exterior observation domain, outside the
lenses and matching disks, the exact transformations give
\[
 p_{N,\alpha,c}(z)
 =e^{Ng_N(z)}\mathscr P_\alpha(s)D_c^{\rm out}(s)q_c(s).
\]
Comparison with \eqref{eq:newpolymodel} therefore yields
\begin{equation}\label{eq:exact-model-quotient}
 \frac{p_{N,\alpha,c}(z)}{p^{\rm mod}_{N,\alpha,c}(z)}
 =\frac{q_c(I_{1,N}(z))}{C_{N,\alpha,c}(I_{1,N}(z))}
\end{equation}
wherever the denominator is nonzero. There is no undetermined scalar
constant: $q_c$ and $C_{N,\alpha,c}$ tend to one at infinity, and
both polynomial expressions have the monic normalization.
For each fixed $c>0$, negative-axis positivity and
Proposition~\ref{prop:observation} give a zero-free neighborhood of
$z=-c$ for both numerator and denominator. Holding $c$ fixed when
differentiating and putting $a=I_{1,N}(-c)$, we obtain
\begin{align}\label{eq:exact-model-logderivative}
 \left.\partial_z\log
 \frac{p_{N,\alpha,c}(z)}{p^{\rm mod}_{N,\alpha,c}(z)}
 \right|_{z=-c}
 &=\frac1{J_N'(a)}
 \left.\partial_s\log\frac{q_c(s)}{C_{N,\alpha,c}(s)}\right|_{s=a}
 \notag\\
 &=\frac{\Psi_{\rm full}(a)-\Psi_N(a)}{J_N'(a)}.
\end{align}
Here $J_N'(a)\ne0$, since $a<-1-\delta_N$ is outside the critical
points. The preceding integrable bounds and Jacobian cancellation
now prove \eqref{eq:newmainerror}, uniformly on compact real
$\alpha$ intervals. No assumption of a uniform modified-weight
full-contour inverse is needed.
\end{proof}

The estimate above controls the global RH comparison. The final
$o(1)$ in Theorem~\ref{thm:newmain} also includes central-model
convergence, whose rate need not equal $N^{-1}\log N$ for a general
admissible $H$.

\section{Ratio asymptotics and field normalization}
\label{sec:ratio-completion}
\begin{proof}[Proof of Theorem~\ref{thm:newmain}]
Proposition~\ref{prop:integrated-comparison} gives
\eqref{eq:newmainerror}.
Subtracting the
exact resolvent identity at $\alpha=0$ cancels its $N/(1+c)$ term. In the
exterior model, the equilibrium exponential, geometric band factor,
$H$ factor, and insertion scalar factor are independent of $\alpha$. The
remaining band integral is exactly $\alpha\chi_N$, and the central
integral is exactly the $\alpha$-derivative of the logarithm of the
Fredholm determinant. Hence
\begin{equation}
 F_N'(\alpha)=\alpha\chi_N+
 \partial_\alpha\log\tau_{N,h_0}(\alpha)
 -\left.\partial_\alpha\log\tau_{N,h_0}(\alpha)\right|_{\alpha=0}
 +O_I(N^{-1}\log N),\label{eq:newfinitebridge}
\end{equation}
where $\tau_{N,h_0}=\det(I-K_{N,\alpha})$ is the exact retained-core
determinant. The evenness of the original integral gives $F_N'(0)=0$.

The finite-$N$ central maps have the expansions
\[
 z_{\pm,N}(t)+1=\mathfrak{s}_Ne^{2t\pm i(2h_0+\phi)}(1+o(1)),\qquad
 p_{\pm,N}(t)+1=\mathfrak{s}_Ne^{2t\pm i(2h_0-\phi)}(1+o(1)).
\]
The phase estimate in the appendix and the scalar band factors give
the previously stated limiting two-contour kernel. Their mixed
derivatives are bounded by
$C_m e^{C_m(|t|+|v|)}e^{-c\cosh t-c\cosh v}$ after balancing.
Local convergence plus this envelope implies trace-norm convergence:
apply $(1+t^2-\partial_t^2)(1+v^2-\partial_v^2)$ to the kernel
difference in $L^2$, then factor through the two Hilbert--Schmidt
inverses of the harmonic oscillator. The same proof applies to
derivatives with respect to $\alpha$. The insertion ratio is uniformly bounded, with the
same local convergence at its effective center, so this comparison is
uniform over the family of modified weights as used above.

Lemma~\ref{lem:limiting-deformation} identifies the limiting
interior-contour determinant with $\tau_\varepsilon$ in
\eqref{eq:newtau}. Its proof is independent of the global RH
comparison. The deformation concerns the limiting determinant;
neither a finite-$N$ cutoff nor a negative-axis observation point is
moved in that argument.

For the unmodified amplitude, evenness of $H$ gives
$\mathcal A_H(-t)=\overline{\mathcal A_H(t)}$, so $v$ is even.
Let $\mathsf Rf(t)=f(-t)$. Then
$\mathsf RQ_\phi\mathsf R=Q_{\pi-\phi}$ and
$u_\alpha(-t)=\overline{u_{-\alpha}(t)}$.
Therefore $\mathsf R\mathcal T_\alpha\mathsf R=\mathcal T_{-\alpha}$,
and $\tau_\varepsilon(\alpha)=\tau_\varepsilon(-\alpha)$.

Uniform convergence of determinants and their derivatives with respect to $\alpha$,
together with their uniform nonvanishing, therefore yields
\[
 F_N'(\alpha)=\alpha\chi_N+
 \partial_\alpha\log\tau_\varepsilon(\alpha)+o(1).
\]
Integrating from zero proves \eqref{eq:newconclusion} under the analytic assumptions of Theorem~\ref{thm:newmain}.
The exact endpoint equation gives
\eqref{eq:newchi}, so the constant
term in $F_N(\alpha)-(\alpha^2/2)\log(NC_b/r_*)$ is
$\log[\tau_\varepsilon(\alpha)/\tau_\varepsilon(0)]$.
All stated estimates concern fixed coupling and positive $r_*$, locally uniform
in $\alpha$. No uniformity as $b\to\infty$ or $r_*\downarrow0$ is asserted.
For the physical correction, Appendix~\ref{sec:tba-strip-proof} supplies
all the analytic input on $H$ and $g$ used in this argument.
\end{proof}

\subsection{The external field-normalization constant}
\label{sec:field-normalization}
The constant $\kappa_a$ in \eqref{eq:intro-physical-cutoff} belongs to
the normalization of the QFT exponential field. It does not occur in
\eqref{eq:newZ}, \eqref{eq:physical-integral-normalization}, or
\eqref{eq:physical-ratio-identification}, and no choice of $\kappa_a$
is needed for our finite-dimensional ratio theorem.
For clarity, the value implied by matching the infinite-volume
normalization can nevertheless be recorded. Set $\alpha=2a$,
$N=2M+1$, and $r_*=R/\sin\phi$. The theorem gives
\[
 \lim_{M\to\infty}\left(\frac{4\pi M}{R}\right)^{-2a^2}
 \frac{I_M(R_{\rm circ},a)}{I_M(R_{\rm circ},0)}
 =\left(\frac{C_b\sin\phi}{2\pi}\right)^{2a^2}
 \frac{\tau_-(2a;H_R)}{\tau_-(0;H_R)}.
\]
As $R\to\infty$, the TBA bounds imply $H_R\to0$ and the central
amplitudes tend to zero in their trace-class estimates, so both
determinants tend to one. Taking this limit after the large-$M$
limit and imposing the conformal-normalization condition
$\langle e^{a\varphi}\rangle_{R_{\rm circ}}\to\mathcal G_a$
of \cite[Eqs.~(24)--(25)]{Lukyanov01} therefore requires
\begin{equation}\label{eq:field-normalization-value}
 \kappa_a=\mathcal G_a
 \left(\frac{2\pi}{C_b\sin\phi}\right)^{2a^2}.
\end{equation}
Here $\mathcal G_a$ is the independently normalized infinite-volume
expectation, initially for $|a|<Q/2$, the convergence range of its
integral formula. Equation~\eqref{eq:field-normalization-value} is the
normalization implied by our ratio asymptotics, not an independent
proof of Lukyanov's QFT identification at finite circumference.
That identification would additionally require
$\langle e^{a\varphi}\rangle_{R_{\rm circ}}/\mathcal G_a
=\tau_-(2a;H_R)/\tau_-(0;H_R)$.

\appendix

\section{An exact principal equilibrium problem and its central phase}
\label{sec:exact-general-equilibrium}
In this appendix only, the potential parameter is denoted by $r=r_*$.
We assume $\theta=b^2\geq1$, put
\[
 A=1+\theta,\qquad B=1+\theta^{-1},\qquad
 p=\frac1{2B},\qquad u=\frac p\theta=\frac1{2A},
 \qquad \phi=\frac\pi A.
\]
In particular $p+u=1/2$. All estimates below are for fixed $r>0$ and
fixed $\theta$; they are uniform when these parameters range over compact
subsets of their indicated domains.
The equilibrium measure minimizes
\begin{equation}\label{eq:gb-energy}
 \mathcal E_N[\mu]=\int V_N(x)\,d\mu(x)
 -\frac12\iint\bigl(\log|x-y|+\log|x^\theta-y^\theta|\bigr)
 \,d\mu(x)d\mu(y),
\end{equation}
where
\[
 V_N(x)=\frac A2\log x+\frac r{2N}(x^p+x^{-p}).
\]
After $x=e^{2B\lambda}$ this differs by a constant from
\[
 \frac rN\int\cosh\lambda\,d\nu(\lambda)
 -\frac12\iint\log\left|
 \sinh A(\lambda-\mu)\sinh B(\lambda-\mu)\right|
 \,d\nu(\lambda)d\nu(\mu).
\]
Thus reflection in rapidity is an exact symmetry of the principal problem.

\subsection{A scalar equation for the endpoints}
We specialize the soft-edge map
$\widetilde J_{c_0,c_1}$ of
\cite[Section~4.3, Lemma~4.3]{Claeys-Romano14} to
$c_0=c_1(1+\delta)$. For $\delta>0$ define
\begin{align}
 c_1(\delta)&=\delta^{-B/2}(1+\delta)^{-(1-1/\theta)/2},
 \label{eq:gb-c1exact}\\
 J_\delta(s)&=c_1(\delta)(s+1+\delta)
 \left(\frac{s+1}{s}\right)^{1/\theta}.
 \label{eq:gb-Jexact}
\end{align}
The power has its cut on $[-1,0]$ and is positive for $s>0$.
The critical points are
\[
 s_\pm=\frac{-(\theta-1)\pm\sqrt{(\theta+1)^2+4\theta\delta}}
 {2\theta},\qquad s_-<-1<0<s_+.
\]
Set $a=J_\delta(s_-)$ and $d=J_\delta(s_+)$; the letter $d$ here
is an endpoint, not the coupling. The conformal domains and inverse
branches are those of this Claeys--Romano map. Its two critical
values are the support endpoints. The two conjugate preimages of $x\in(a,d)$
are denoted by $I_+(x)$ and $I_-(x)$, with $\Im I_+(x)>0$.

The normalization in \eqref{eq:gb-c1exact} gives the exact identities
\begin{equation}\label{eq:gb-involution}
 ad=1,\qquad
 \mathfrak r_\delta(s)=-\frac{(1+\delta)(s+1)}{s+1+\delta},\qquad
 J_\delta(\mathfrak r_\delta(s))=J_\delta(s)^{-1}.
\end{equation}
The last identity uses the corresponding analytic branches. Indeed, at a
critical point $s+1+\delta=\theta s(s+1)$, and hence
\[
 J_\delta(s_-)J_\delta(s_+)
 =c_1^2\delta^{1+1/\theta}(1+\delta)^{1-1/\theta}=1.
\]

Define the explicit hypergeometric functions
\begin{equation}\label{eq:gb-Iq}
 I_q(\delta)=(1+\delta)^q\,
 {}_2F_1\left(-q,-\frac q\theta;1;\frac1{1+\delta}\right).
\end{equation}
The endpoint parameter is the positive solution of
\begin{equation}\label{eq:gb-mass-exact}
 \boxed{\displaystyle
 N=\frac{rp}{A}\left\{c_1(\delta)^p I_p(\delta)
                  -c_1(\delta)^{-p}I_{-p}(\delta)\right\}.}
\end{equation}
This is an exact equation, not a limiting mass condition.

\begin{proposition}\label{prop:gb-exact-equilibrium}
For every $N,r>0$ and $\theta\geq1$, equation
\eqref{eq:gb-mass-exact} has exactly one positive solution $\delta_N$.
The principal equilibrium measure is supported on
\[
 [x_{-,N},x_{+,N}]=[e^{-2BL_N},e^{2BL_N}],\qquad
 e^{2BL_N}=J_{\delta_N}(s_+).
\]
It has a strictly positive density on the open interval and nonzero
square-root vanishing at both endpoints. In particular there is no
finite-$N$ one-cut/two-cut transition in this principal equilibrium
problem.
\end{proposition}

\begin{proof}
We supply the construction because a regularity hypothesis for an
unspecified fixed potential would not suffice here. Let $\gamma$ be
the counterclockwise boundary of the conformal domain of $J_\delta$,
and put
\[
 U(s)=J_\delta(s)V_N'(J_\delta(s))
 =\frac A2+\frac{rp}{2N}\{J_\delta(s)^p-J_\delta(s)^{-p}\}.
\]
For $q=\pm p$, a binomial expansion on a circle
$1<|s|<1+\delta$ gives
\[
 \frac1{2\pi i}\oint_\gamma\frac{J_\delta(s)^q}{s}\,ds
 =c_1^q I_q(\delta).
\]
The circle may replace $\gamma$ because it encloses $[-1,0]$ and
excludes the zero $-1-\delta$. The coefficient of $s^0$ in
$(s+1+\delta)^q(1+1/s)^{q/\theta}$ is precisely the
hypergeometric series in \eqref{eq:gb-Iq}.
Thus \eqref{eq:gb-mass-exact} is the first mass condition
\[
 \frac1{2\pi i}\oint_\gamma\frac{U(s)}s\,ds=A.
\]
The second endpoint condition in the scalar construction of
\cite[Section~4.4]{Claeys-Romano14} follows automatically. To see
this, write $M_0,M_1,M_\delta$ for the contour integrals of $U$ divided
by $s,s+1,s+1+\delta$, respectively. Since $U(\mathfrak r_\delta(s))=A-U(s)$,
change of variables by $\mathfrak r_\delta$ gives
$M_0=A-M_1+M_\delta$. Also
\[
 0=\frac1{2\pi i}\oint_\gamma d\bigl[V_N(J_\delta(s))\bigr]
   =M_\delta+\frac{M_1-M_0}\theta.
\]
Consequently $M_0=A$ implies $M_1=1$.

\paragraph{Scalar RH problem for $\mathcal N$.}
The spectral function solves the scalar Riemann--Hilbert problem
of \cite[Section~4.4]{Claeys-Romano14}, with the present $U$:
\begin{enumerate}
\item $\mathcal N$ is analytic in the interior and exterior of
$\gamma$.
\item Its interior and exterior boundary values satisfy
\[
 \mathcal N_{\rm out}(s)+\mathcal N_{\rm in}(s)=U(s),
 \qquad s\in\gamma.
\]
\item At infinity, $\mathcal N(s)=1+O(s^{-1})$. The interior
values satisfy $\mathcal N(0)=\theta$ and $\mathcal N(-1)=0$.
\end{enumerate}
The last two values are the endpoint conditions verified above.
Let $\mathcal C U(s)=(2\pi i)^{-1}\oint_\gamma
 U(t)/(t-s)\,dt$, and set
\[
 \mathcal N(s)=
 \begin{cases}1-\mathcal C U(s),&s\text{ outside }\gamma,\\
 \mathcal C U(s)-1,&s\text{ inside }\gamma.
 \end{cases}
\]
Then $\mathcal N_{\rm out}+\mathcal N_{\rm in}=U$ on $\gamma$,
$\mathcal N(\infty)=1$, $\mathcal N(0)=\theta$, and
$\mathcal N(-1)=0$. These are exactly the scalar conditions in
\cite[Section 4.4]{Claeys-Romano14}. The involution moreover gives
\[
 \mathcal N_{\rm out}(s)+\mathcal N_{\rm out}(\mathfrak r_\delta(s))=1,
 \qquad
 \mathcal N_{\rm in}(s)+\mathcal N_{\rm in}(\mathfrak r_\delta(s))=\theta.
\]
For example, the sums solve an additive constant-jump problem; the
values at $0,-1$ determine its two constants. It follows that
$\mathcal N_{\rm out}(-1-\delta)=0$.
Thus the functions $G(z)=\mathcal N_{\rm out}(I_1(z))/z$ and
$\widetilde G(z)=\mathcal N_{\rm in}(I_2(z))/z$ have the required
regularity at zero, the respective normalizations $1/z$ and
$\theta/z$ at infinity, and the coupled boundary condition
$G_\pm+\widetilde G_\mp=V_N'$.
Here $I_1,I_2$ are the exterior and interior inverse branches.
This also explains why the singularity of $V_N$ at zero does not prevent
using the scalar construction: all its contour data are away from zero,
and regularity of $G$ at zero has just been checked explicitly.

For clarity, positivity can be checked without an appeal to a formal
one-cut ansatz. Collapsing the Cauchy contour onto $[-1,0]$ gives
\begin{equation}\label{eq:gb-Nout-cut}
 \mathcal N_{\rm out}(s)=1+
 \frac{rp\sin(\pi u)}{2\pi N}\int_0^1
 \frac{\mathcal W_\delta(v)}{v+s}\,dv,
\end{equation}
where
\[
 \mathcal W_\delta(v)=
 c_1^p(1+\delta-v)^p\left(\frac{1-v}{v}\right)^u
 +c_1^{-p}(1+\delta-v)^{-p}\left(\frac{1-v}{v}\right)^{-u}>0.
\]
The corresponding density is
\begin{equation}\label{eq:gb-positive-density}
 N\rho^{\mathrm{eq}}_N(x)=\frac{rp\sin(\pi u)}{2\pi^2x}
 \Im I_+(x)\int_0^1
 \frac{\mathcal W_{\delta_N}(v)}{|v+I_+(x)|^2}\,dv,
 \qquad x_{-,N}<x<x_{+,N}.
\end{equation}
All integrals converge since $0<u\leq1/4$.
The density is positive; its mass is one by $G(z)=1/z+O(z^{-2})$.
At either endpoint the inverse map has a simple square-root branch,
while the integral in \eqref{eq:gb-positive-density} has a finite,
strictly positive limit. This proves the two nonzero soft-edge
coefficients.

The coupled boundary condition gives the Euler--Lagrange equality.
For the inequality outside the support, return to rapidity and write
\[
 \mathcal U(\lambda)=\frac rN\cosh\lambda-
 \int\log|\sinh A(\lambda-\mu)\sinh B(\lambda-\mu)|\,d\nu(\mu).
\]
Off the support,
\[
 \mathcal U''(\lambda)=\frac rN\cosh\lambda+
 \int\left\{\frac{A^2}{\sinh^2 A(\lambda-\mu)}
             +\frac{B^2}{\sinh^2 B(\lambda-\mu)}\right\}d\nu(\mu)>0.
\]
Its derivative is zero at either endpoint by the soft-edge behavior and
the equality on the band. Hence $\mathcal U$ is strictly above its
band value outside it. The constructed positive measure is therefore
the equilibrium measure. Uniqueness follows from strict convexity of
\eqref{eq:gb-energy}: on signed measures of total mass zero each of
the two negative logarithmic energies is positive, by pushforward
under $x\mapsto x^\theta$ for the second one.

It remains to verify existence of a root. The right side of
\eqref{eq:gb-mass-exact} is continuous for $\delta>0$, tends to infinity
as $\delta\downarrow0$, and tends to zero as $\delta\to\infty$.
For the last assertion put $z=(1+\delta)^{-1}$: the expression in braces is
\[
 (1-z)^{-1/4}{}_2F_1(-p,-u;1;z)
 -(1-z)^{1/4}{}_2F_1(p,u;1;z)=\frac z2+O(z^2).
\]
The intermediate value theorem gives a root. Every root gives an
equilibrium measure by the preceding construction. Uniqueness of that
measure fixes its support. Moreover
\[
 \frac{d}{d\delta}\log\frac d a
 =\frac1{s_++1+\delta}-\frac1{s_-+1+\delta}<0,
\]
where the derivatives of the critical points drop out. Thus the support
fixes $\delta$, proving uniqueness of the root.
\end{proof}

\subsection{Endpoint and inverse-map asymptotics with errors}
Set
\[
 C_b=\frac{\Gamma(u)\Gamma(p)}{\sqrt\pi},\qquad
 \gamma_\theta=A\theta^{-\theta/A},\qquad
 j_\theta=\frac{A^B}{\theta}.
\]
Gauss's hypergeometric evaluation and its first derivative imply
\begin{align*}
 I_p(\delta)&=\frac{\Gamma(3/2)}{\Gamma(1+p)\Gamma(1+u)}+O(\delta),\\
 I_{-p}(\delta)&=\frac{\Gamma(1/2)}{\Gamma(1-p)\Gamma(1-u)}
                  +O(\delta^{1/2}).
\end{align*}
Using $c_1^p=\delta^{-1/4}(1+\delta)^{-(\theta-1)/(4A)}$
in the exact mass equation therefore gives
\begin{align}
 c_1^p&=\frac{NC_b}{r}\bigl(1+O(N^{-2})\bigr),
 &\delta_N&=\left(\frac r{NC_b}\right)^4\bigl(1+O(N^{-2})\bigr),
 \label{eq:gb-mass-asymptotic}\\
 e^{L_N}&=\sqrt{\gamma_\theta}\frac{NC_b}{r}
                  \bigl(1+O(N^{-2})\bigr).
 \label{eq:gb-L-asymptotic}
\end{align}
Indeed $J_{\delta}(s_+)=c_1j_\theta(1+O(\delta))$ and
$j_\theta^p=\sqrt{\gamma_\theta}$. No substitution of an asymptotic
mass equation is required for these error bounds.
At $\theta=1$ this construction gives $c_1=\delta^{-1}$,
$\delta=\operatorname{csch}^2(2L_N)$, and
$c_1^{\pm p}I_{\pm p}=P_{\pm p}(\cosh4L_N)$ by Pfaff's
hypergeometric transformation. Thus \eqref{eq:gb-mass-exact} reduces
exactly to the Legendre mass equation of the self-dual calculation.

Introduce the useful exact scale
\[
 \mathfrak{s}_N=c_1^{-1/B},\qquad
 \mathfrak{s}_N=\left(\frac r{NC_b}\right)^2(1+O(N^{-2})),\qquad
 \delta_N=\mathfrak{s}_N^2(1+O(\mathfrak{s}_N^2)).
\]
Let $0<c<1$ and $0<h<\phi/2$. The inverse branches continued from
the upper and lower sides of the band satisfy, uniformly for
$|\Re\lambda|\leq cL_N$, $|\Im\lambda|\leq h$,
\begin{equation}\label{eq:gb-inverse-uniform}
 I_\pm(e^{2B\lambda})+1
 =\mathfrak{s}_N e^{2\lambda\pm i\phi}
 \left(1+O\bigl(\mathfrak{s}_N\cosh(2\Re\lambda)\bigr)\right).
\end{equation}
To verify this, put $w=s+1$ in the equation $J_\delta(s)=e^{2B\lambda}$.
For $|w|\ll1$ and $\delta/|w|\ll1$ it becomes
\[
 e^{2B\lambda}=c_1w^B e^{\mp i\pi/\theta}
 (1+\delta/w)(1-w)^{-1/\theta}.
\]
The leading root is $w_0=\mathfrak{s}_Ne^{2\lambda\pm i\phi}$.
The analytic implicit-function theorem, on a circle of radius a fixed
multiple of $|w_0|(|w_0|+\delta/|w_0|)$ around $w_0$, gives
\eqref{eq:gb-inverse-uniform}. The errors tend to zero uniformly on
the displayed rectangles. The same bound holds after every fixed
number of rapidity derivatives, on slightly smaller rectangles.

\subsection{An exact integral and a uniform complex central phase}
Let $\rho^{\mathrm{eq}}_{\lambda,N}(\lambda)=2Be^{2B\lambda}\rho^{\mathrm{eq}}_N(e^{2B\lambda})$
for real $\lambda\in(-L_N,L_N)$, and use its analytic continuation
in the following proposition. Put
\[
 w_{\pm,N}(\lambda)=\frac{I_\pm(e^{2B\lambda})+1}{\mathfrak{s}_N},
 \qquad \mathfrak d_{N}=\frac{\delta_N}{\mathfrak{s}_N},
\]
where $\mathfrak d_{N}$ is unrelated to the right endpoint. Define
\[
 W_N(y)=(y+\mathfrak d_{N})^p y^u(1-\mathfrak{s}_Ny)^{-u}
       +(y+\mathfrak d_{N})^{-p}y^{-u}(1-\mathfrak{s}_Ny)^u.
\]
Changing variables $v=1-\mathfrak{s}_Ny$ in \eqref{eq:gb-Nout-cut}
yields the exact representation
\begin{equation}\label{eq:gb-density-central-exact}
 N\rho^{\mathrm{eq}}_{\lambda,N}(\lambda)=
 -\frac{r\sin(\pi u)}{4\pi^2 i}
 \int_0^{1/\mathfrak{s}_N}W_N(y)
 \left\{\frac1{w_{+,N}(\lambda)-y}
       -\frac1{w_{-,N}(\lambda)-y}\right\}\,dy.
\end{equation}
This expression supplies the required analytic continuation: for the
rectangles in \eqref{eq:gb-inverse-uniform}, the two poles stay strictly
above and below the integration ray, respectively.

\begin{proposition}[Uniform central density and phase]\label{prop:gb-central-phase}
For $0<c<1$, $0<h<\phi/2$, and every fixed nonnegative integer $j$,
\begin{equation}\label{eq:gb-central-density-bound}
 \partial_\lambda^j\left\{
 N\rho^{\mathrm{eq}}_{\lambda,N}(\lambda)
 -\frac{r\sin\phi}{2\pi}\cosh\lambda\right\}
 =O\left(N^{-1}\cosh(2\Re\lambda)\right)
\end{equation}
uniformly on $|\Re\lambda|\leq cL_N$, $|\Im\lambda|\leq h$.
Let $\Phi_{+,N}$ denote the analytic continuation of the upper band
phase, normalized by $N\Phi_{+,N}(0)=i\pi N$. Then
\begin{equation}\label{eq:gb-central-phase-bound}
 N\Phi_{+,N}(\lambda)=i\pi N-ir\sin\phi\sinh\lambda
 +O\left(N^{-1}\cosh(2\Re\lambda)\right).
\end{equation}
The same error bound holds for its fixed-order derivatives after the
corresponding differentiation of the leading expression.
\end{proposition}

\begin{proof}
We give bounds that remain useful on growing central contours. Write
$\mathfrak{s}=\mathfrak{s}_N$ and $R=e^{2\Re\lambda}$. By
\eqref{eq:gb-inverse-uniform}, $|w_{\pm,N}|$ is comparable to $R$
and their arguments stay a fixed positive distance from the real
integration ray. Consequently
\[
 \left|\frac1{w_{+,N}-y}-\frac1{w_{-,N}-y}\right|
 \leq C\frac R{(y+R)^2},\qquad y>0.
\]
For large $N$, the relevant range of $R$ lies between $\mathfrak{s}$ and
$\mathfrak{s}^{-1}$. The elementary weighted estimate
\begin{align}\label{eq:gb-W-error}
 &\int_0^{1/\mathfrak{s}}|W_N(y)-y^{1/2}-y^{-1/2}|
                 \frac R{(y+R)^2}\,dy\notag\\
 &\hspace{8mm}+\int_{1/\mathfrak{s}}^\infty(y^{1/2}+y^{-1/2})
                 \frac R{(y+R)^2}\,dy
 \leq C\sqrt{\mathfrak{s}}\,(R+R^{-1})
\end{align}
follows by splitting at $\mathfrak{s}$ and $1/(2\mathfrak{s})$.
On $(0,\mathfrak{s})$ the integral of the potentially largest term
$(y+\mathfrak d_{N})^{-p}y^{-u}$ is $O(\sqrt{\mathfrak{s}})$, since $p+u=1/2$
and $\mathfrak d_{N}$ is comparable to $\mathfrak{s}$.
On the middle interval the weight difference is bounded by
\[
 C\mathfrak{s}\bigl(y^{3/2}+y^{1/2}+y^{-1/2}+y^{-3/2}\bigr).
\]
Integration against $R/(y+R)^2$ gives the bound in
\eqref{eq:gb-W-error}. On the upper interval the factor
$(1-\mathfrak{s} y)^{-u}$ remains integrable because $u<1$; scaling
$y$ by $1/\mathfrak{s}$ gives $O(\sqrt{\mathfrak{s}} R)$.
The omitted infinite tail has the same bound.

The change from $w_{\pm,N}$ to $w_\pm^0=e^{2\lambda\pm i\phi}$
costs at most
\[
 C\mathfrak{s}(R^2+1)(R^{-1/2}+R^{-3/2})
 \leq C\sqrt{\mathfrak{s}}(R+R^{-1}),
\]
by the resolvent identity and
$\int_0^\infty(y^{1/2}+y^{-1/2})/(y+R)^2\,dy
=O(R^{-1/2}+R^{-3/2})$.
Finally, elementary beta integrals give
\[
 \int_0^\infty(y^{1/2}+y^{-1/2})
 \left\{\frac1{e^{2\lambda+i\phi}-y}
       -\frac1{e^{2\lambda-i\phi}-y}\right\}\,dy
 =-4\pi i\cos(\phi/2)\cosh\lambda.
\]
The difference makes the integral convergent at infinity; branches in
its beta-integral evaluation are fixed by the upper and lower poles.
As $\sin(\pi u)=\sin(\phi/2)$ and $\sqrt{\mathfrak{s}}=O(N^{-1})$,
this proves \eqref{eq:gb-central-density-bound} for $j=0$.
Cauchy estimates on a slightly larger rectangle prove every fixed
$j$. The identity
$\partial_\lambda(N\Phi_{+,N})=-2\pi iN\rho^{\mathrm{eq}}_{\lambda,N}$,
integrated from zero first vertically and then horizontally, proves
\eqref{eq:gb-central-phase-bound}; evenness of the equilibrium measure
fixes $N\Phi_{+,N}(0)=i\pi N$.
\end{proof}

In particular, on an upper rapidity lip $\lambda=t+ih$,
\[
 e^{-N\Phi_{+,N}(t+ih)}
 =(-1)^N e^{ir\sin\phi\sinh(t+ih)}
   \exp\left[O\bigl(N^{-1}\cosh2t\bigr)\right].
\]
For $|t|\leq L_N/3+1$ the phase error is $O(N^{-1/3})$.
More generally, throughout $|t|\leq cL_N$, the real part of
$-N\Phi_{+,N}(t+ih)$ is bounded above by $-c_0\cosh t$
for some $c_0>0$ and large $N$.
Thus the surviving central jumps and their double-exponential tails
follow from an exact spectral calculation. This result does not alone
prove trace-norm convergence of the full central operator: the outer
factors, partner-sheet maps and their Jacobians must also be controlled.

The endpoint notation $x_{-,N},x_{+,N}$ will be used throughout. Also $\Phi_{+,N}(\lambda)$ is the upper-band continuation of $\varphi_N(e^{2B\lambda})$ defined above.

\section{Verification of \texorpdfstring{Assumption~\ref{ass:H}}{Assumption on H} for the physical TBA correction}
\label{sec:physical-TBA-Schur-proof}\label{sec:tba-strip-proof}

This appendix verifies both parts of Assumption~\ref{ass:H} for the
physical TBA correction. Proposition~\ref{prop:TBA-bounded} supplies
the regularity and decay in part~(i); Theorem~\ref{thm:TBA-amplitude}
supplies the amplitude representation and strip bounds in part~(ii).

The bounded-correction existence and uniqueness theorem used in this
appendix is a specialization of Fring--Korff--Schulz
\cite[Section~5]{Fring-Korff-Schul99}. We reproduce that argument in the normalization
of the present ensemble, then derive the regularity estimates required
by the RH analysis. Two further steps are given separately: reduction
of real finite-convolution solutions to the bounded class, and
identification of the central amplitude with an analytic strip factor.
The analytic reconstruction is related to the established
$Q$-function constructions in \cite{Lukyanov01,Zamolodchikov06,Bytsko-Teschner06,Teschner08}; the
particular amplitude identity below is verified directly.

\subsection{Normalization of the physical equation}
Fix $0<\phi\leq\pi/2$ and $r_*>0$, and put
\[
 R=r_*\sin\phi,\qquad a_0=\frac\pi2,\qquad h=\frac\phi2.
\]
All convolutions in this section mean
$(f*g)(t)=\int_{\mathbb R}f(t-v)g(v)\,dv$.
Use the kernels
\[
 S_0(t)=\frac1{\pi\cosh t},\qquad
 S_\phi(t)=\frac{\sin\phi\cosh t}
 {\pi(\sinh^2t+\sin^2\phi)}.
\]
The physical TBA equations in this normalization are
\begin{equation}
 G=S_\phi*L,\qquad
 \epsilon(t)=R\cosh t-G(t),\qquad
 L(t)=\log(1+e^{-\epsilon(t)}),\qquad H=S_0*L.
 \label{eq:TBA-Schur-definitions}
\end{equation}
This normalization follows from Lukyanov's
\cite[Eqs.~(19)--(21)]{Lukyanov01}: if $R=mR_{\rm circ}$, then
$\Phi/(2\pi)=S_\phi$ and
\[
 \log\mathcal Q(t)=-\frac{R}{2\sin\phi}\cosh t
 +\frac1{2\pi}\int_{\mathbb R}\frac{L(v)}{\cosh(t-v)}\,dv.
\]
Thus $\mathcal Q(t)^2=\exp[-r_*\cosh t+H(t)]$ with
$r_*=R/\sin\phi$.  Here $\sin(\pi b/Q)=\sin\phi$.
This fixes both the sign and the scale of the physical correction;
$R$ is the TBA driving coefficient and $r_*$ is the coefficient in
our one-body weight.\footnote{We use the rapidity conversion
$t=\gamma/(bq)$, with $q=Q$ in Lukyanov's notation. Under
$b\mapsto b^{-1}$ and $\gamma\mapsto\gamma/b^2$, this rapidity
and the measure and pair interaction in \cite[Eq.~(13)]{Lukyanov01}
are invariant. The products $bq\gamma$ printed in
\cite[Eqs.~(14), (16)]{Lukyanov01} are not, and conflict with the
self-duality asserted after Eq.~(21) there. We interpret those
products as typographical inversions; this is not an author-confirmed
erratum. The TBA and $\mathcal Q$ formulas \cite[Eqs.~(19)--(21)]{Lukyanov01}
are used without alteration.}

Fring--Korff--Schulz define convolution with the measure $dv/(2\pi)$.
Their sinh-Gordon kernel, divided by $2\pi$, is our $S_\phi$.
Thus their bounded unknown $\xi=R\cosh-\epsilon$ is precisely $G$.
The full pseudoenergy $\epsilon$ is unbounded; all assertions of
boundedness below refer to the correction $G$, not to $\epsilon$.

\subsection{The bounded fixed point and its regularity}
The existence--uniqueness part of the next proposition is
\cite[Section~5, Eqs.~(127)--(130)]{Fring-Korff-Schul99}. Their invariant set is
\[
 D_{q,R}=\left\{f:\|f\|_\infty
 \leq R+\log\frac{q}{1-q}\right\},\qquad e^{-R}\leq q<1.
\]
Choosing $q=e^{-R}$ gives the bound
$M=-\log(1-e^{-R})$ and contraction constant used here.
Positivity allows us to restrict to $0\leq G\leq M$.
We include the proof for the explicit estimates and append the
regularity consequences needed later.

\begin{proposition}[FKS fixed point and strip regularity]
\label{prop:TBA-bounded}
The equations \eqref{eq:TBA-Schur-definitions} have a unique solution
with bounded $G$.  It is real and even, and with
\[
 M=-\log(1-e^{-R}),\qquad \rho=e^{-R}<1,
\]
one has
\begin{equation}
 0\leq G,L\leq M,\qquad
 L(t)\leq e^{M-R\cosh t},\qquad
 0<n(t):=\frac1{1+e^{\epsilon(t)}}\leq\rho.
 \label{eq:TBA-Schur-bounds}
\end{equation}
The function $H$ is real and even on the real axis, holomorphic on
$|\Im z|<\pi/2$, and, on every closed substrip, $H$ and each fixed
derivative are $O(e^{-|\Re z|})$.
Writing $D=G-L$, one has $D\in C^\infty\cap L^1$,
$|D|\leq M$, exponential decrease through fixed derivatives, and
\begin{equation}
 \int_{\mathbb R}D(t)\,dt=0,
 \qquad \log n(t)=-R\cosh t+D(t).
 \label{eq:TBA-Schur-D}
\end{equation}
\end{proposition}

\begin{proof}
We first specialize the contraction argument of \cite{Fring-Korff-Schul99}.
The kernel $S_\phi$ is positive and has mass one: the substitution
$u=\sinh t$ gives
\[
 \int_{\mathbb R}S_\phi(t)\,dt
 =\frac{\sin\phi}{\pi}\int_{\mathbb R}
   \frac{du}{u^2+\sin^2\phi}=1.
\]
On the closed subset $0\leq G\leq M$ of $C_b(\mathbb R)$ consider
\[
 \mathcal F(G)=S_\phi*\log(1+e^{-R\cosh+G}).
\]
It maps this set into itself, since
$\log(1+e^{-R+M})=M$.  The derivative of the scalar nonlinearity
with respect to $G$ is at most
$1/(1+e^{R-M})=e^{-R}$.  Thus $\mathcal F$ is a strict contraction
in the supremum norm.  Banach's theorem gives its fixed point.
Any bounded fixed point is nonnegative; its supremum $M'$ satisfies
$M'\leq\log(1+e^{-R+M'})$, which rearranges to $M'\leq M$.
This proves uniqueness among all bounded corrections.  Reflection
commutes with the map, giving evenness.

The same scalar estimates give \eqref{eq:TBA-Schur-bounds}.
Every derivative of $S_\phi$ decreases exponentially.  Convolving
with the displayed double-exponential envelope of $L$ first gives
all real derivatives of $G$, and differentiation of the scalar
nonlinearity then gives the corresponding smoothness and
double-exponential envelopes for fixed derivatives of $L$.
On every closed substrip of $|\Im z|<\pi/2$, the kernel $S_0(z)$
and all its fixed derivatives are bounded by a constant times
$e^{-|\Re z|}$.  Holomorphic dominated convergence proves the
claims about $H$.  In particular this strip is strictly wider than
the required $|\Im z|\leq\phi/2$.
Finally $G,L$ are integrable and $\int S_\phi=1$, so Fubini gives
$\int(G-L)=0$.  The last identity in \eqref{eq:TBA-Schur-D} follows
from $\log n=-\epsilon-\log(1+e^{-\epsilon})$.
\end{proof}

\subsection{Reduction from finite convolution to bounded correction}
The next argument supplies a reduction not contained in the bounded
fixed-point proof of \cite[Section~5]{Fring-Korff-Schul99}. It uses the exponential
comparison for this particular kernel to show that a weak finiteness
assumption already places a real solution in the preceding uniqueness
class.

\begin{proposition}[Automatic boundedness for ordinary real solutions]
\label{prop:TBA-automatic}
Suppose real measurable functions satisfy
\eqref{eq:TBA-Schur-definitions} almost everywhere, and the
nonnegative integral defining $G$ is finite at one real point.
Then, after choosing the continuous convolution representative of $G$,
they coincide with the solution just constructed.  In particular no
separate boundedness or finite-free-energy assumption is needed within
this real, finite-convolution class.
\end{proposition}

\begin{proof}
For fixed $\phi$ there are positive constants such that
\[
 c_\phi e^{-|s|}\leq S_\phi(s)\leq C_\phi e^{-|s|}.
\]
Finiteness of the convolution at one point therefore implies
$\int e^{-|s|}L(s)\,ds<\infty$.  The same comparison shows that
the convolution is finite and locally bounded at every real $t$;
dominated convergence also makes it continuous.  Moreover
\[
 e^{-|t|}G(t)=\int e^{-|t|}S_\phi(t-s)L(s)\,ds\longrightarrow0
 \qquad(|t|\longrightarrow\infty).
\]
For each fixed $s$ the integrand tends to zero, and it is dominated by
$C_\phi e^{-|s|}L(s)$ because $|s|\leq|t|+|t-s|$.
Consequently, outside a compact interval,
\[
 R\cosh t-G(t)\geq\frac R4e^{|t|},\qquad
 L(t)\leq\exp\left(-\frac R4e^{|t|}\right).
\]
The local bound on $G$ bounds $L$ locally as well.  Thus $L\in L^1$
and $\|G\|_\infty\leq\|S_\phi\|_\infty\|L\|_1<\infty$.
The supremum estimate and contraction uniqueness proved above now
apply.  This argument concerns the displayed real TBA system with its
ordinary logarithm and nonnegative convolution; it makes no assertion
about different complex branches or other TBA equations.
\end{proof}

\subsection{Identification and bounds for the central strip factor}
Reconstructing strip-holomorphic functions from TBA data is an
established part of the $Q$-function approach; see
\cite[Section~2, Eqs.~(2.9)--(2.14)]{Zamolodchikov06},
\cite[Section~6.2]{Bytsko-Teschner06}, and \cite[Appendix~A]{Teschner08}.
Here the function to be reconstructed is the scalar amplitude
$\mathcal A_H$ of \eqref{eq:newamp}. Its precise phase must be matched,
not only its modulus. We give that calculation explicitly and then
apply the classical Cauchy and strip Poisson formulas. These steps
are additional to the bounded TBA existence theorem.

Define, for $0<\Im z<a_0$,
\begin{equation}
 g(z)=\exp\left\{
 iR\sinh z+\frac1{\pi i}\int_{\mathbb R}
                  \tanh(z-v)D(v)\,dv\right\}.
 \label{eq:TBA-Schur-g}
\end{equation}
The integral is locally holomorphic and absolutely convergent,
because its poles are on the strip boundary or outside it and
$D\in L^1$.

\begin{theorem}[Exact physical amplitude and strip contractivity]
\label{thm:TBA-amplitude}
Let $\mathcal A_H$ be the amplitude defined by the scalar Cauchy
formula in \eqref{eq:newamp}, with the correction
$H=S_0*L$ from \eqref{eq:TBA-Schur-definitions}.  Then
\begin{equation}
 \mathcal A_H(t)=g(t+ih)\qquad(t\in\mathbb R),
 \label{eq:TBA-Schur-amplitude}
\end{equation}
with equality of the full complex amplitudes, not just their moduli.
For $0<y<a_0$,
\begin{align}
 \log|g(t+iy)|
 &= (J_y*\log n)(t),
 &J_y(s)&=\frac{\sin(2y)}{\pi(\cosh(2s)+\cos(2y))},
 \label{eq:TBA-Schur-modulus}\\
 |g(t+iy)|&\leq e^{-2Ry/\pi}<1,
 \label{eq:TBA-Schur-interior}\\
 |g(t+iy)|&\leq
 \exp\left\{\frac{2My}{\pi}-R\sin y\cosh t\right\}.
 \label{eq:TBA-Schur-envelope}
\end{align}
Its boundary values satisfy
\begin{equation}
 |g(t)|=1,\qquad |g(t+i\pi/2)|=n(t)\leq e^{-R}<1.
 \label{eq:TBA-Schur-boundaries}
\end{equation}
In particular the physical correction satisfies the amplitude and
strip bounds in Assumption~\ref{ass:H}(ii), including
\eqref{eq:newschur}, with
$\rho=e^{-r_*\sin\phi}$.  On every closed interior substrip, $g$
and each fixed holomorphic derivative have a double-exponential
envelope in $|\Re z|$.
\end{theorem}

\begin{proof}[Proof of the full complex identification]
Use the Fourier convention
$\widehat f(k)=\int e^{-ikt}f(t)\,dt$, and abbreviate
$X=\pi k/2$, $Y=\phi k/2=hk$.
The elementary transform of $S_0$ is $1/\cosh X$.
The identity
\[
 S_\phi(t)=\tfrac12\left[
 S_0(t+i(\pi/2-\phi))+S_0(t-i(\pi/2-\phi))\right]
\]
and a contour shift inside the poles of $S_0$ therefore give
\[
 \widehat S_\phi(k)=\frac{\cosh(X-2Y)}{\cosh X}.
\]
It follows that
\begin{equation}
 \widehat H=\frac{\widehat L}{\cosh X},\qquad
 \widehat D=-\frac{2\sinh(X-Y)\sinh Y}{\cosh X}\widehat L.
 \label{eq:TBA-Schur-Fourier-TBA}
\end{equation}

Direct substitution in the scalar Cauchy formula
\eqref{eq:newCH} gives
\[
 \mathcal C_H(e^{2t})=(P_\phi*H)(t),\qquad
 \mathcal C_H(e^{2t+2i\phi})=(R_\phi*H)(t),
\]
where
\[
 P_\phi(s)=\frac{\sin\phi}{\pi(\cosh2s-\cos\phi)},\qquad
 R_\phi(s)=\frac{\coth(s+ih)-\coth(s+3ih)}{2\pi i}.
\]
These signs follow, for example, by expressing
$2e^{2v-i\phi}/(e^{2v-i\phi}-e^{2t})$
as $1+\coth(v-t-ih)$.
Their transforms are
\begin{equation}
 \widehat P_\phi=\frac{\sinh(X-Y)}{\sinh X},\qquad
 \widehat R_\phi=-\frac{e^{X-2Y}\sinh Y}{\sinh X}.
 \label{eq:TBA-Schur-PR}
\end{equation}
The values at zero are understood continuously.  One verification
uses the distributional transform
\[
 \widehat{\coth(\cdot+ia)}(k)
 =-i\pi\frac{e^{(\pi/2-a)k}}{\sinh(\pi k/2)},
 \qquad 0<a<\pi,
\]
which follows by shifting $\tanh t$; the singularity at zero is a
principal value.  In the differences defining $P_\phi,R_\phi$
that singularity cancels.  For $P_\phi$ use also
$\coth(s-ih)=\coth(s+i(\pi-h))$.

Since $H$ is holomorphic and exponentially decreasing on a wider
strip, translation by $ih$ has Fourier multiplier $e^{-Y}$.
Thus the correction in the exponent of \eqref{eq:newamp} has
transform
\begin{align*}
 \left(\widehat P_\phi-\widehat R_\phi-e^{-Y}\right)\widehat H
 &=\frac{2e^{-Y}\sinh(X-Y)\sinh Y}{\sinh X\cosh X}\widehat L\\
 &=-\frac{e^{-Y}}{\sinh X}\widehat D.
\end{align*}
On the other hand the tempered transform of
$K_h(t)=(\pi i)^{-1}\tanh(t+ih)$ is
\[
 \widehat K_h(k)=-\frac{e^{-hk}}{\sinh(\pi k/2)}.
\]
At $h=0$ this is the odd principal-value distribution, with no
delta term; translation gives the displayed expression at $h>0$.
Multiplication by $\widehat D$ is unambiguous because
$\widehat D(0)=\int D=0$.  Consequently the exponent correction
is exactly $K_h*D$, proving \eqref{eq:TBA-Schur-amplitude}.
Equivalently, equality after differentiating in $t$ determines the
correction up to a constant; both corrections tend to zero at
infinity because $H$ decays and $\int D=0$.  This also verifies
that no constant phase has been omitted.
\end{proof}

\begin{proof}[Proof of contractivity and the boundary statements]
The real part of $(\pi i)^{-1}\tanh(s+iy)$ is exactly $J_y(s)$.
Moreover elementary substitutions give
\[
 J_y>0,\qquad \int_{\mathbb R}J_y(s)\,ds=\frac{2y}{\pi},
 \qquad (J_y*\cosh)(t)=\sin y\cosh t.
\]
For example $u=\tanh s$ proves the mass identity, and
$u=\sinh s$ proves the last identity after oddness removes the
$\sinh t$ term.  Taking the real part of
\eqref{eq:TBA-Schur-g} and using \eqref{eq:TBA-Schur-D} now gives
\eqref{eq:TBA-Schur-modulus}.  The bounds
$\log n\leq-R$ and $D\leq M$ give
\eqref{eq:TBA-Schur-interior} and
\eqref{eq:TBA-Schur-envelope}, respectively.

At the lower boundary the exponent correction is purely imaginary,
giving modulus one directly.  As $y\uparrow\pi/2$, $J_y$ is an
approximate identity of mass tending to one; hence its convolution
with $D$ tends to $D$.  The full upper boundary trace also exists.
Indeed $\tanh(s+i\pi/2-i0)=\coth(s-i0)$, and the Plemelj formula
gives
\[
 \int\tanh(t-v+i\pi/2-i0)D(v)\,dv
 =\operatorname{PV}\int\coth(t-v)D(v)\,dv+i\pi D(t).
\]
The principal value exists because $D$ is smooth and decreases
exponentially.  Therefore
\[
 g(t+i\pi/2)=
 \exp\left\{-R\cosh t+D(t)
 +\frac1{\pi i}\operatorname{PV}\int\coth(t-v)D(v)\,dv\right\},
\]
whose modulus is $n(t)$.  Finally the interior envelope and Cauchy's
derivative estimate on small disks inside the strip give the claimed
double-exponential estimates for every fixed derivative.
\end{proof}

\paragraph{Consequences for the Hardy-space contraction estimates and the modified weights.}
Proposition~\ref{prop:TBA-bounded} verifies part~(i) of
Assumption~\ref{ass:H}, since $\phi/2\leq\pi/4<\pi/2$.
Theorem~\ref{thm:TBA-amplitude} verifies part~(ii).
Hence the physical correction $H=S_0*L$ satisfies
Assumption~\ref{ass:H} for every fixed $b\geq1$ and $r_*>0$.
The construction supplies the exact analytic phase required by the
Hardy factorization.  In particular the contraction constant there can
be taken to satisfy
\[
 c_{\phi,R}^2\leq
 \sup_{k\in\mathbb R}
 \frac{\cosh((\pi-\phi)k)+e^{-2R}\cosh(\phi k)}
      {1+\cosh(\pi k)}<1.
\]
Multiplication of $g$ by the strip Blaschke factor associated with $w(x)/(x+c)$
preserves both boundary moduli, so the same bound is uniform in the
insertion position.  No uniformity as $b\to\infty$ or $r_*\downarrow0$
is asserted.  On compact subsets of $1\leq b<\infty$ and $r_*>0$,
the bounds remain uniform.

\section*{Acknowledgments}

The author was partially supported by the National Natural Science Foundation of China under grant number 12271502, and the University of Chinese Academy of Sciences start-up grant 118900M043. The author is grateful to Alexander Its for suggesting the connection between the Lukyanov integrals and Muttalib--Borodin ensemble.

GPT-6 Astra (OpenAI) was used as a writing tool to resolve some technicalities. 
The authors provided the mathematical content and verified all AI-assisted output against independent checks. The authors take full responsibility for the accuracy and integrity of all content, including AI-assisted portions.

\def\cydot{\leavevmode\raise.4ex\hbox{.}}
  \newcommand{\ssf}[1]{\mathsf{#1}}\renewcommand{\Bbb}[1]{\mathbb{#1}}

\end{document}